\documentclass[aps,prb,twocolumn,floatfix,showpacs,article
]{revtex4-1}

\usepackage{graphicx}
\usepackage{amsmath}
\usepackage{amssymb}
\usepackage{epsfig}
\usepackage{color}
\usepackage{amsfonts}
\usepackage{wrapfig}
\usepackage{pstricks}
\usepackage{pst-node}
\usepackage{dblfnote}
\begin{document}
\title{Rare-earth mononitrides}

\author{F.~Natali}
  \affiliation{The MacDiarmid Institute for Advanced Materials and Nanotechnology, School of Chemical and Physical Sciences, Victoria University of Wellington, PO Box 600, Wellington 6140, New Zealand}
\author{S.~Granville} 
  \affiliation{The MacDiarmid Institute for Advanced Materials and Nanotechnology, Industrial Research Ltd., PO Box 31-310, Lower Hutt, New Zealand}
\author{B.~J.~Ruck} 
  \affiliation{The MacDiarmid Institute for Advanced Materials and Nanotechnology, School of Chemical and Physical Sciences, Victoria University of Wellington, PO Box 600, Wellington 6140, New Zealand}
\author{N.~O.~V.~Plank}
  \affiliation{The MacDiarmid Institute for Advanced Materials and Nanotechnology, School of Chemical and Physical Sciences, Victoria University of Wellington, PO Box 600, Wellington 6140, New Zealand}
\author{H.~J.~Trodahl}
  \affiliation{The MacDiarmid Institute for Advanced Materials and Nanotechnology, School of Chemical and Physical Sciences, Victoria University of Wellington, PO Box 600, Wellington 6140, New Zealand}
\author{C.~Meyer}
  \affiliation{Institut N\'{e}el, CNRS/UJF, BP 166, 38042 Grenoble Cedex 9, France}
\author{W.~R.~L.~Lambrecht}
  \affiliation{Department of Physics, Case Western Reserve University, Cleveland, Ohio 44106-7079, USA}

\begin{abstract}

When the rare earth mononitrides (RENs) first burst onto the scientific scene in the middle of last century, there were feverish dreams that their strong magnetic moment would afford a wide range of applications. For decades research was frustrated by poor stoichiometry and the ready reaction of the materials in ambient conditions, and only recently have these impediments finally been  overcome by advances in thin film fabrication with ultra-high vacuum based growth technology. Currently, the field of research into the RENs is growing rapidly, motivated by the materials demands of proposed electronic and spintronic devices.  Both semiconducting and ferromagnetic properties have been established in some of the RENs which thus attract interest for the potential to exploit the spin of charge carriers in semiconductor technologies for both fundamental and applied science.  In this review, we take stock of where progress has occurred within the last decade in both theoretical and experimental fields, and which has led to the point where a proof-of-concept spintronic device based on RENs has already been demonstrated. The article is organized into three major parts. First, we describe the epitaxial growth of REN thin films and their structural properties, with an emphasis on their prospective spintronic applications. Then, we conduct a critical review of the different advanced theoretical calculations utilised to determine both the electronic structure and the origins of the magnetism in these compounds. The rest of the review is devoted to the recent experimental results on optical, electrical and magnetic properties and their relation to current theoretical descriptions. These results are discussed particularly with regard to the controversy about the exact nature of the magnetic state and conduction processes in the RENs.

\end{abstract}

\date{\today}
\maketitle

\section{Introduction}

The rare-earth mononitrides were first investigated in the 1960s, when technological developments overcame the problems faced in separating the chemically similar members of the lanthanide series. That early work was discussed in reviews from the 1960s and 1970s.\cite{hullig78,hullig79} In the following period, from 1980 to 2005, there was a low level of interest, with further reviews appearing in the 1990s and in the early years of this century.\cite{VogtMatten93,ruck09,duanreview} In the past few years the literature on the subject has grown rapidly, based on a combination of breakthroughs in theoretical developments and the facility to grow stoichiometric epitaxial films. The rare-earth nitrides (RENs) show promise in applications as diverse as spintronics, infrared (IR) detectors and as contacts to III-V compounds, with now about a dozen laboratories worldwide reporting the growth and study of REN thin films.

The rare-earths, with atomic numbers from 57 (La) to 71 (Lu), comprise the elements across which the $4f$ orbitals are filled. They have atomic configurations $[Xe] 6s^24f^n$, with $n$ varying from 1 for Ce (0 for La) to 14 for Lu, and certain of the series have an additional 5$d$ electron (La, Ce, Gd, Lu). Their most common ionic charge state is 3+, with the $4f$ levels spanning the Fermi energy. They are the only stable elements with more than marginally filled $f$-shell electronic orbitals, and as a consequence they are the elements with the largest spin and orbital moments. In ordered solids they contribute to the most strongly ferromagnetic materials, a contribution that has ensured their utility in technologies that require strong permanent magnets. Despite their name they are by no means rare, with the exception of promethium, which has no stable nuclear isotope. They are found in the Earth's crust at concentrations exceeding that of Ag, Cd and Hg, and similar to Ge and As.\cite{usgs}

The magnetic states of most rare-earth monopnictides (RE-V) have been known for some decades, after studies undertaken already in the 1960s. The heavier pnictides were found to be antiferromagnetic, but in contrast the nitrides are almost all ferromagnetic; the magnetic properties are reported in previous reviews.\cite{hullig78,hullig79,VogtMatten93} The RE-V adopt magnetic order only at cryogenic temperatures, with the highest Curie temperature ($T_C$), for GdN, of 70 K.\cite{busch,schumacher65}
The current interest in the potential of spintronic devices has raised the level of urgency in the exploration of intrinsic ferromagnetic semiconductors, of which the RENs offer a rich set of examples.  Yet their potential has been poorly explored until recently, despite EuO having become a well-known intrinsic magnetic semiconductor.\cite{EuO} Unlike the dilute magnetic semiconductors (DMS) the RENs do not rely on the presence of foreign ions, nor on a huge hole concentration that prevents independent doping control. Thus in principle this permits controlled doping independent of the ferromagnetism opening new possibilities for spintronic devices and fundamental spin-transport research. However even within the DMS scenario the RENs hold some promise, with reports that Gd leads to room-temperature ferromagnetism with an unexpectedly large moment when it is introduced as a dilute impurity in GaN.\cite{GaGdN} A thorough understanding of the corresponding GdN is a prerequisite to understand this proposed DMS system.

In contrast with the magnetic properties, the transport properties and electronic band structures of the RENs were until recently much less certain. Even for ostensibly identical compounds one can find reports of the conductivity ranging from insulating to metallic. The source of the uncertainty lies in two issues that impact strongly on their achievable stoichiometry: a propensity for rapid oxidation when exposed to air and for the formation of N vacancies ($V_N$). The former requires that any thin film must be protected by a passivation capping layer, while the latter is related to the small formation energy of $V_N$ that ensures these will be present at the level of at least 1\% in any film grown substantially above ambient temperature.

Among the earliest of the recent theoretical discussions there was a prediction that the RENs would display a series of contrasting conductive states, ranging from semiconductors to semimetals, and including half metals.\cite {aerts} This range of interrelated magnetic and conducting properties is then an obvious testing ground for prototypical spintronics structures. Following that prediction there have been a number of laboratories that have initiated programmes to grow polycrystalline and epitaxial films, so there is the hope that some spintronic structures can ultimately be explored. That hope is still some way off, waiting for improved techniques for the growth and capping of thin films of these severely reactive materials, although already now there is a report of a GdN-based spin filter.\cite{GdNspinfilter}

Against this background there has been for some time a theoretical interest in the rare-earth monopnictides. The localized atomic-like properties of the open-shell $4f$ electrons defy the standard density functional band structure theory approach. Because the RENs form a family of materials with the same simple
rocksalt structure, they have become a useful testing ground for new theoretical developments dealing with the strongly correlated $4f$ electrons. That work is impeded by the paucity of reliable experimental data, especially about the electronic state of the nitrides; there is relatively little too inform the theoretical work. The recent rapid expansion of experimental studies, especially on well-ordered films, has begun to provide data with which to tune the treatment of strong correlation.

In this review
we will focus on the recent developments on the rare-earth (RE) nitrides. However, occasionally, we will also include closely related materials,
such as other RE-V,
as well as some RE chalcogenides and oxides, such as
EuO. For example, because the RE shell is half-filled in both EuO and GdN,
it is of interest to compare them.
We should mention that the theoretical properties of RE-V were reviewed not so long ago
by Duan et al.~\cite{duanreview}
Nonetheless, the present review, focused on RENs will be
complementary, particularly because of their contrasting magnetic behaviour the recent experimental studies based on advances in epitaxial thin-film growth.

We have prepared the present review with an eye to the potential exploitation of this class of material in various applications. Their potential depends on specific aspects of their crystal structures and electronic properties, and although these will be discussed in detail in later sections, in the following section we give a brief description of the most obvious applications. Specific attention will be paid to issues such as their epitaxial compatibility. Following that we will cover recent advances of, in Section \ref{growth}, epitaxial thin-film growth and in Section \ref{theory} theoretical descriptions. The magnetic and electronic/magnetoelectronic advances will be discussed in Sections \ref{magnetics} and \ref{optics}, respectively, and Section \ref{summary} is a summary.

\section{ The basis of potential exploitation}

The RENs form in the face-centered (FCC) cubic NaCl structure with lattice constants ranging from 5.305 \AA~for LaN to 4.76 \AA~for LuN, in total a 5\% difference across the series and less than 0.5\% between nitrides of neighbouring atomic species. There is clearly potential for epitaxial growth of custom-designed heterostructures, including superlattices, and even for controlled strains to be introduced. In later sections we will discuss the complementary electronic properties and strongly contrasting magnetic behaviours of the RENs, traits which immediately suggest them as the basis for a variety of spin-dependent devices. It is essential for any such devices that techniques are developed for the growth of well-ordered epitaxial structures and that a thorough understanding is reached of the electronic band structures and magnetic behaviours of the RENs, both in bulk and in thin epitaxial layer forms.

The strong exchange interaction results in a significant spin splitting of a few hundred meV in both band edges, with the majority spin having the lower energy in the conduction band and the higher energy in the valence band. Thus carriers at both edges, electrons and holes, are in majority-spin bands; the minority spin edges are unoccupied at ambient temperature. Any device, such as a diode, transistor or filter that requires doping or accumulating carriers into the band edges will involve transport of carriers with only majority spin state. Based on this strong exchange splitting, a REN-based spin-filter Josephson junction has been recently demonstrated.\cite{GdNspinfilter} A thin layer (5 nm) of GdN acting as a spin-dependent tunneling barrier is placed between two NbN superconductor contacts, with the resulting spin-filter efficiency estimated to be about 75\% at 4.2 K.  
With a strong magnetoresistance over a broad field range in its ferromagnetic state, as well as a small coercive field, GdN is an appropriate material to explore for use in magnetic field sensors. The recent enhancement of the Curie temperature up to 200~K in N-deficient GdN films would allow operation of the sensor well above the boiling point of liquid nitrogen.\cite{plank10,senapati11a}

With magnetic states that vary strongly across the series and coercive fields depending strongly on the growth conditions, the RENs are of specific interest for non-volatile magnetic memory elements such as MRAMs. In particular, SmN is the only known near-zero-moment ferromagnetic semiconductor, with an enormous coercive field,~\cite{Mey2008} and, combined with GdN which has a coercive field some three orders of magnitude smaller, they may form an ideal hard- and soft-ferromagnetic pair with potential in memory elements.

The ongoing effort for the epitaxial growth of RENs is further driven by the potential to integrate REN-based devices with group-III nitrides to develop new functionalities combining both families. The narrow band gaps of the RENs, with optical absorption edges lying near 1 eV and absolute gaps on the order of one half of that, are interesting for IR detectors. Thus the properties of the RENs are complementary with those of the wide band-gap group-III nitrides, and a heterojunction involving the two semiconductors might have very attractive properties for multi-wavelength photonic devices. In addition, as spin-polarised carrier injection cannot be accomplished efficiently from metals into a semiconductor, a GdN layer could be regarded as a spin injector in GaN-based transistors or diodes.

Although the Si-RE reaction is so rapid that it prevents the epitaxial growth of RENs directly onto Si, they can be grown as polycrystalline films at ambient
temperature, which may provide useful injection of spin selected electrons. The use of AlN and GaN buffer layers will be seen to permit epitaxial REN/Si integration; other buffers have yet to be explored.~\cite{natali10}

Very recently there has been an \textit{ab initio}-based suggestion that GdN will have a ferroelectric ground state under 3\% compressive in-plane strain, and a similar result can be expected for the other RENs.\cite{liu11} The lattice constant variation across the REN series permits a thorough investigation of the possibility. If the prediction is verified they would form interesting electromagnetic multiferroics.

Thus among the most interesting aspects of the series is their epitaxial compatibility coupled with their contrasting magnetic and complimentary electronic properties, showing promise for a wide range of possible spintronic and electronic structures. It is then of special interest to provide a much fuller description of the magnetic and electronic properties of the entire series. To date it is only GdN that has been subjected to very thorough experimental investigation, as will be realised very quickly on reading this review.

Finally, there are reports of catalytic\cite{imamura06} and large magneto-caloric\cite{yamamoto04,plaza04,nakagawa04a,nakagawa06a,nakagawa06b,nishio06,hirayama08} effects in the RENs, suggesting them as promising candidates for magnetic refrigeration. Thin films are unlikely to contribute to these technologies.

\section{Crystal structure and epitaxial growth of rare-earth nitrides}
\label{growth}

One of the major hurdles in growing REN thin films, either epitaxial or in polycrystalline form, is their propensity to form $V_N$ and to decompose in air into RE oxide/hydroxides.~\cite{dismukes70,Eick1956} This is the reason for some of the long standing controversies concerning their electronic structure, transport properties and magnetic properties. The growth of RENs in both thin film and bulk form dates back more than seventy years and quite a number of processes have been tested, but none was found fully satisfactory at that time.\cite{hullig79} It is only recently that high quality epitaxial thin films have been achieved, mainly thanks to the continuous development of film growth techniques, such as ultrahigh vacuum (UHV)-based methods. In this section we describe the recent developments in the field of epitaxial growth of REN thin films. The growth of bulk materials will not be treated in this review. Polycrystalline thin films will be mentioned briefly, primarily because their lower growth temperature offers some control over the formation of $V_N$.\cite{ruck09}

\subsection{Crystal structure}

\begin{figure}
\includegraphics[width=\columnwidth]{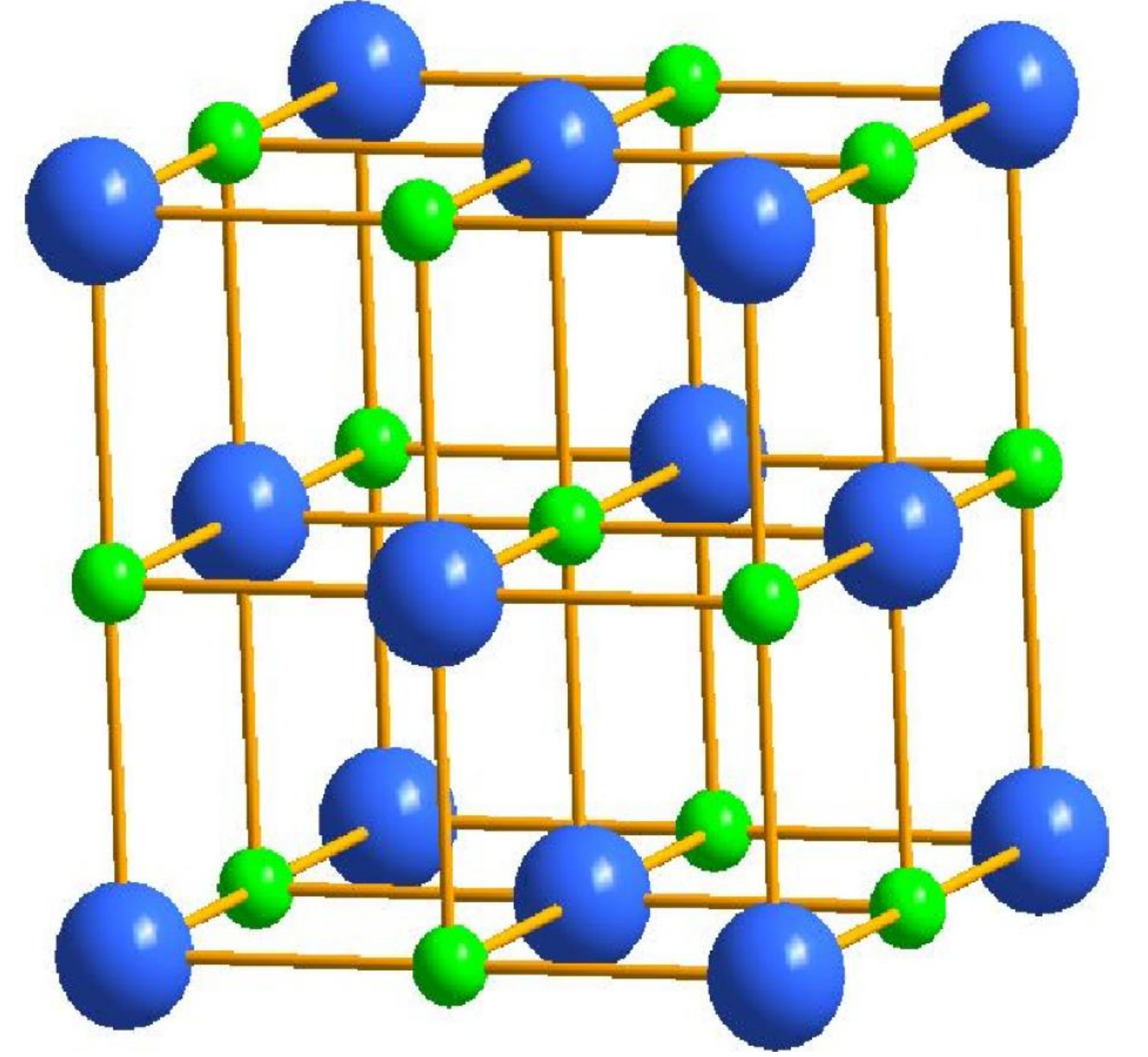}

\caption{The common rocksalt crystal structure of the RENs.  The large spheres represent the RE cations while the small spheres are the N anions.
\label{NaCl}}
\end{figure}

The RENs adopt a FCC NaCl structure (space group \textit{Fm-3m} (225)) as shown in Figure ~\ref{NaCl} and in all NaCl-type RENs with the exception of CeN (tetravalent) the cation is trivalent. Each RE atom is coordinated by 6 nitrogen atoms and, conversely, each nitrogen atom is coordinated by 6 RE atoms. The large difference in electronegativity between nitrogen (3.0) and RE (1.1 to 1.5) leads to a strong affinity and a predominantly ionic character (more than 50\%) of the RE-N bonds.\cite{Marchand98} High quality single crystals have proven to be exceedingly difficult to prepare and as mentioned above there remains much controversy about various experimental results, and many fundamental material parameters have still not been reliably measured. 

Table~\ref{tabgaps} shows some of the crystal lattice parameters for the RENs available in the literature, which range from about 5.305 \AA~ to 4.76 \AA~ for LaN to LuN, respectively. The lattice constants of the RENs decrease with increasing $4f$ occupancy as expected from the decrease of cation sizes across the series. The theoretical lattice constants for the entire REN series determined from first-principles calculations are shown in Table~\ref{tabgaps} and differ from most of the experimental lattice constants by about 1-2\%.\cite{Larson07} The bulk moduli calculated following the same computational approaches are also displayed in Table~\ref{tabgaps} with in this case larger discrepancy among the calculation methods. However, the calculated values are still comparable to experimental values determined for the $5f$ nitrides UN (203GPa) and ThN (175GPa),\cite{gerward85} and for the group III-nitrides AlN (209GPa), GaN (183GPa) and InN (133GPa).\cite{bechstedt02} No experimental data for the thermal expansion, thermal conductivity or elastic stiffness coefficient, including Young's modulus, have yet been obtained, with the exception of CeN. For the latter, the bulk modulus of 153 GPa has been determined by X-ray diffraction\cite{jakobsen02} while the hardness and elastic modulus of epitaxial CeN(001) films were determined from nanoindentation measurements to be $15.0\pm 0.9$~GPa and $330\pm16$~GPa, respectively.\cite{lee03} A full computational study of the elastic properties of CeN has been performed by Kanchana et al.~\cite{kanchana11} and the results are in close agreement with the experimental values. More recently, several groups have theoretically explored the elastic properties and hardnesses of PrN,\cite{kocak10} TbN~\cite{ciftci12} and HoN,\cite{bhajanker12} and Yang et al.~have studied the entire REN series\cite{yang10} and compared their results with the literature. Furthermore, linear dielectric constants of the RENs derived from the dielectric theory of chemical bonds for solids are also displayed in Table~\ref{tabgaps}.\cite{xue00}

\begin{table}
\caption{Experimental lattice constants, calculated lattice constants, bulk moduli ($B$) and dielectric constants ($\epsilon$) of rare-earth nitrides.}
\begin{ruledtabular}
\begin{tabular}{lc|ccc}
REN & Expt.&\multicolumn{3}{c}{Calc.}\\
    & $a(\AA)$ & $a(A)$\cite{Larson07} & $B(GPa)$\cite{Larson07,yang10,kocak10,bhajanker12,ciftci12} &$\epsilon$\cite{xue00} \\
LaN & 5.305\cite{olcese}&5.38 &130, 122&7.07\\
CeN &5.022\cite{olcese}&4.90 &210, 121&	7.01 \\
PrN & 5.135\cite{ettmayer79}&5.29 &140, 121, 129.14&6.96\\
NdN& 5.132\cite{olcese}&5.24 &140, 197&6.90 \\
PmN & &5.19 &150, 114&6.85 \\
SmN &5.035\cite{hullig79}&5.10 &180, 127&6.79 \\
EuN &5.017\cite{jacobs78}&5.14 &110, 114&6.74\\
GdN &4.974\cite{dxli97}&5.08 &150, 111&6.68\\
TbN &4.922\cite{hullig79}&5.05 &150, 241, 155.53&6.63\\
DyN &4.895\cite{hullig79}&5.03 &160, 121&6.57\\
HoN &4.865\cite{hullig79}&4.98 &170, 138, 137.9&6.51 \\
ErN &4.842\cite{olcese}&5.00 &160, 56&6.46 \\
TmN &4.80\cite{hullig79}&4.90 &190, 138&6.40\\
YbN &4.781\cite{degiorgi90}&4.79 &190, 136&6.35\\
LuN &4.76\cite{suehiro04}&4.87 &170, 183&6.29\\
\end{tabular}
\label{tabgaps}
\end{ruledtabular}
\end{table}

\subsection{Epitaxial growth of REN}

The progress in studies of the RENs over the last few years was achieved using UHV-based methods such as molecular beam epitaxy (MBE),\cite{gerlach07,scarpulla09,natali10,Ruck2011,richter11,natali13,kent12} pulsed-laser deposition (PLD),\cite{shimomoto09,ludbrook09,natali13,Ruck2011} and dc/rf magnetron sputtering.\cite{osgood98,yoshitomi11} These are the techniques of choice for deposition of RE-based materials because the high-vacuum and inert-gas environment helps to ensure material purity and interface quality and these methods are subsequently the most common used in this field. Among these UHV-based techniques, the recent advances in the growth of GdN and EuN by MBE tend to show that this is probably the best growth method to achieve rapid improvement in the quality of REN thin films.\cite{scarpulla09,natali10,richter11,natali13} It is worth mentioning that the purity of the as-received RE charges or targets is far from the one commonly used for the growth of conventional III-V semiconductors; typically the purity of the RE source is about 99.99\% by weight with the main impurities being oxygen, carbon, nitrogen, calcium and iron. 

While solid sources are commonly used for the RE elements, a wide range of options are available for nitrogen sources/precursors. The most straightforward is to use pure N$_2$ gas by taking advantage of the catalytic breakdown effect of the RE atoms on the molecular nitrogen. So far epitaxial growths of GdN\cite{natali10,natali13} and SmN\cite{natali13} as well as polycrystalline films of DyN, ErN, LuN\cite{Granville09} have been demonstrated. The growth of EuN must be achieved under the presence of radical nitrogen, for example using low energy nitrogen ions from a Kaufmann ion source or N$_2$ plasma.\cite{natali13,richter11,Ruck2011,shimomoto09} The other nitrogen precursors used for the growth of REN are those commonly used for the growth of GaN: ammonia (NH$_3$)\cite{scarpulla09}, which is decomposed on the surface of the substrate by pyrolysis, and nitrogen plasma obtained either by radio-frequency or electron cyclotron resonance.\cite{scarpulla09,shimomoto09,Ruck2011,ludbrook09} Epitaxial growth of the RENs is still at too early a stage to conclude on the best nitrogen source in terms of material quality.

The growth conditions of CeN, GdN and SmN under NH$_3$, pure N$_2$ and low energy nitrogen ions are in fact close to those currently used for the III-V nitrides. They are grown with an excess of nitrogen species with respect to the RE flux to avoid $V_N$ formation and/or metallic RE clusters. Nitrogen ions to Ce ratio of 15\cite{lee03} and pure N$_2$\cite{natali10} or NH$_3$\cite{scarpulla09} to Gd/Sm
ratios larger than 100 have been reported. Using these conditions, stoichiometric films are achieved and the material properties are of high quality. Epitaxial growth of GdN has also been performed using an N$_2$ plasma cell with a slight excess of Gd by MBE\cite{scarpulla09} and in a large excess of N$_2$ by PLD.\cite{natali13,ludbrook09}  The growth of EuN is somewhat intriguing as epitaxial films have been obtained when the growth occurs in an adsorption-controlled growth regime, where the Eu flux is set to be much higher than the N$_2$ flux and a high substrate temperature ($\sim$800$^{\circ}$C) is maintained to re-evaporate any excess of Eu.\cite{natali13} EuN has also been grown by PLD but the authors do not report on the RE:N ratio while they point out that high growth temperatures, up to 860$^{\circ}$C, are required to achieve epitaxial films.\cite{shimomoto09} An optimised growth temperature window around 600$\pm$50$^{\circ}$C was found for SmN,\cite{natali13} again in a range of growth temperatures and pressures where any excess of Sm can be re-evaporated from the substrate. 

While it seems there is a consensus about the growth temperature for EuN and SmN thin films, the situation is far from clear about the one to use for GdN. Gerlach et al.~\cite{gerlach07}  reported a growth temperature of 750$^{\circ}$C, comparable with those used later on by Natali et al.~\cite{natali10} for the growth of GdN under pure N$_2$ (650-750$^{\circ}$C) or by Ludbrook et al.~(700-850$^{\circ}$C) by PLD using a N$_2$ plasma source.\cite{ludbrook09} No clear significant temperature dependence of the structural quality of the films was found between 700 and 850$^{\circ}$C. Temperatures much lower, such as 500$^{\circ}$C, have been used for the growth of epitaxial GdN films by reactive radio frequency magnetron sputtering.\cite{yoshitomi11} GdN layers have also been grown at 450$^{\circ}$C by MBE using an N$_2$ plasma cell and NH$_3$ as the nitrogen precursor.\cite{scarpulla09} The authors mentioned that deposition at higher temperatures did not yield GdN films. It is worth mentioning that Gd has a vapor pressure of 1.3 10$^{-4}$ mbar at 1175$^{\circ}$C, and, therefore, the growth process does not involve the re-evaporation of the excess Gd. Only one paper reports on the epitaxial growth of CeN, for which the growth temperature was 700$^{\circ}$C.\cite{lee03}

Although metal-organic chemical vapor deposition (MOCVD) is the technique of choice for the growth of group-III nitrides, for the RENs the weak point of the technique has been for a long time related to the absence of an efficient RE precursor. Only recently have GdN and DyN polycrystalline films been achieved using guanidinato-complexes of Gd and Dy as precursors.\cite{milanov09,thiede11} Physical vapor deposition\cite{Leuenberger05,shalaan06} and more recently chemical vapor deposition\cite{brewer10} as well as plasma-enhanced atomic layer deposition\cite{fang12} have been used to grow polycrystalline GdN films.

\subsection{Substrates and capping layers}

One of the major difficulties which has hindered REN epitaxial growth is the lack of native substrates. A plethora of suitable material that is more or less matched with the RENs has been employed. Below we give a brief description of each of these substrates.

\begin{figure}
\includegraphics[width=6cm]{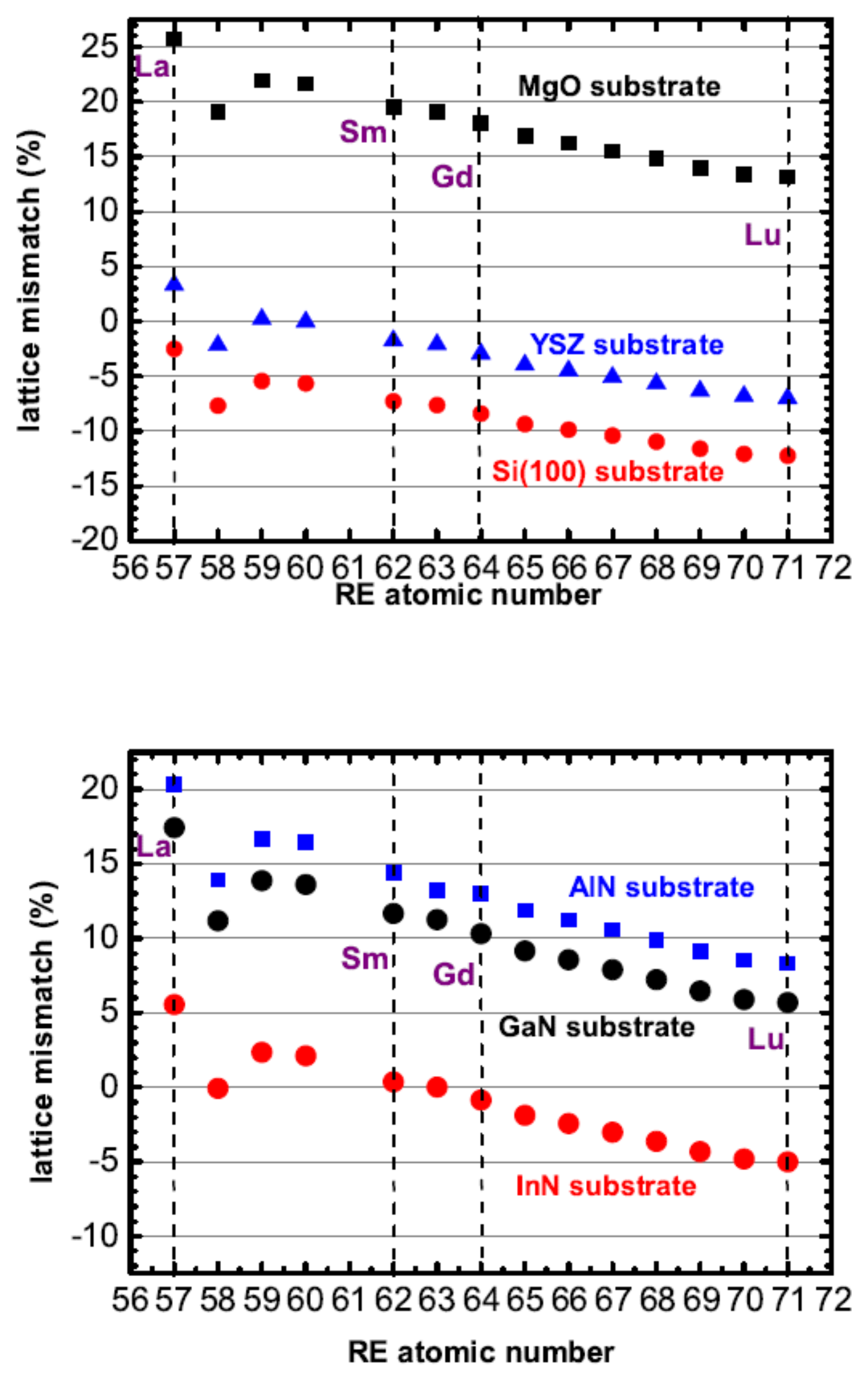}
\caption{Calculated lattice mismatch between RENs and substrate materials. Top: 	(001) REN and MgO(001), YSZ (001) and Si(001). Bottom: (111) REN c-plane (0001) wurtzite AlN, GaN and InN lattices as a function the RE atomic number. From Reference \onlinecite{natali13}.
\label{X.2.}}
\end{figure}

Most research in the past has selected (100) oriented substrate surfaces for REN epitaxy. This is supported by the fact that the RENs adopt a FCC (NaCl) structure. Historically the first epitaxial growths of RENs, CeN by T.-Y. Lee et al.~\cite{lee03} in 2003 and then GdN by Gerlach et al.~\cite{gerlach07} in 2007, have been performed on MgO(100) substrates with a lattice mismatch of about +18.7\% and +19.2\% for GdN and CeN, respectively (Figure \ref{X.2.}). The film/substrate epitaxial relationships were demonstrated to be (001)REN$\parallel$(001)MgO and [100]REN$\parallel$[100]MgO. In both cases the films were shown to be of high crystalline quality, but in the case of GdN little was reported on the magnetic or transport properties and it has been observed that films thicker than 60~nm have very rough surfaces. The most closely lattice-matched substrate for REN epitaxy would be YSZ (100) with a lattice parameter of 5.125~\r{A}, nearly matching the lighter RENs as shown in Figure \ref{X.2.}. Epitaxial growth on YSZ substrates of GdN,\cite{natali13,ludbrook09} SmN,\cite{natali13} and EuN\cite{natali13} using PLD and then later EuN by MBE\cite{richter11} have been reported. It is worth mentioning that an oxide layer, RE$_2$O$_3$, is formed at the interface between the substrate and the REN films, likely due to the strong affinity of the RE for oxygen and the mobility of oxygen in YSZ.\cite{natali13} For economic and technological reasons it would be a real advantage to demonstrate the growth of REN films on silicon substrates.  The REN lattice constants are within about 10\% of that of Si (Figure \ref{X.2.}), but, as mentioned previously, silicide formation at the Si/REN interface is still a major issue to overcome and has so far prevented epitaxial growth.

The possibility to take advantage of the hexagonal symmetry of the \{111\} plane of the REN rocksalt structure to investigate growth on c-plane (0001) wurtzite GaN surfaces has been proposed by Scarpulla et al.,\cite{scarpulla09} and then later on AlN surfaces.\cite{natali10,yoshitomi11} Earlier work by Shimomoto et al.~reported the growth of EuN as a buffer layer on MgO(111) and Al$_2$O$_3$ substrates for the subsequent growth of InN.\cite{shimomoto09} Figure \ref{X.2.} shows the lattice mismatch between (111) REN lattice and c-plane (0001) wurtzite AlN, GaN and InN lattices as a function of the RE atomic number.\cite{natali13} This graph shows that RENs can be grown nearly lattice-matched on an InN substrate, but its lack of commercial availability make GaN and AlN more attractive for the moment. The 7.5-15\% lattice mismatch between the RENs and either AlN or GaN, though relatively severe, is comparable to that in heteroepitaxial systems.\cite{liu99} The growth of the REN (111) plane on a (0001) surface resulted in two rotational variants of the grains due to the lower 3-fold symmetry of GdN compared to the 6-fold symmetry of the group-III nitride surface.\cite{scarpulla09} The in-plane epitaxial relationship for the two rotational variants are [1-10]GdN$\parallel$[10-10]GaN and [-1010]GaN, in addition to (111)GdN$\parallel$(0001)GaN.\cite{scarpulla09} The quality of the films of Ref.~\cite{scarpulla09} was high enough to address the properties of the RENs with relative confidence. To date most of the laboratories have selected group-III nitride substrates for REN epitaxial growth. This choice is supported by the fact that these substrates are widely available with a good crystal quality, due to their transparent nature, and ease of handling and pre-growth cleaning. Their use is also driven by the possibility to develop new functionalities combining nitride semiconductors and REN materials. Additionally, the possibility to grow GdN on a 100-nm thick AlN buffer layer on top of a silicon substrate has been demonstrated\cite{natali10} and may pave the way for the integration of RENs into mainstream silicon technology.

Due to their decomposition in air, REN films must be passivated with an effective capping layer to avoid reaction with the atmosphere. A series of polycrystalline or amorphous capping layers have been tried successfully in the past, including metallic layers such as W,\cite{Leuenberger05} Cr,\cite{Leuenberger05} Cu,\cite{thiede11} TaN\cite{fang12} and NbN\cite{senapati11a,GdNspinfilter} and insulator films such as YSZ,\cite{ludbrook09} GaN,\cite{gerlach07,scarpulla09,trodahl07,Granville06} AlN\cite{lee03,natali10,yoshitomi11} and MgF$_2$.\cite{Granville06}  Polycrystalline AlN and GaN are the most commonly used capping layers which can be attributed to their transparency allowing optical measurements, their ease of growth, and good chemical stability over time.

\subsection{Structural Properties}

In spite of efforts to improve crystal quality, only little is known about the structural defects in the RENs. This is mainly due to their instability in air which makes high resolution transmission electron microscopy challenging. However, for some of the RENs, such as GdN, the oxidation rate is sufficiently slow to allow cross-section scanning electron microscopy.\cite{mckenzie06,scarpulla09, yoshitomi11} As most of the RENs are prepared by heteroepitaxy on III-V nitride substrates, the most prevalent structural defects should be threading dislocations. In addition, a misfit type of dislocations lying at the interface to relieve the misfit strain in the large lattice mismatched growths is expected whatever the growth orientation, (001) or (111). Reflection high energy electron diffraction (RHEED) has been used for measuring the plastic relaxation during the growth of GdN on AlN.\cite{natali10} It has been observed that after $\sim$2.5MLs the lattice constant of GdN starts to increase, while the RHEED pattern remains streaky, and a fully relaxed GdN layer is obtained after only 6MLs. RHEED has also been useful to monitor the presence of RE$_2$O$_3$ at the interface during growth on a YSZ substrate.\cite{ludbrook09,natali13} Interestingly it has also signalled the presence of a (2x2) surface reconstruction after the growth of (001) EuN,\cite{richter11} while no surface reconstruction was observed for (111)-oriented films. 
 In the case of growth onto group III-nitride surfaces, planar defects in the form of twin boundaries rotated by 60 degrees with respect to each other are present in the films.\cite{scarpulla09} These are separated by grain boundaries and/or antiphase domains, meaning that single-crystalline films will only be possible when one variant can be suppressed. 

The traditional figure of merit used to assess the quality of the films is the X-ray diffraction (XRD) coherence length determined from $\theta$-2$\theta$ and rocking curve scans. Full-width half maximums (FWHMs) for the symmetric (002) rocking curves of 2.5$^{\circ}$ and 1.4$^{\circ}$  have been reported for (001)-oriented GdN grown on YSZ(001)\cite{ludbrook09} and MgO(001)\cite{gerlach07} respectively, while a larger FWHM, 2.16$^{\circ}$, for the symmetric (002) rocking curve of (001)-oriented CeN grown on MgO (001)\cite{lee03} has been measured. Narrower linewidths are found for films grown in the (111) direction. Scarpulla et al.~\cite{scarpulla09} reported values with a FWHM for the symmetric (111) peak of 0.251$^{\circ}$ and 0.321$^{\circ}$ for GdN grown on GaN using an N$_2$ plasma or with NH$_3$ as a nitrogen source. Such values lead to in-plane X-ray coherence lengths up to 100~nm. While the structural quality of these films is higher than for any other reported films, further optimisation of the growth parameters is still required to reduce the degree of twinning and mosaic spread.

\section{Recent theoretical advances}\label{theory}

 The theoretical discussion starts with a review
of some basic textbook notions about open-shell $4f$ systems, such as Hund's rules, in
Sec. \ref{basic}.
Next, we will briefly review the shortcomings of standard density
functional theory for $4f$ systems. In Sec. \ref{early} we will review the
early band structure work
on RE-V, which used the simplest
possible approach, namely, treating the $4f$ electrons as a partially filled
core state. Then we will turn to each of the main new theoretical
approaches for dealing with the strongly correlated $4f$ electrons:
the self-interaction correction approach (Sec. \ref{sic}), the LSDA+$U$ method (Sec. \ref{lda+u}),
the hybrid functionals (Sec. \ref{hybrid}), GW theory (Sec. \ref{GW}), and the dynamical mean field theory (DMFT) approach (Sec. \ref{dmft}).
All of this will mainly be focused on the electronic structure.
One of the main questions to be answered is whether the RENs
are semimetals or semiconductors.

A large fraction of the existing theory work is devoted to understanding the
magnetic properties, in particular why GdN is ferromagnetic
while the other pnictides are antiferromagnetic. On the other RENs, there are far fewer detailed studies of the magnetic properties
from a band structure point of view, although there is a large body
of older work dealing with the complex magnetic ordering phenomena in these
materials. Besides the magnetism there is also some interesting theory work on the optical and
lattice dynamical properties of the RENs.
Therefore, we discuss the magnetism (Sec. \ref{magnetics}), electronic and optical properties (Sec. \ref{optics})
and vibrational properties (Sec. \ref{Raman}) in separate sections.

\subsection{Basic properties of $4f$ electrons.}
\label{basic}
The atomic-like properties of $4f$-electrons arise from their localized
nature. The high angular momentum $l=3$ of an $f$ electron results in
a strong centrifugal barrier $l(l+1)/r^2$ which keeps the electrons
separated from the other valence shells, like the $5d$
and $6s$ electrons. Because the $4f$ radial wave function has no node
it is rather localized. However, it increases as $r^3$ near the origin and is
thus peaked away from the nucleus,
but at the same time at a much smaller radius than the outer $6s$ and $5d$ electrons
which need to stay orthogonal to the lower $s$  and $d$ shells.
In contrast, when we consider the actinides, the $5f$ shell is more spread out and
behaves less atomic-like. Of course, how localized or band-like
the electrons are depends on the crystal structure and on the separation
of the atoms. Hence, there is some interest in early lanthanides, whether
under pressure they can convert from an atomic-like to a band-like behavior of the 4$f$ electrons.

The atomic localized character
implies a strong Coulomb interaction between the electrons as we gradually
fill the $4f$ shell. The basic atomic physics is thus dominated by the
electron-electron interaction. In atomic physics, the basic question
to answer is then, what are the quantum numbers that determine the ground
state of the multi-electron system consisting of an $f^N$ system?

This question has been studied long ago in terms of the different coupling
schemes: ${\bf L}\cdot{\bf S}$ coupling, ${\bf J}\cdot{\bf J}$ coupling,
etc., and is for example summarized in the well-known book by
Condon and Shortley.\cite{CondonShortley} If the spin-orbit coupling is smaller than the
electron-electron interaction, one expects the  ${\bf L}\cdot{\bf S}$
coupling scheme to be valid. The many-electron wave function is a linear
combination of Slater determinants with a given total orbital and
spin angular momentum ${\bf L}$ and ${\bf S}$ and these then couple to give a
total angular momentum ${\bf J}$. In that case, the good quantum numbers
are $L,S,J,M_J$. Here $L,S,J$ define the eigenvalues of $L(L+1)\hbar^2$,
etc., of the angular momentum operators $\hat{L^2}$, $\hat{S^2}$, $\hat{J^2}$. These
determine the magnitude of the corresponding angular momenta. The
$M_J$ determine the eigenvalues of the total angular momentum along
one chosen quantization axis, $J_z$. As is well known, quantum mechanics
allows us only to measure the magnitude and one component of the angular
momentum, but not the various components simultaneously.
Determining all possible combinations of these quantum numbers for
a given number of $f$ electrons determines the so-called {\sl  multiplet splitting terms}.

The above theory for free atoms results in the well-known Hund's rules,
which determine which of the angular momentum quantum numbers result in the lowest energy. These rules are: (1) first the total spin $S$ should be maximized,
(2) then, for a given maximal spin, the total orbital angular momentum
$L$ should be maximized, (3) if the shell is less than half filled, then the
lowest energy is found for  $J=|L-S|$, while for a more than half-filled
shell, $J=L+S$. In other words, the third rule results from the
spin-orbit coupling and says that the orbital and spin momenta oppose each other
for less than half-filling and are in the same direction for more than
half-filling.   As an example, take the Eu$^{3+}$ ion with 6 $f$ electrons.
The total spin will be $S=3$ if the spins are all parallel. But
since by Pauli's principle we cannot have more than one electron with
the same orbital and spin quantum numbers $m_l$ and $m_s$ and we need
to keep the angular momenta opposite to the spin,
we must have a total angular momentum of $M_L=\sum_{-3}^2 m_l=-3$. So, the
maximum orbital angular momentum in this case is $L=3$. The total
$J=L-S=0$ in this case. We denote this ground state multiplet term
by $^{2S+1}L_J$ or in this case: $^7F_0$. As usual in the
spectroscopic notation, one here replaces $L=0,1,2,3\dots$
by $S,P,D,F\dots$.  On the other hand, a Gd$^{3+}$
ion or a Eu$^{2+}$ ion would have 7 $f$ electrons (half-filled shell) and
lead to a $^8S_{7/2}$ ground state.

Now, when we place the ion in a solid environment, we need to worry
about how the symmetry breaking of the surrounding ions will split
these atomic levels. While for $d$ electrons, this splitting is strong
and leads to so-called quenching of the orbital angular momentum, this
is not the case for $4f$ electrons. As mentioned earlier, they are shielded from the surrounding ions and stay atomic-like.
The theory of how the crystal field splits the multiplet terms was
largely worked out by Racah\cite{Racah49} in terms of group theory.

This gives a very brief summary of the basic atomic concepts of $f$ electrons
in free atoms and in isolated ions as would occur for example for a RE impurity
in an ionic solid. These atomic multiplet splittings largely determine the optical and paramagnetic properties of RE ions.
While optical transitions between $f$ electrons are strictly speaking
dipole forbidden in the free atom, the small admixture with neighboring ligand
orbitals and the resulting symmetry breaking for these ions in a solid
environment allows weak optical transitions between the ground state
and the various excited state multiplets. These optical transitions
result in sharp luminescence lines  largely unaffected by the host and
form the basis of many solid-state lasers, such as Nd:YAG lasers etc.
They are also of great interest for RE impurities in semiconductors
and for optical amplifiers. Because wide band gap semiconductors like GaN have
an advantage for the excitation of the RE optical transitions, it has spawned
a good deal of work on RE-doped GaN.\cite{steckl02}

As far as paramagnetic centers, the magnetic dipole moment of an ion is
determined by the expectation value of ${\bf L}+g_0{\bf S}$ in the particular
ground state $|LSJM_J\rangle$ resulting from Hund's rules. As is well known
from textbooks, this is determined by the Wigner-Eckardt theorem and
leads to the Land\'e $g$ factor.

\subsection{Early band structure work.}\label{early}

The earliest approach to RE-V from a band structure point of view is to treat the $4f$ electrons as core electrons. In a spin-density functional method, this leads to a different potential for spin-up and spin-down valence electrons. To the best of our knowledge, the first applications of this approach
to RE-V were by Hasegawa and Yanase.\cite{hasegawa77,hasegawa80}
They successfully predicted already the essential elements of the band structure.
The valence bands are formed from the group V-$p$ orbitals, and the conduction bands by the RE-$d$ bands. In all pnictides, except for the nitrides, the RE-$d$ band
dips below the valence band maximum (VBM), which  itself is located at $\Gamma$,
at the X-point in the Brillouin zone. These materials are thus semimetals,
except that the nitrides were found to have almost zero gap, so possibly
semiconductors, in view of the usual underestimation of the band gap
by the local density approximation (LDA). The drawback of this method is that it does not
show the $4f$ states as bands at all. Their location for the occupied spin
is in fact above the N-$2s$ bands but they are nonetheless not included in the band structure picture. The location of the empty spin-down states is not calculated. It is as if the $4f$ electrons live in a different world. 

The same approach was also used by Petukhov et al.~\cite{Petukhov96} in
a rather complete survey of the pnictides, including the nitrides.
Detailed studies were made of ErAs using this approach, in particular because
Shubnikov-de-Haas measurements were available for the Fermi surface.\cite{Petukhov94,Lambrecht97}
The approach was quite successful in determining the Fermi surface
properties, in particular when spin-orbit coupling was included.

However, it does not accurately describe the spin-splitting of the valence
bands. If we call the majority spin of the $f$ electrons spin-up, then
one finds as expected that the RE-$d$ spin-up states are also lower than the spin-down.
This results from the interatomic $p-d$ coupling. However, one also finds in
these calculations that the group V-$p$ VBM also is
higher for spin-down than spin-up states. This is because the difference between the spin-up and spin-down potential simply acts more or less uniformly on all electrons
in this approach and there is no explicit hybridization of the RE-$4f$ with the
V-$p$ electrons. In approaches that do take this interaction into account explicitly, as we will see in later sections, the N-$p$ spin-up electrons are
pushed up by their interaction with the RE spin-up $4f$ electrons lying below them. The N-$p$ spin-up states thus form antibonding combinations with the RE-$4f$ spin-up states. The spin-down states of the
N-$p$, on the other hand, are pushed down by the empty $4f$ levels above them.
Thus, the ordering of the VBM spin states is opposite
to that of the RE-$f$ electrons. This has a profound effect on the band gaps
in the nitrides, as we will see below.

Let us here mention that the ErAs Fermi surface was recently recalculated
using a much more involved dynamical mean field theory,\cite{Pourovskii09} which
reproduced the Fermi surface only slightly more accurately, but now also obtains
the $4f$ levels in the band structure picture in the right place.

The main difficulty with $4f$ electrons, as already mentioned, is their
strongly localized character. As such, the local spin density approximation (LSDA),
which is derived from a free electron gas is no longer valid.
If one were to do a straightforward LSDA calculation, one would find the
$f$ levels at the Fermi level, except for the Gd case where the
half filling will result in a large enough splitting to push the levels
away from the Fermi level even in LSDA.  Leaving the core levels
out of the picture altogether avoids this problem.

\subsection{Self-interaction correction (SIC)}\label{sic}
\subsubsection{Background on the SIC theory}
In Hartree-Fock theory, the terms representing the Coulomb interaction of
an electron in a given orbital with itself are exactly cancelled by the corresponding exchange term. In approximate schemes such as LDA, this is no longer the case, and thus there is a residual
self-interaction error. The self-interaction correction (SIC) approach to LDA was first
introduced by Perdew and Zunger.\cite{PerdewZunger81} It subtracts the
self-energy correction for each orbital as follows
\begin{equation}
\delta_{\alpha\sigma}=\int \frac{n_{\alpha\sigma}({\bf r})n_{\alpha\sigma}({\bf r}')}{|{\bf r}-{\bf r}'|}d^3rd^3r'+E_{xc}^{LSD}[n_{\alpha\sigma},0]
\end{equation}
The SIC formalism however has some unusual properties. The Hamiltonians
determining the one-electron states are in principle different for each orbital.
So, the eigenstates are no longer automatically orthogonal because they derive
from different Hamiltonians. While for a finite atomic system it is
straightforward to apply the correction,
in an extended system with only extended orbitals, the correction can
be shown to go to zero. However, one may find solutions which are localized, i.e., break the periodic symmetry,
by searching for solutions different from Bloch functions with an explicit
energy minimization approach. This approach was introduced by Svane and Gunnarsson\cite{Svane88,Svane90} and was first applied to a RE system in a study of Ce by Svane.\cite{Svane94}

SIC is somewhat cumbersome because it departs from the
usual Bloch band structure picture. On the other hand, Heaton et al.~\cite{Heaton83} introduced the idea of a {\sl unified Hamiltonian} which,
using a projector technique, restored
a Hamiltonian applicable to periodic Bloch sums of such localized solutions
on each site. This technique was combined with the linearized muffin-tin
orbital (LMTO) approach and further simplified by Temmerman et al.~\cite{Temmerman93}

\subsubsection{Applications of SIC to RENs}

The SIC approach was applied to various RE elements: Pr,\cite{Temmerman93}
hcp Gd,\cite{Heinemann94} and Ce.\cite{Szotek94,Svane94}
It was applied to Ce chalcogenides, \cite{Svane99} Eu-chalcogenides and pnictides,\cite{Horne}
and Yb-compounds (including YbN).\cite{Temmerman99,Svane2000} The particular choices of Ce, Eu and
Yb were because
these RE elements exhibit a competition between different valencies and one of the strengths of SIC
is that it can determine which $f^N$ configuration has the lowest energy.

\begin{figure}
\vskip 1 cm
\includegraphics[width=\columnwidth]{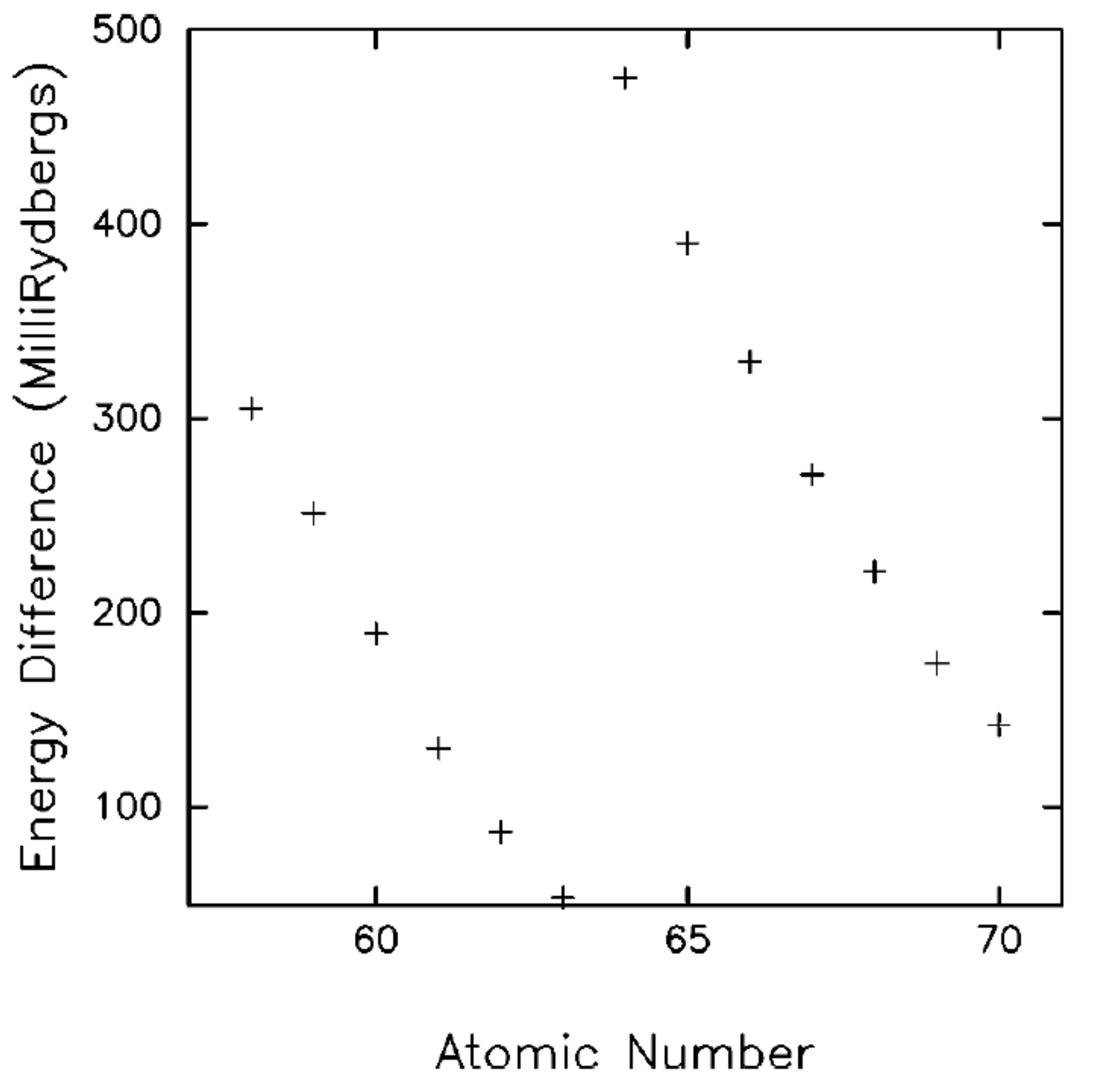}

\vskip 1 cm
\caption{Energy difference between divalent and trivalent RE ions in RENs, from Ref.~\onlinecite{Aerts2004}. \label{figvalency}}
\end{figure}

The SIC approach was applied to the entire series of RENs by
Aerts et al.~\cite{Aerts2004} The paper focuses on determining self-consistently
how many $f$ electrons need to be treated as localized states  and hence
on determining the effective valency. Their results clearly show that all
RENs prefer the trivalent over the divalent states. As shown
in Figure \ref{figvalency}, there is a systematic trend of the divalent minus
trivalent total energy: it decreases in the first part of the series with
a minimum at Eu and then jumps to its maximum value in Gd and decreases again
toward Yb. We will later see that this potential
competition between trivalent and
divalent behavior for Eu and Yb will also show up in other approaches.

This study nicely reproduced the decreasing trend in lattice constant with
atomic number throughout the series, with the exception of Ce. The authors
explain the trends in bonding and emphasize the hybridization effects of
the RE-$f$ with the N-$p$ states.

In terms of magnetic properties, they find that a small magnetic
moment opposite to that of the RE is induced on the N
in the first half of the series and a moment
with equal sign is induced in the second half.  This follows the same trend as
the RE-$4f$ orbital moments, which are opposite to the spin moment according
to Hund's third rule. The total spin magnetic moments
are integer with the exceptions of ErN, TmN and YbN. This implies insulating
(Tb, Dy, Ho) behavior or half-metallic (Pr to Gd) behavior. CeN is found to
be non-magnetic and metallic.
We note that they find the lighter RENs all to have a zero gap for
majority spin. The minority spin gaps gradually increase from PrN (0.53~eV)
to EuN (1.46~eV), then decrease to 1.11~eV in GdN. The next three, TbN, DyN and HoN, are found
to have a small gap in both spin channels with the minimum
gap ranging from 0.05~eV in TbN to 0.11~eV in TbN and 0.24~eV in DyN.
The last three, ErN, TmN and YbN, are found to have zero gap.
All of these results are based on densities of states. No band structure
dispersions are reported in this work.

As an example, in TbN, Aerts et al.~find the occupied
majority spin $4f$ bands to lie below the N-$2s$ band, while the minority
spin filled $4f$ bands lie just above it. The empty $4f$ bands lie just above
the conduction band minimum (CBM). This differs from the LDA+$U$ results discussed
in Section \ref{lda+u}, which generally find all the occupied $4f$ bands above
the N-2$s$ bands. This indicates a very strong effect of the SIC on pushing the
occupied $4f$ bands down. The splitting between occupied and empty $f$ bands
of the same spin also appears to be significantly larger than in the LDA+$U$ results discussed
in Sec. \ref{lda+u}.

In a closely related paper, Svane et al.~\cite{Svane05} also studied some of the other
RE-V and chalcogenides with the SIC technique, in particular,
the Sm compounds. In the case of SmS, SmSe and SmTe, they
find a transition from trivalent to divalent behavior under pressure.
In other words, depending on the pressure, the number of $f$ electrons behaving as
localized or delocalized may change.
In that paper,
they also applied the Hubbard-I approach\cite{Lichtenstein98}
to the spectral functions,
which allows one to see the atomic multiplet splittings of the $4f$ electrons.
Unfortunately, they did not apply this technique to the nitrides.  We will discuss this
approach in more detail in Sec. \ref{dmft}.

\subsection{LDA+$U$}\label{lda+u}
\subsubsection{Theoretical background on LDA+$U$}
An alternative approach for dealing with the localized nature of $f$ electrons is the
so-called LDA+$U$ approach, introduced by Anisimov et al.~\cite{Anisimov91,Anisimov93,Liechtenstein95}
In this approach, the orbital dependence of the Coulomb and exchange interactions for
a set of localized orbitals is grafted onto the DFT-LDA framework by adding Hubbard-like
terms to the Hamiltonian and total energy functional:
\begin{equation}
E_{LDA+U}=E_{LDA}+E_U-E_{dc}
\end{equation}
The Hubbard terms $E_U$ depend on the occupation numbers of the localized orbitals, or, in a more
general formulation, on their density matrix. In the most general open shell case, they
are written in terms of Slater's $F_k$ Coulomb and exchange integrals\cite{CondonShortley}
and Clebsch-Gordan coefficients and
correspond to the configuration-averaged Hartree-Fock expression for the Coulomb energy
but using an empirically determined
screened direct Coulomb interation $F_0=U$. The exchange terms are usually
taken unscreened and written in terms of one effective exchange interaction $J$.
The total energy is then minimized not only
as a function of the spatial density function $n({\bf r})$ but separately as a function of the occupation numbers for specific orbitals $n_{m\sigma}$, or more generally their density matrix $\rho^\sigma_{mm'}$.
 This leads to an orbital dependent potential.

However, because these Coulomb interactions are already
included in an approximate way in the LDA, a double counting correction must be added.
There has been considerable discussion of this double counting
correction\cite{Anisimov91,Anisimov93,Czyzyk,Petukhov03}
and there are several slightly different variants of the LDA+$U$ approach.
One of these is the around mean field (AMF) approach while the other is the
fully localized limit (FLL).
The latter is clearly the most logical approach for strongly localized states such as $f$ electrons.
It assumes that in the atomic limit of integer occupations, the LSDA and LSDA+$U$
will give the same total energy.

While in the earliest formulation, the spin effects are
supposed to result completely from the $U$ and $J$ terms, the most common approach is to start already
from a spin-polarized LSDA and then the $U$ and $J$ terms merely need to add the orbital dependence
of the Coulomb and exchange interactions. The latter are in principle dependent on the
$m$ quantum numbers of the orbitals in an open-shell case. This can thus in principle lead to
orbital as well as spin ordering. Sometimes one spherically averages away all these effects, \cite{Dudarev} while
in the most complete formulation of Liechtenstein et al.~\cite{Liechtenstein95} these orbital effects are included in
detail.

However, the theory still remains a single Slater determinant Hartree-Fock like theory
and thus does not deal explicitly with the multiplet splittings discussed in the introduction.
One may go beyond the Hartree-Fock solution for the localized electrons in a so-called
dynamical mean field theory (DMFT) framework.\cite{Lichtenstein98} This
will be discussed in Sec. \ref{dmft}.
The advantage of the LSDA+$U$ approach
is that it fits more easily into a standard band structure approach than the SIC approach.

In a broad sense, the effect is pretty similar to that of the SIC potential, in that it tends to push
occupied states to lower energy and empty states to higher energy. However, unlike
SIC where the strength of this interaction is determined self-consistently within the
approach in LDA+$U$ it is determined by an adjustable parameter $U$. In the original work
this screened interaction $U$ was determined independently from impurity like
constrained DFT calculations.\cite{AnisimovGunnarsson91} Recently, a closely related
linear response approach was proposed by Cococcioni and de Gironcoli.\cite{Cococcioni}
For the most part, however, it has been customary to treat the $U$ parameters as
adjustable parameters and to either study the behavior as a function of $U$ or to determine the
values of $U$ based on experimental inputs.

In a formal sense, the relation to a Hartree-Fock treatment for the localized orbitals with
a screened Coulomb interaction makes the method closely related to the GW theory
and also to the recently developed screened exchange and hybrid functionals.

\subsubsection{Applications of LSDA+$U$ to RENs}
Because the application of LSDA+$U$ to RENs, other RE-V and chalcogenides
is straightforward, several groups have applied this method. In spite of the similarity of
the approach used by these different groups, there are significant differences in the results.
As usual, the devil is in the details.

One of the earlier attempts to use this method by Lambrecht\cite{Lambrecht2000}
did not yet use a full-fledged LSDA+$U$ formulation but rather added energy shifts directly to the
diagonal Hamiltonian matrix elements in the LMTO basis set. The effect on the band
structure is the same as in the LSDA+$U$ method but without the underlying justifying theoretical
framework. In this work, the focus was on the question whether ScN and GdN
were semiconductors or semimetals and shifts were added for both the $f$ and $d$ states. These shifts of the $f$ levels were based on the experimental
X-ray photoemission spectroscopy (XPS) and Bremsstrahlung isochromat spectroscopy (BIS)
for the $f$ levels in Gd-pnictides.\cite{Yamada96}
For the $d$ shifts the GW theory was used as a guide. It would predict
an inverse proportionality of the shifts to the dielectric constant. The latter can
itself be obtained from the interband transitions in the band structure, so the circle can be closed.
As a starting point, one needs to know the shift in one material, and ScAs
was used for that purpose. That material is a semimetal and, as already mentioned above,
detailed knowledge on the size of the Fermi surface, related to the anion-$p$ metal-$d$ band overlap
allowed the author to determine the required shifts in ScAs. This work successfully
concluded that ScN was a semiconductor and was the first to provide predictions of the
red shift of the gap in GdN due to spin polarization.
However, this shift approach was not sufficiently flexible, too heavily based on experimental
input and restricted to half-filled $f$ shells.

\begin{figure}
\includegraphics[width=8cm]{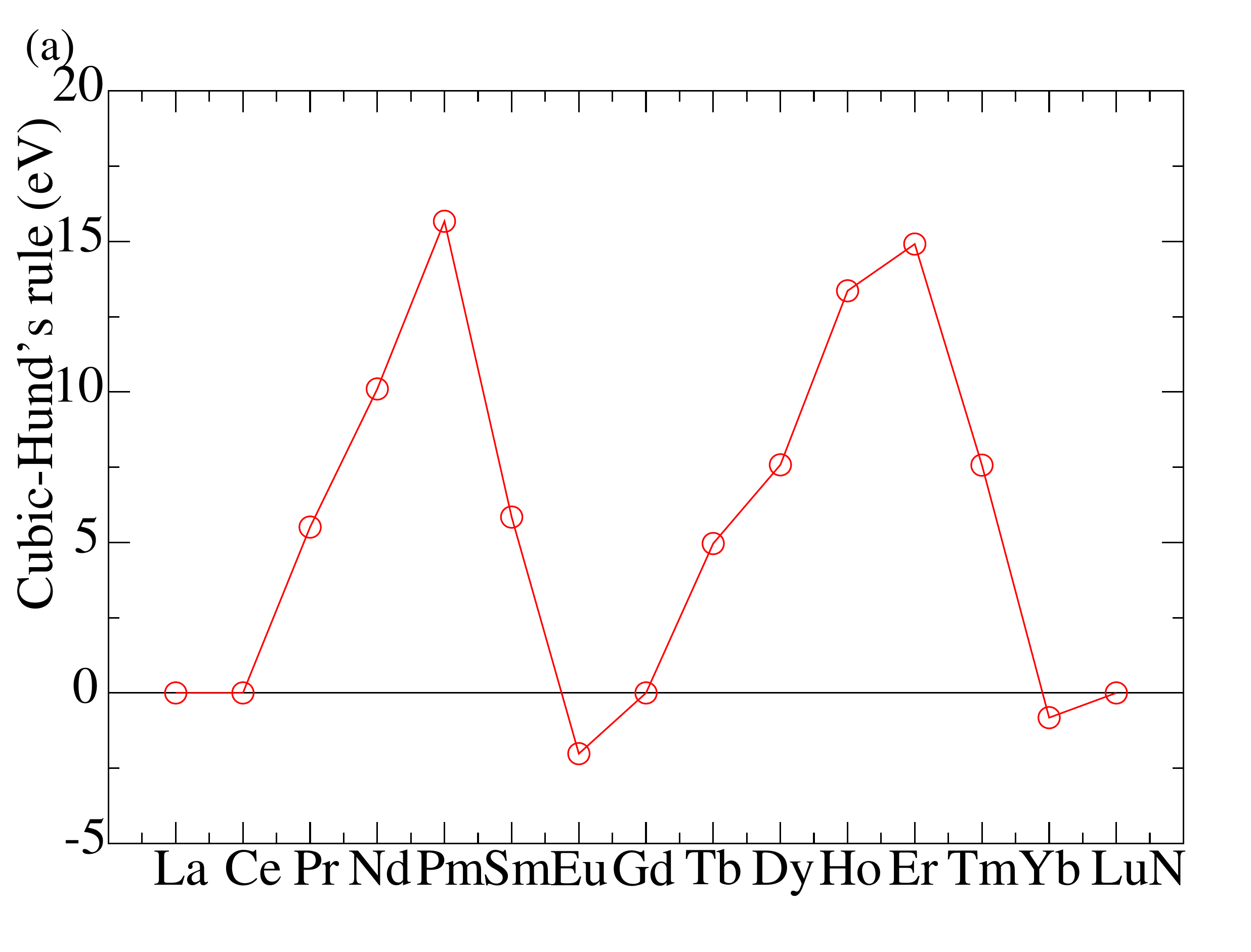}
\includegraphics[width=8cm]{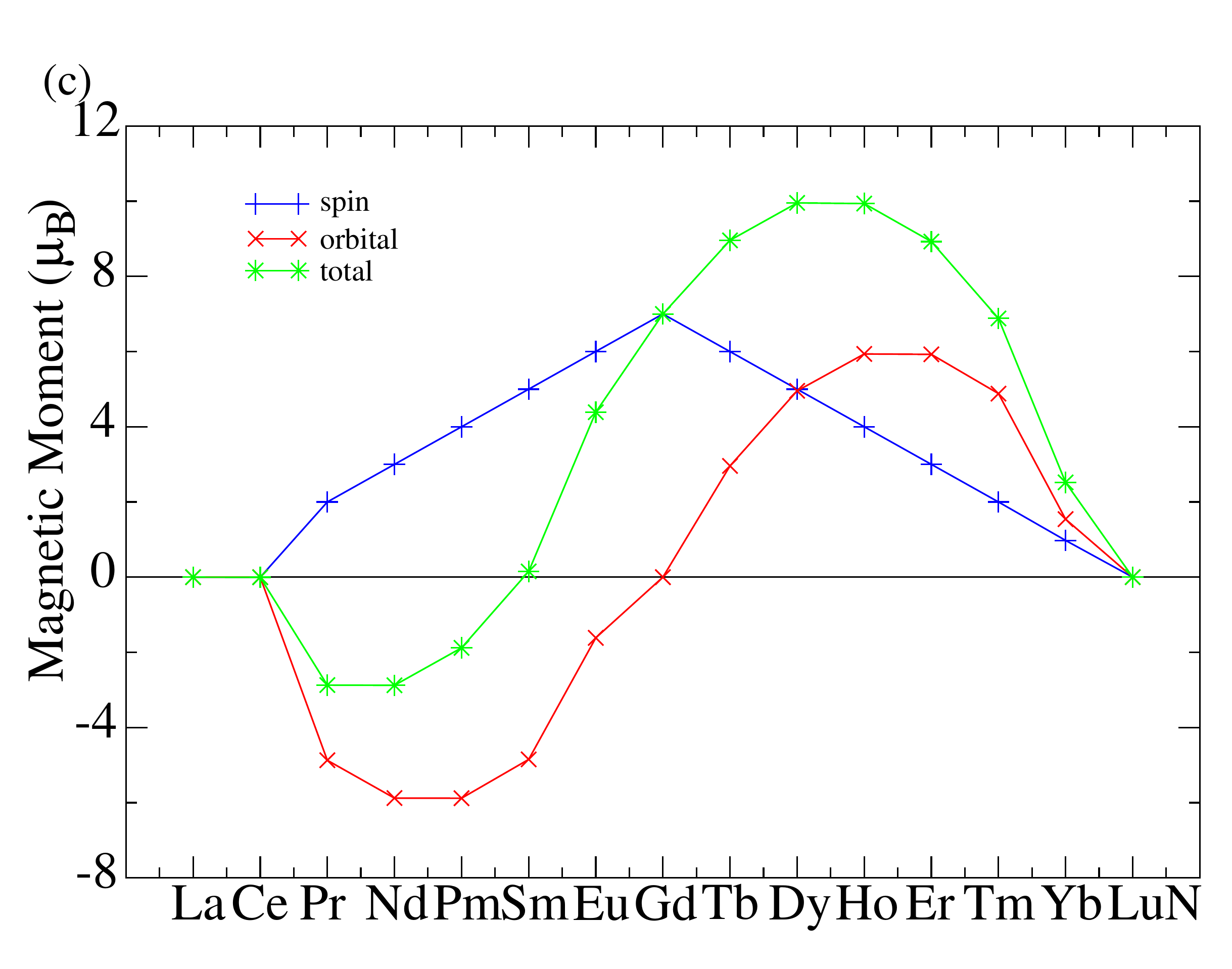}
\caption{Top: Energy difference between cubic and Hund's rule solutions to the LSDA+$U$ results.
Bottom: Magnetic moments in the RENs decomposed into spin and orbital contribution.
From Ref.~\onlinecite{Larson07}.\label{fighundsrule}}
\end{figure}

We begin our discussion of the LSDA+$U$ results
with the most comprehensive study of the RENs using this approach,
by Larson et al.~\cite{Larson07} This paper systematically explores the band structures of the
entire REN series and discusses the magnetic moments. The emphasis of the paper is on the
question of how the $f$ electron shell is occupied. In fact, as the LSDA+$U$ formalism for open shells
can lead to orbital ordering, one must determine self-consistently not only the number of
localized $f$ electrons (as in the SIC approach), but specifically for which $m$-quantum numbers
of the $f$ orbitals the states are occupied or empty. In other words, one must make the
$f$ electron density matrix self-consistent. The trouble here is that in principle there can
be multiple minima and one must find the lowest energy minimum.  In particular, the authors
discuss two separate plausible starting points and find which of the two gives lower energy.
In the first one, one assumes cubic symmetry is fully maintained. In that case, the energy minimization is dominated by the desire to move all the occupied states well below the Fermi level.
This depends on the specific filling and on the crystal field splitting of the $f$ electrons
in the octahedral environment. In the second approach, one assumes a slight modification of
Hund's rules. As in Hund's rules, the spin is maximized first, but then instead of maximizing
$L$, one maximizes $L_z$ since the moments are supposed to stay fully parallel to each other in
a periodic ferromagnetic solution.  This Hund's rule solution explicitly breaks the
cubic symmetry by the orbital polarization. It was found that the Hund's rule solution
had lower energy in all cases, except possibly for EuN and YbN as is shown in Figure \ref{fighundsrule}.
The difficulty with EuN and YbN, as already mentioned in the SIC section, is that in these cases
the divalent solution is close by in energy and during self-consistent iterations, the solution
may evolve toward an erroneous local minimum.

In contrast to the SIC calculations, the RE-$4f$ bands are found to be always between the
N-$2s$ and N-$2p$ bands and a gap is obtained in most RENs. Of course, these results depend
on the choice of $U$ parameters.  In fact, the origin of the gap in these calculations, as opposed
to an almost zero gap, results from the inclusion of both a $U_d$ and $U_f$ parameter.
While the $U_f$ is well justified by the localized character of the $f$ electrons, the
$U_d$ requires some separate discussion. The $U_f$ parameters determine the splitting of the
empty and filled $f$-states. In the Larson paper these were determined rather carefully
based on experimental data.\cite{Larson07} For GdN, they are based on the experimental data on
this splitting from X-ray photoemission and inverse photoemission for the entire series
of Gd pnictides, (GdP, GdAs, GdSb, GdBi).\cite{Yamada96}
They also give good results for the occupied $f$ states
in GdN. The values for the other RENs were then based on the assumption that the
$U_f$ should scale like the atomic Coulomb Slater parameter $F_0$ and exhibit similar
screening. In view of the overall similar electronic structure of the outer shells this
is an excellent approximation and at the same time it incorporates the correct atomic
trends.

Within this approach, the same authors had earlier determined that GdN would still be a
semimetal with the Gd-$d$ band at $X$ dipping below the N-$p$ like VBM at $\Gamma$.\cite{Larson06}
To account for the experimental observation of a gap in GdN, they added a $U_d$ which
shifts up the empty states relative to the filled states. This parameter was adjusted to the best
available data at the time for the spin averaged gap. It does lead to a prediction of the
ferromagnetic red shift of the gap, or the difference in gap between majority and minority
spins. Although the RE-$d$ like bands are not really narrow localized bands, they in fact
show strong band dispersion. This extension of the LDA+$U$ formalism to open up gaps
should be viewed as an empirical correction for the well-known LDA underestimate of band gaps
in most semiconductors.

\begin{figure}
\includegraphics[width=8cm]{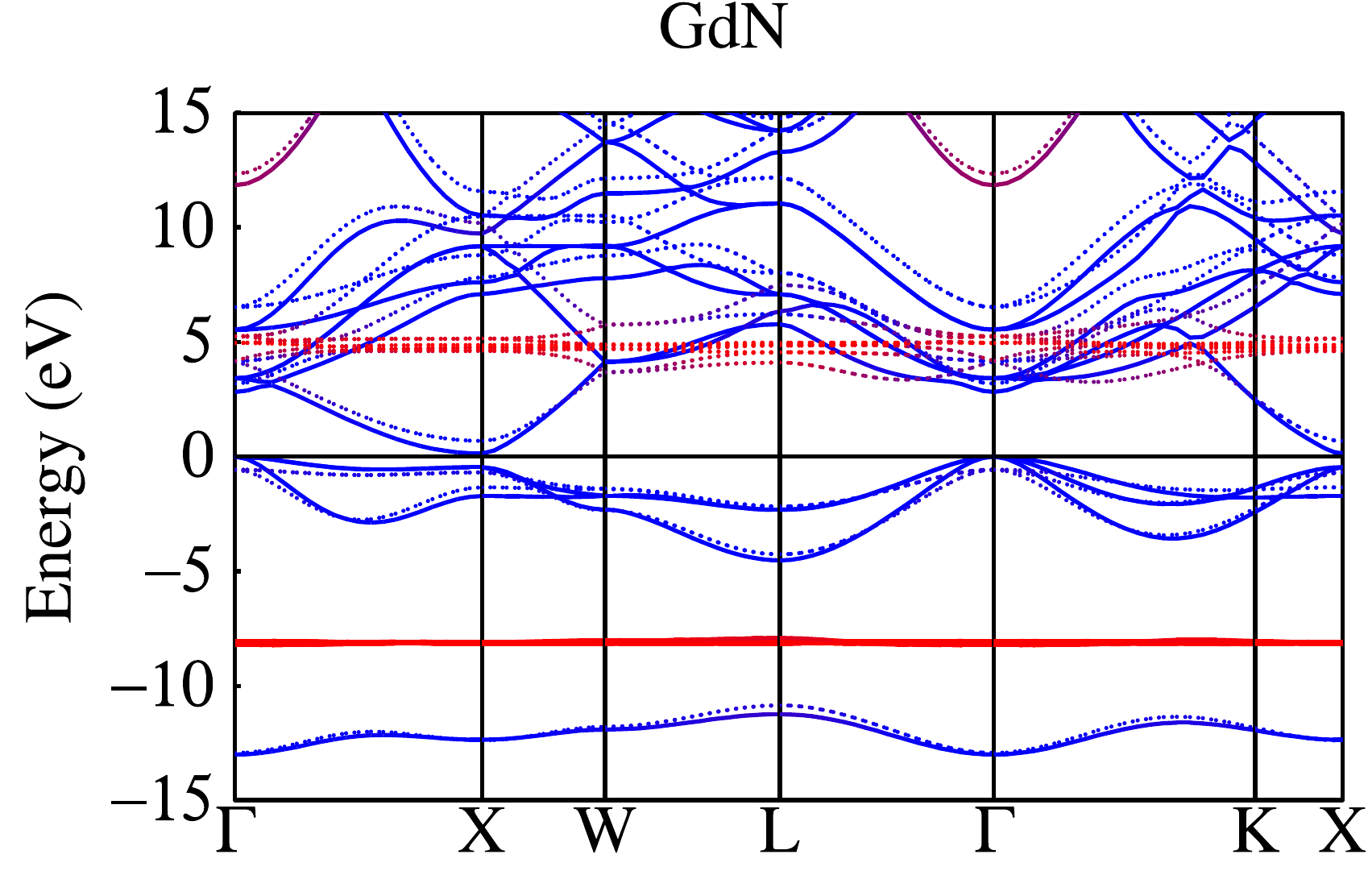}
\includegraphics[width=4cm]{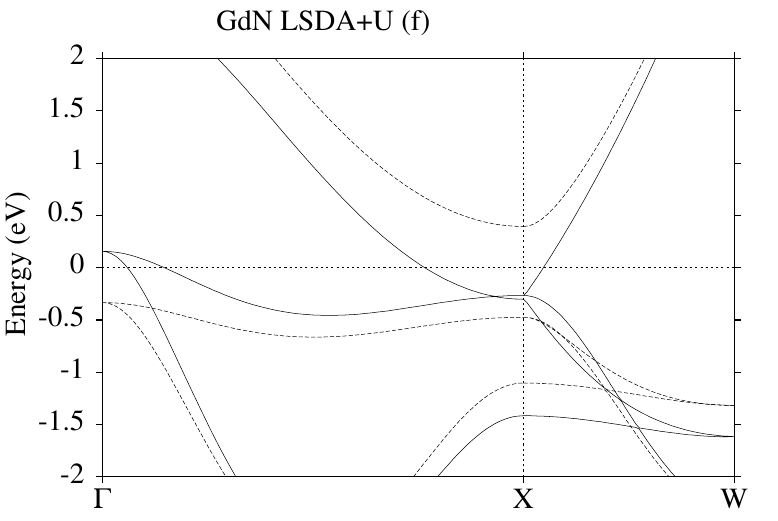}\includegraphics[width=4cm]{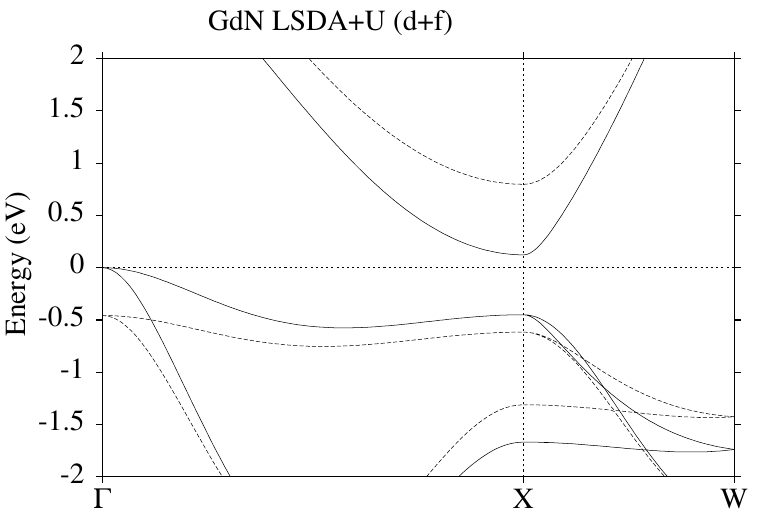}
\caption{Band structure of GdN in LSDA+$U$. Top panel, from Ref.~\onlinecite{Larson07}:
The $f$-bands are colored red, the blue and purple indicate mixed $f$ and other orbitals
character. Solid and dashed lines indicate majority and minority spin. Second and third panel,
from Ref.~\onlinecite{Larson06}: Detail of the bands near the gap with and without $U_d$.
\label{figgdn}}
\end{figure}

As an example, the band structure of GdN is shown in Figure \ref{figgdn}.
The top panel shows the bands occupied and empty $4f$ bands in red. The lower panels
show the details along $\Gamma-X$ with and without the inclusion of the $U_d$.

Basically, the importance of Ref.~\onlinecite{Larson07} is that it clearly establishes that Hund's rules
stay valid even for periodic RE compounds.  This is born out in practice by the fact
that the RE in RE-V definitely have an orbital moment and hence break the cubic symmetry.
The slight twist on Hund's rules of making it
compatible with a periodic ferromagnetic solution, however, results in an interesting
situation of approximately zero magnetic moment for Sm rather than for Eu.
As we will see in Section \ref{magnetics} this is confirmed by the experiments for SmN in the ferromagnetic state.

Among the RENs, GdN and the related pnictides have
received more attention than any of the others.
In part, this is because these are the easiest to deal with theoretically because they have a half-filled
shell so even LSDA gives reasonable results.
The LSDA+$U$ approach was applied to these compounds by Duan et al.,~\cite{duanprl05,duanjap05,duanapl06}
Ghosh et al.,~\cite{ghosh05} Larson and Lambrecht, \cite{Larson06,Larson07,Larsonjpcm06} Mitra and Lambrecht \cite{mitra08mag,mitra08opt} and Abdelouahed and Alouani.\cite{alouani07,alouani09}
We will discuss the differences between these papers in later sections focused on the
magnetism (Section \ref{magnetics}) and the optical properties (Section \ref{optics}). First, we complete our tour of the various computational
approaches to deal with the correlation of the $f$ electrons.

\subsection{Hartree-Fock, Hybrid functionals}\label{hybrid}
As was emphasized in the two previous sections, the main error of LSDA for $f$ electrons
is the self-interaction error. By treating the $f$ electron Coulomb terms separately  at the
Hartree-Fock level, LSDA+$U$ corrects the self-interaction error, and SIC of course explicitly removes it. The drawback of LSDA+$U$ is that it is somewhat empirical and {\sl ad-hoc} as the $f$ Coulomb
interactions are just added onto the theory. One might want to use a more rigorous first-principles
approach which treats all electron orbitals on the same footing.  Hartree-Fock theory is self-interaction
free but misses correlation entirely. It is known to strongly overestimate band gaps in
semiconductors, and to tend to overestimate spin-splittings and magnetic moments. Its application
to $f$ electron systems has been delayed by technical difficulties. Nonetheless, recently,
it has become possible to apply Hartree-Fock to periodic solids with Gaussian basis sets and
recently also to include $f$ electrons in the CRYSTAL code.\cite{CRYSTAL06}

This also opened the way toward hybrid
functionals. The hybrid functional approach, i.e., mixing some fraction of Hartree-Fock with
LDA, has long been popular in the quantum chemistry community but has recently gained considerable
ground in solid state physics applications. Two functionals have been commonly used, B3LYP\cite{becke:5648,becke:1372,becke:1040}
and more recently the HSE\cite{HSE03,HSE06} (Heyd-Scuseria-Ernzerhof) functional.
The latter mixes in about
25 \% exact exhange, but cuts off the long range part of the exact exchange, and thereby
uses effectively a screened exchange. This screening is done with an error-function type cut-off.
The method was recently implemented in the VASP (Vienna Ab-Initio Simulation Package)
code\cite{paier:154709,paier:249901}
 and has since gained a lot of popularity as it was found to be promising to significantly improve the band gaps.  It has not yet
been widely applied to $f$ electron systems.
B3LYP and Hartree-Fock were both applied to GdN by Doll et al.~\cite{Doll} using the
CRYSTAL code. Very recently Schlipf et al.~\cite{Schlipf2011} implemented the HSE functional
within an FLAPW code and applied it to GdN.

As expected, Hartree-Fock gives a large band gap of about 5 eV in  GdN
and also overestimates the spin-splitting
of the $f$ electrons. Doll et al.~\cite{Doll} find the occupied $f$ levels at about 15 eV
below the VBM and the empty spin-down levels at about 22 eV above the VBM.
Interestingly, the B3LYP gave different solutions with somewhat different total energy.
The band structures in the lowest two of these solutions show a strange extra band near the
Fermi level with rather different dispersion to the well-known N-$p$ or Gd-$d$ like bands.
In one case, it occurs in the majority spin and in the other in the minority spin.
In a third solution, which however, they find to have a 2.4 eV higher total energy, they find
an insulating band structure with the usual  indirect ($\Gamma-X$) gap of about 0.7 eV
for majority spin. The empty and occupied $f$ levels lie about $\pm$7 eV relative to $E_F$ and are in reasonable
agreement with experiment and LSDA+$U$ calculations.  The origin of the strange additional
solutions in B3LYP is not clear.

We should mention one other Hartree-Fock plus correlation energy study of GdN by
Kalvoda et al.~\cite{Kalvoda} This paper however focuses entirely on the cohesive energy
and does not give any information on the band structures. It treated the $f$ electrons as
core states.

Schlipf et al.~\cite{Schlipf2011} used the HSE functional and obtained results very close to the
LSDA+$U$ results. They find the $f$-levels at about $\pm$6 eV relative to the VBM.
They find a very nearly zero band gap, with a slight positive gap of 0.01 eV
at the experimental lattice constant (4.988 \AA) and a slight negative gap of $-0.06$ eV at their
theoretical equilibrium lattice constant (4.963 \AA). Thus, like Duan et al., \cite{duanprl05}
they obtain a solution at the brink of a metal-insulator transition.  Their smallest
direct gap at $X$ is 0.90 eV for the ferromagnetic state and 1.17 eV (average of spin-up and
spin-down gap) above $T_C$. This red-shift of the gap  of 0.27 eV is quite close to the
results of Larson et al.~\cite{Larson06} or Trodahl et al.~\cite{trodahl07} of about 0.4 eV.

\subsection{The GW method}\label{GW}
The most accurate and rigorously first-principles computational method available today for band structures
is Hedin's GW approach.\cite{Hedin65} Here $G$ and $W$ stand for the one-electron Green's
function and screened Coulomb interaction respecively. While the band structure eigenvalues
of the Kohn-Sham equations in density
functional theory are in principle only Lagrange parameters related to the one-electron
wavefunctions' orthogonality and normalization, used in minimizing the total energy as a function
of density, they do not strictly speaking represent excitation energies. The theory for quasiparticle
excitations, i.e., extracting or adding an electron to the system, is conceptually the correct theory
for addressing photoemission and inverse photoemission theory. Within this many-body theoretical
framework, GW is the first order approximation for the self-energy operator in an expansion of the screened Coulomb interaction.
The equations of the GW method can be summarized as follows. The quasiparticle excitation
energies $E_i$ are given by
\begin{align}\label{quasiparticle}
&\left[-\frac{1}{2}\nabla^2 +v_{H}({\bf r})\right]\psi_i({\bf r}) & \nonumber \\
+&\int d^3r' \Sigma_{xc}({\bf r},{\bf r}',E_i)\psi_i({\bf r}')=E_i\psi_i({\bf r}),
\end{align}
where $v_H({\bf r})$ is the Hartree potential representing the classical interaction with the nucleus
and all electrons, $\Sigma_{xc}({\bf r},{\bf r}',\omega)$ is the self-energy operator which is non-local and energy dependent and $\psi_i({\bf r})$ is the quasiparticle wavefunction.
The self-energy operator is given by
\begin{equation}
\Sigma_{xc}({\bf r},{\bf r}',\omega)= \frac{i}{2\pi}\int d\omega'G_0({\bf r},{\bf r'},\omega-\omega')W({\bf r},{\bf r}',\omega')e^{-i\delta\omega'}
\end{equation}
in which $G_0$ is the one-electron Green's function of the corresponding
independent particle approximation, which is usually the Kohn-Sham equation,
\begin{equation}
\left[-\frac{1}{2}\nabla^2 +v_{H}({\bf r})+v_{xc}({\bf r})\right]\psi_i({\bf r})=\epsilon_i\psi_i({\bf r}),
\end{equation}
and with the screened Coulomb interaction,
\begin{equation}
W({\bf r},{\bf r}',\omega)=\int \varepsilon^{-1}({\bf r},{\bf r}'',\omega)v({\bf r}'',{\bf r})d^3r''
\end{equation}
in which $\varepsilon$ is the dynamic dielectric response function and $v$ the bare Coulomb interaction
($1/|{\bf r}-{\bf r}'|$). Schematically, one can write
$W=(1-v\Pi_0)^{-1}v$ in which $\Pi_0$ is the independent particle polarizability, which itself is
given in terms of the Green's functions as
\begin{equation}
\Pi_0(1,2)=-iG_0(1,2)G_0(2,1). \label{eqrpa}
\end{equation}
Here we used  a short-hand notation where $1$ stands for $\{{\bf r}_1, \sigma_1, t_1\}$, i.e.
position, spin and time of particle 1.  In practice, these equations are Fourier transformed over time
and lead to a convolution in the frequency domain and, instead of directly using the
position dependent functions, the corresponding operators  are expanded in suitable basis functions.

While this formalism has been available since Hedin's work of 1965,
applications to real systems have lagged because of the considerable
technical challenges in implementing it, in particular because the
frequency dependent dielectric function is needed.
  The first applications in the 1980s were restricted by
pseudopotentials and applied only to standard semiconductors but led to much
more accurate band gaps than LDA.  Implementing it to systems
with localized $d$ and $f$ electrons proved a further
challenge.\cite{Aryasetiawan94,Aryasetiawan95,Aryasetiawan96}
Typically, the equations are solved only in first order perturbation theory starting from the
LDA, which one calls a ``one shot $G_0W_0$'' approximation.
In the most recent incarnation, which is called the quasiparticle self-consistent GW (or QSGW) approach,
a new exchange correlation potential is extracted from the $\Sigma_{xc}$ and leads to new
one-electron Green's functions, which in turn lead to a new $\Sigma_{xc}$. These equations
are then solved self-consistently and the quasiparticle energies $E_i$ then become equal to
the $\epsilon_i$.

This method
was implemented in the FP-LMTO framework by van Schilfgaarde et al.~\cite{vanSchilfgaarde06,Kotani07}
It is important to note that this is an all-electron method without the need for pseudopotential
approximations, so that core-valence exchange is properly included.
It has already been shown to give accurate band gaps for all standard semiconductors and some
transition metal compounds.
The method was applied to $f$ electron systems including GdN and ErAs by Chantis  et al.~\cite{Chantis07}

The QSGW method gives a band structure that supports the results of prior LSDA+$U$ calculations
in general terms. That is, the filled and empty $f$ orbitals are moved far away from the
band gap region. Looking in more detail at the gap, the direct application of the
QSGW method gives a gap of 0.22 eV for GdN. Because in most semiconductors, QSGW is found to overestimate
the gaps slightly, a hybrid approach between LDA and QSGW using 80\% QSGW and 20 \% LDA as exchange
correlation was used and gives a gap of only 0.05 eV. This method thus once again gives a gap
very close to the metal insulator transition in GdN. The direct gaps at $X$ are
0.46 eV for majority and 1.48 eV for minority spin and thus give an average of 0.97 eV, close to the
experimental absorption edge.\cite{hullig79} Recent measurements place this gap slightly higher
at 1.3 eV\cite{trodahl07,yoshitomi11} which suggests that the uncorrected QSGW may be closer to the experiment
in this case than with the correction factor of 0.8. So clearly GW does quite well on the band gaps
and also supports the HSE calculations discussed in the previous section.

The main problem with the QSGW approach at present is that it appears to overestimate the
majority and minority spin splitting of the $f$ levels. In particular, the empty $f$
levels lie significantly higher than in the LSDA+$U$ and HSE results. This is also obvious
for other 4$f$ systems like ErAs and metallic Gd as shown by Chantis et al.~\cite{Chantis07}
These authors blame this discrepancy, as well as the remaining overestimate of the QSGW gaps
in most cases, to the underestimate of the screening by the random phase approximation
used in Eq. \ref{eqrpa}.  On the other hand, this seems to not affect the quality of the
energy levels near the gap.

\subsection{Dynamical Mean Field Theory} \label{dmft}
In all previous methods, the $f$ electrons are still taken into account at the Hartree-Fock
level. In other words, in a many-body theory framework, they correspond to a single
determinant solution. These methods thus do not do justice to the intricate splittings of the $f$ levels
one observes in atomic systems and also in solid state systems for any but the simplest
half-filled case. The problem is that a multi-electron framework in which the
multiplet splittings appear seems conceptually totally
separate from a one-electron band structure picture.  The dynamical mean field theory (DMFT) provides
a way to overcome this difficulty. The key is to realize that what a band structure really represents
is the one-particle excitation spectrum. One may then calculate the one-particle excitations
between different multi-electron configurations (or multiplet terms, to use the terminology
established in the introductory section) corresponding to specific quantum numbers $L,S,M_L,M_S$
for the isolated $f$ electrons first. These define a well defined one-particle Green's function
for the isolated $f$ systems without interaction with the other electrons in the system.
They define an on-site self-energy for the $f$ electrons.
The DMFT approximation then consists in assuming that the self-energy is not \textbf{k}-dependent
and can be solved by a so-called impurity solver.\cite{Kotliar}
The interaction with the remaining electrons can then be switched on assuming these
local interactions occur at each site and this way the broadening effects on the $f$ levels
due to their interaction with other (band electrons) in the system are included as well
as the indirect effect of the $f$ electrons on the rest of the system. Typically, the
quantities compared  with experiment in this case are the spectral functions which can be
directly compared to photoemission and inverse photoemission. 

The DMFT
has many applications beyond $4f$ electron systems, including any system where more complex
many-body effects occur in the localized electron system.
The dynamical aspects of the problem are important for example in Kondo systems, where they can
give rise to both upper and lower Hubbard bands, but also a peak at the Fermi level.
It is also important for the actinides, as shown by Savrasov et al.~\cite{Savrasov06}	
For $4f$ electrons its main importance is that it provides a way to
put the multiplet splittings and band structures on the same footing in a single comprehensive
picture. This particular application of DMFT was previously introduced by
Lichtenstein et al.~\cite{Lichtenstein98} and called the Hubbard-I approximation. It goes
a step beyond LSDA+$U$ but is in the same general spirit of including the relevant atomic physics,
but grafted onto a DFT formalism.

The DMFT approach was applied to Ce-pnictides by L{\ae}gsgaard and Svane.\cite{Laegsgaard}
The approach was recently applied to ErAs as already  mentioned and led to  results
in good agreement with the Fermi surface Shubnikov-de Haas experiments.\cite{Pourovskii09}
At the same time it gives quite different $f$-level splittings than LSDA+$U$.
Experimentally, from X-ray absorption and emission spectra for $M_{4,5}$ edges, i.e.,
$3d\rightarrow4f$ transitions, it is clear that the spectral shapes are indeed dominated by
these atomic multiplet effects.

Interesting hybridization effects of the localized multiplet split levels with other bands were observed in photoemission spectra of
EuNi$_2$P$_2$ by Danzenb\"acher et al.~\cite{Danzenbacher} These effects are also typical for
heavy fermion systems involving light RE such as Ce.
In the LSDA+$U$ band structure studies by Larson et al.,\cite{Larson07} EuN
stood out as a possibly interesting case from this perspective. In fact, it was found that
this case was one of the exceptions to the Hund's rules in which a solution of the equations
maintaining the cubic symmetry had lower energy. This however, turns out to be due to
the proximity for Eu of the divalent and trivalent solutions. This problem was previously
pointed out by Johannes and Pickett.\cite{Johannes05} In the work of Larson et al.,\cite{Larson07} it led to a prediction
of a rather unusual band stucture with a band of mixed $d$ and $f$ character which changes
its character as a function of {\bf k} in the Brillouin zone crossing the Fermi level.
EuN was thus targeted as one of the RENs worthy of a more detailed investigation.
Recently this came to fruition in a paper on EuN combining various theoretical approaches
compared to various X-ray spectroscopies.\cite{richter11}
It was found that the LSDA+$U$ fails for this system, while GW gives a correct prediction of
the existence of a band gap but only the DMFT in the Hubbard-I approximation can
satisfactorily describe the details of the $f$ electrons and their hybridization effects on the
Eu $d$ bands.

Even though the DMFT comes closest to combining the atomic multiplet aspects of the $f$ electrons
with the band structure picture, it is still limited in that it is geared toward studying the
one-electron excitation spectra, as occurs most directly in photoemission and inverse photoemission,
and to some extent in X-ray absorption and emission. In other words, it can treat
excitations in which the number of $f$-electrons changes by 1: $f^N\rightarrow f^{N-1}$
or $f^N\rightarrow f^{N+1}$.
However, it does not yet allow
a proper treatment of the optical excitations. In that case, one is often interested in
excitations between multiplet terms within a single $f^N$ configuration. These types
of excitations are mostly important for RE impurities but may also become important to fully
understand the optical properties of RENs.

Recently, a constrained DFT approach was proposed by Hourahine\cite{Hourahine}
which could potentially deal with such
situations. It minimizes the DFT total energy under the constraint of a particular set of
$L,S,M_L,M_S$ quantum numbers. In other words, it constrains the energy optimization
by constraining the expectation values of $\hat{L}^2$, $\hat{S}^2$, $\hat{L}_z$ and $\hat{S}_z$.
It has not yet been applied to RENs and was only implemented in a tight-binding type of
band structure approach.  For the sake of completeness, however, we mention it here as an approach
which has potential to provide further progress on this difficult problem.
It is in some sense a generalization of earlier work by Eriksson et al.~\cite{Erickson90} to
incorporate orbital polarization so as to impose Hund's rules, also
discussed by Solovyev et al.~\cite{Solovyev94}

Having done a tour of the main approaches to treat the electronic structure of
the $4f$ electrons within a band structure framework and their applications to RENs,
we will discuss in the following sections some details on how these methods give us results to be compared to measurements concerning magnetism and interband transitions. But first we complete this section with a discussion of  lattice dynamics.

\subsection{Pressure effects and phase transitions}\label{phonon}
The sensitivity of the GdN band structure to the lattice constants was already pointed out
by various authors and in particular emphasized by Duan et al.~\cite{duanprl05} As we mentioned earlier,
GW and HSE also find GdN to be at the brink of a metal insulator transition. Hence, studies as a
function of pressure or for films under tensile stress, as can be realized in epitaxial layers, are important. Abdelouahed and Alouani\cite{alouani07} took this a step further and investigated the possibility of
phase transitions between different crystal structures. They show that under tensile stress
there would be a transition to zinc blende GdN. It is somewhat doubtful that such a transition
can ever be reached because it requires a negative pressure,
but it is nonetheless of interest as a hypothetical compound in the context
of Gd-doped GaN.  On the other hand, Abdelouahed and Alouani\cite{alouani07} predict also a phase transition
to the wurtzite structure under applied pressure. This is rather surprising given that the
wurtzite phase has lower coordination that the rocksalt phase and remains to be confirmed by
experiment or other calculations. They predict a transition pressure of 68.3 GPa in GGA+$U$
and 19 GPA in GGA.

\section{Magnetic studies}\label{magnetics}

The REN series was originally a goldmine for neutron scientists, who found ferromagnetic materials with a simple cubic structure very easy to investigate through neutron scattering. This outcome was in contrast to work on the pure RE metals which have more complex crystallographic structures. Through such studies, the ferromagnetic nature of most of the RENs had been found already in the early 1960s.\cite{hullig79,VogtMatten93}  The easy axis of magnetisation, as controlled by the intrinsic anisotropy of the RE, was also determined. In a few cases, the lack of
single crystals left some uncertainties in the magnetic structures,
because powder samples only give the direction of the ordered
moments with respect to the scattering vector.\cite{schumacher66,child63} However,
the general findings were in agreement with theoretical predictions
based on an atomic approach to the RE magnetic properties, resulting from
crystal field calculations performed in the ground state multiplet
$J$.\cite{Tram63}

Since less than ten years ago the magnetic properties of the RENs have been
reinvestigated owing to the combined progress in the growth of
thin films and in \textit{ab initio} calculation techniques. The RENs are
indeed now considered as model systems for band calculation and used to
check the ability of different models to describe the localised RE states (see Section \ref{theory}). The central question of the
exchange mechanism in the RENs has also been addressed by these theoretical
calculations.

In the following we will give some examples of the more recent
studies on the magnetic properties of REN thin films, which can be
considered as real advances after so many years of limited progress. First, in Section \ref{magnetic_expt} we recall the experimental tools which
are used to establish magnetic properties, with emphasis on the techniques for thin films. Next, we give details about some of
the more significant data, obtained predominantly on epitaxial thin
films and mostly on GdN (Section \ref{exchange}). A great step forward in understanding
the magnetism of the RENs was taken when the prominent role of
$V_N$ as carrier dopants was experimentally
demonstrated. In Section \ref{other_ren} we describe the recent experimental
results obtained on other RENs, in particular SmN and
EuN. Finally, in Section \ref{magnetism} we discuss the different theoretical
models that have been proposed to explain the magnetic properties of
the RENs.

\subsection{Experimental details}
\label{magnetic_expt}

Standard magnetometry with a Superconducting Quantum Interference Device (SQUID) or Vibrating Sample Magnetometer (VSM) is the first step needed to determine
the magnetic properties of the REN films, and these techniques have been the basis of
all experimental studies. The magnetic moment, the magnetic ordering temperature and
the magnetic anisotropy can all be derived, however the diamagnetic or
paramagnetic contributions from the substrate or the buffer and cap
layers may introduce some uncertainties inherent in thin
films. For weakly anisotropic ferromagnets like GdN, the correction
for parasitic magnetic signals is easy, while for the other RENs
with non-collinear ferromagnetic arrangements or strong
anisotropies, the correction can be problematic.\cite{Mey2008}
Having recourse to complementary techniques that avoid the substrate
contribution entirely is becoming the rule, mainly Ferromagnetic
Resonance (FMR) and X-ray Magnetic Circular Dichroism (XMCD). In
addition, each of these techniques provides its own specific information.

FMR brings valuable information about both the magnetisation and the
anisotropy constants, even on polycrystalline films, through the application of a
static magnetic field along different directions with respect to the
film plane.\cite{Kha2006,ohta11}

X-ray absorption spectroscopy (XAS) is becoming an essential
technique for probing the magnetic properties of materials, in particular with XMCD.
The element selectivity of XMCD ensures that only the magnetic element of
interest will be analysed. The energy selectivity permits access
to the magnetic polarisation of each electronic level separately; 4$f$
and 5$d$ for the RE and 2$p$ for the N. There are two
energy ranges, which probe different aspects of the RE magnetism.
The $M_{4,5}$ absorption edges in the soft X-ray spectral range
probe transitions to the 4$f$ orbital empty states. Therefore, the
XMCD difference signal intensity is proportional to the 4$f$ magnetic
moment. With the help of specific sum rules, it is in principle
possible to separate the spin and orbital contributions to the
magnetic moment.~\cite{Tho92,Tera96} The $L_{2,3}$ absorption edges
in the hard X-ray spectral range probe the transitions to the 5$d$
band empty states, which may be polarised through the intra-atomic
$f$-$d$ exchange interaction. In metallic RE compounds the 5$d$ states are conduction electron states and contribute to the
exchange mechanism between the localised 4$f$ spins. In the case of
the RENs, the $L$-edge XMCD signal provides indirect
information about the 4$f$ magnetic moment and also about the other
sources of polarisation in the lattice. In addition, the RE XAS peak
positions are well separated for different valence states, for both
spectral edges. The energy separation is important in the case of anomalous REs which may exhibit mixed valence states (Ce,  Sm, Eu, Tm, Yb). We will show in Section \ref{other_ren}
an illustration of this situation in the case of EuN.

The existence of magnetic polarisation at the N site can be
obtained independently by measuring the XMCD signal at the N $K$-edge.
The spectra involve the $1s \rightarrow 2p$ electronic transition and probe the
2$p$ character of the unoccupied states. Leuenberger et
al.~observed a very large XMCD signal at the N
site in GdN.~\cite{Leuenberger05} The corresponding hysteresis loop from SQUID measurements exactly reflects the
Gd 4$f$ magnetisation derived from the Gd $M_{4}$ edge
XMCD signal. The same authors have investigated the XMCD spectra at
the Gd $L_{2,3}$ edges in the case of a
strained film of GdN (see details in Section \ref{strain}).~\cite{Leuenberger06}

A full XMCD investigation of EuN at the $M_{4,5}$ and
$L_{2,3}$ edges of the Eu ion has given a fundamental set of information~\cite{Ruck2011,richter11} which we will
discuss later. Theoretical calculations have been of great support to
understand the experimental results and have in turn benefitted from the data obtained by measurements.

Antonov et al.~have provided an in-depth theoretical study of XMCD in GdN for various absorption edges.~\cite{Antonov07} They used an
ASA-LMTO Hamiltonian and applied both LDA and LDA+$U$ methods. In
addition, they investigated the effects of the core-hole on the
spectra and on the surface states. While they find good general
agreement in the main features for all edges, they point out the need to include multiplet splitting effects to fully understand the high energy
shoulders in the fine structure of the $M_{4,5}$ edges. The
$L_{2,3}$ edges are explained in detail in terms of the various
spin-orbit split states. In addition, Antonov et al.~study the effect of electric quadrupole
and magnetic dipole transitions on these spectra. Abdelouahed
and Alouani have also studied the GdN $L_{2,3}$ edge spectra.~\cite{alouani07}  For the N
$K$-edge, the XMCD clearly requires the inclusion of core-hole effects
and even then Antonov et al.~could not fully explain the fine
structure observed in the experiments of Leuenberger et
al.\cite{Leuenberger06} Finally, the
surface states make a sizable contribution to the first peak at
400~eV in the N $K$-edge of GdN.

\subsection{Impact on the theory of the exchange mechanism in GdN}
\label{exchange}

In the following we will give an overview of the main parameters that
show a significant impact on the magnetic properties of the RENs. We
show how a better control of these parameters in the recent studies on thin films has
permitted some advance in the understanding of the mechanism of
exchange interactions in the RENs. These data all focus on GdN as the
model material. The compound GdN was indeed the first to be
reinvestigated by several groups. There soon appeared some
differences in the results published mainly by two groups,
in G\"{o}ttingen~\cite{Leuenberger05} and in
Wellington.\cite{Granville06,ludbrook09} More recently, a group in Kobe has reported on the magnetic properties of GdN thin films~\cite{yoshitomi11} and another group in
Cambridge has reported peculiar properties due to the existence of
another Gd-N phase depending on the preparation
conditions.\cite{senapati11a} Other important differences between these studies concern
the transport properties, which will be described in the dedicated
Section~\ref{optics} below.

\subsubsection{Control of the impurity}

In the older preparations of bulk material, a level of oxygen
contamination was unavoidable. Even in thin films where the
growth atmosphere can be controlled very accurately in ultra-high
vacuum systems, a small amount of contamination cannot be excluded.
It is well known that O levels below 1\% cannot be detected
through ion beam analysis techniques such as Rutherford backscattering or secondary ion mass spectroscopies. However, thanks to these older experimental
reports, the effect of oxygen contamination on the magnetism has been analysed and understood.
The ferromagnetic ordering is cancelled with 5\% O in the
lattice and short range antiferromagnetic order sets in.\cite{Gam70}
Therefore, any increase of Curie temperature $T_C$ cannot be accounted for by the
presence of oxygen impurities. 

\subsubsection{Crystallite size}\label{coercivity}

\begin{figure}
\includegraphics[width=8cm]{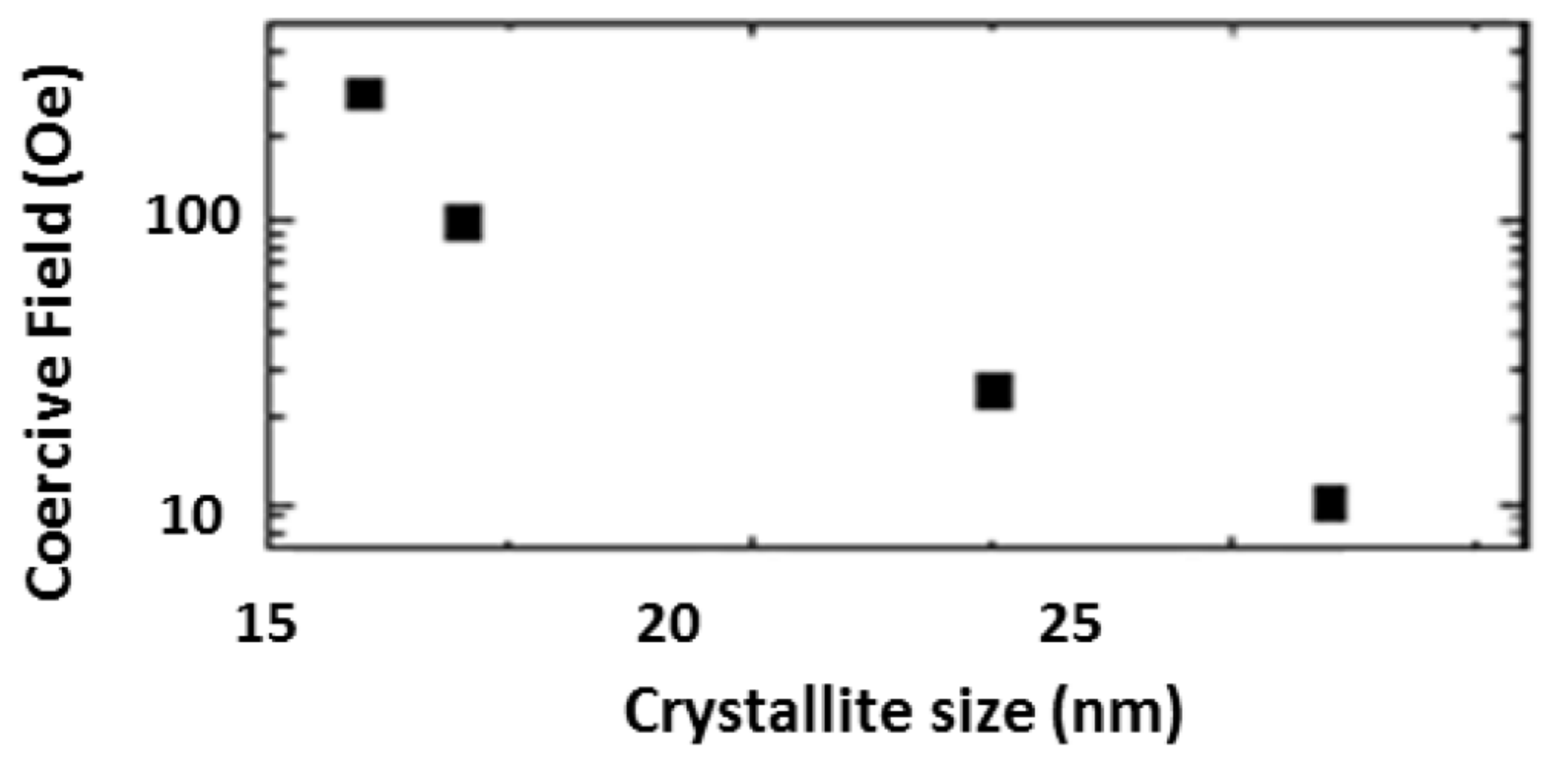}

\caption{The measured coercive field against the inverse of crystallite volume in GdN thin films. From Ref.~\onlinecite{ludbrook09}
\label{GdN_coercivity}}
\end{figure}

GdN is a soft ferromagnet, with the low temperature in-plane coercive fields of recent thin films reported as ranging from approximately 10 Oe to 220 Oe.\cite{Leuenberger05,Granville06,ludbrook09,scarpulla09,yoshitomi11}  Senapati et al.~have reported an enhanced coercive field of several hundred Oe in GdN grown by reactive dc magnetron sputtering at high power, as well as the presence of a small exchange bias in samples that may containg a strained phase of GdN.\cite{senapati10b,senapati11a}  For films with the relaxed GdN structure, Figure~\ref{GdN_coercivity} shows a linear dependence of the coercive field on crystallite size, suggesting that the magnetic reversal mechanism may take place through nucleation of domains at defect sites and that the intrinsic magnetic anisotropy of defect-free GdN is near 15 Oe.\cite{ludbrook09}

\subsubsection{Strain effects}\label{strain}

In  2005, Duan et al.~\cite{duanprl05} investigated the effect of strain
on the band structure of GdN with a LSDA+\textit{U} calculation process.
They demonstrated that the system exhibits a half-metallic band
structure at the equilibrium lattice constant and that a
semimetallic or semiconducting character develops with increasing
lattice constant. They show that the magnetic properties are
extremely sensitive to the volume variation. The calculated
magnitude of the exchange parameter is reduced with increasing
volume, and so is $T_C$.

There are a few experimental works which report on the effect of an increased volume
on the magnetic properties of GdN. Leuenberger et al.~investigated the influence of a lattice expansion on the Gd $L_{2,3}$
XMCD spectra.~\cite{Leuenberger06} $T_C$ is significantly reduced
(30~K instead of 60~K), suggesting a reduction of the exchange
interaction. The film deposited by N$^+$ plasma-assisted reactive
ion beam sputtering at room temperature shows a 9\% larger unit cell
volume with respect to the bulk parameter when the film is deposited
at 450$^{\circ}$C. It is shown that the 5$d$ states are polarised and
carry a magnetic moment. Apart from this reduction in $T_C$, the authors
observe another interesting effect. The XMCD signal amplitude ratio
$\left|L_3/L_2\right|$ = 3 is unexpectedly high in the volume
expanded film, while statistically a ratio of 1 is expected and
observed in most Gd systems, being imposed by the degeneracy of the
2$p_{1/2}$ and 2$p_{3/2}$ core states. The authors attribute this
effect to a reduction of the $L_2$ signal amplitude. The influence
of the lattice constant enlargement is not called on directly, but a
more fundamental reason is proposed. A change in the 5$d$
polarisation is due to a change in the relative occupation of the
spin up and spin down 5$d$ bands, predominantly of majority spin Gd 5$d$ $t_{2g}$
character. The $L_2$ edge mainly probes the Gd
5$d_{3/2}$ and the 5$d$ $t_{2g}$ states have a stronger $j$=3/2
character. XMCD is thus a very accurate technique to give reliable
information about the polarisation of the 5$d$ Gd states, and
consequently about the effect of changes in the exchange
interactions. As a comparison, Preston et al.~measured the $L_{2,3}$ XAS spectra of a GdN (001) epitaxial film with
a bulk-like lattice parameter and a $T_C$ of 70~K.~\cite{preston:032101}  A
$\left|L_3/L_2\right|$ ratio of 1 was measured as expected, which
provides another proof that the 9\% volume expansion of the Leuenberger et al.~sample affected the exchange interaction. Preston et al.~claim that
the anomalies detected in the $L_2$ edge XAS and XMCD amplitudes are
related to a modification of the 5$d$ $t_{2g}$ states, which are
involved in the bottom of the conduction band and experience
exchange splitting below $T_C$. This is in agreement with the
theoretical calculation by Duan et al.~\cite{duanapl06}

The Wellington group also investigated a volume expanded sample of
GdN.~\cite{Kha2006} The sample was a polycrystalline film grown by ion beam
assisted deposition which exhibited a lattice parameter of
5.12~$\r{A}$, corresponding to a 2.4\% tensile strain. $T_C$ was drastically reduced down to 20~K and the
magnetisation measured in-plane saturated only above 4~T, suggesting
some disorder-induced anisotropy. The grain size was indeed quite
reduced (3~nm) compared to the microstructure of the bulk-like films
(10~nm) reported earlier.~\cite{Granville06} This was confirmed through FMR experiments, performed on both
types of film.~\cite{Kha2006} In both films, a strong uniaxial
anisotropy perpendicular to the film plane was found, which is
attributed to strain effects. The magnetoelastic anisotropy
contribution was smaller in the lattice expanded sample. The origin
of this contribution is not yet understood.

\begin{figure}
\includegraphics[width=8cm]{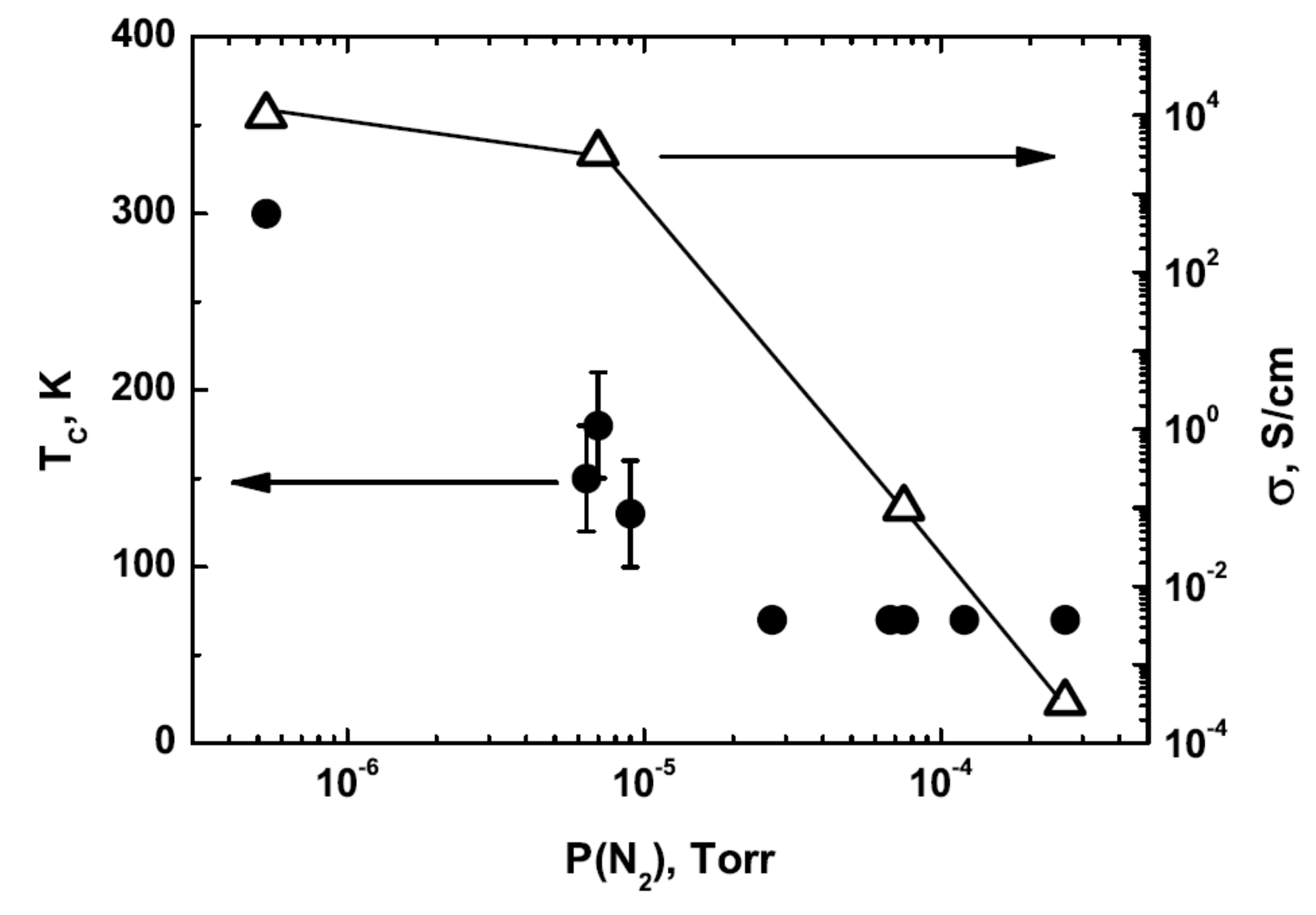}

\caption{Curie temperature $T_C$ and conductivity $\sigma$ just above $T_C$ vs N$_2$ growth pressure. The error bars result from the less than 100\% contribution from the Gd ion moments to the moment in the paramagnetic phase. From Ref.~\onlinecite{plank10}.
\label{GdN_Tc_carriers}}
\end{figure}

In summary, the strong strain effect observed in the magnetic
properties of GdN shows that the exchange interaction is sensitive
to the relative occupation of the spin states in the 5$d$ band, and that the polarisation of the
conduction band is indeed involved in the Gd 4$f$-4$f$ exchange,
suggesting a carrier-mediated exchange interaction brought about by
doping. We will discuss this issue in the next section.

\subsubsection{Stoichiometry - Effect of N vacancies}

A general agreement about the Curie temperature of GdN has been found at
near 70~K (see Table~\ref{tabcomparegd}), but all authors of such studies agree that the RE/N stoichiometry plays a
major role. Similar to the effect of unwanted oxygen contamination, a lack of nitrogen was shown to decrease $T_C$.\cite{WachterKaldis,cutler75}
Very recently, Plank et al.~\cite{plank10} carried out a systematic study
of N deficient polycrystalline thin films of GdN, resulting in the
important discovery that a N deficit enhances the average T$_C$ values and is equivalent to carrier doping.
These carriers may be polarised by the localised magnetic moments
and participate in an 
exchange interaction process of the RKKY type.

In the earliest study of the Wellington group it was reported that the rate of N$_2$ flow during growth
and the sample conductivity were intimately linked,
suggesting that the importance of the right stoichiometry is the
key to understanding the transport and magnetic properties of GdN.~\cite{Granville06} It was
only in the last couple of years that a systematic study of the
$V_N$ became possible, owing to progress in optimising the
preparation conditions and a strict control of the nitrogen
partial pressure.~\cite{plank10} The results of Figure~\ref{GdN_Tc_carriers} delivered through the
influence of the growth pressure are spectacular. It was found that
an increase of the conductivity by eight orders of magnitude
occurred by reducing the N$_2$ growth pressure from 2.6$\times$10$^{-4}$
Torr to 7$\times$10$^{-6}$ Torr, resulting also in a near tripling of the
$T_C$ value to $\sim$200~K. This also demonstrates that some hope exists to utilise GdN as an effective material for spin-polarising devices by driving $T_C$ even higher through proper management of the $V_N$ concentration.

More experimental work has established a link between the magnetic
behaviour and the $V_N$ concentration in the same way.~\cite{senapati10b,senapati11a} The authors claim the presence of another phase which may be antiferromagnetic, but
this conclusion needs some more experimental work.

\subsection{Magnetic properties of other RENs}
\label{other_ren}

Whilst the magnetic behaviour of the reference compound GdN may be near a full understanding, some of the other RENs have not even been reinvestigated since the 1970s. In recent years significant progress has
been made towards understanding a few heavy RENs (DyN and ErN) and light
RENs, in particular two special members of the series, SmN
and EuN. The results are all derived from thin films, either
polycrystalline or epitaxial. 

\begin{widetext}
\begin{center}
\begin{table}
\caption{Structural and magnetic properties of RENs. FM =
ferromagnetic, AFM = antiferromagnetic, PM = paramagnetic, VV PM = Van Vleck paramagnetic. Moment is
the maximum value of the moment at low temperatures and
$\mu_{\textrm{eff}}$ is the paramagnetic moment derived from a Curie-Weiss law fit above $\theta_p$.}
\begin{ruledtabular}
\begin{tabular}{l |c c c c | c c c c c}
& \multicolumn{4}{c}{pre 1994} & \multicolumn{4}{c}{1994-present} \\

Nitride & Magnetic & $T_{\textrm{C}}$, $T_{\textrm{N}}$ or $\theta_p$ & Moment & $\mu_{\textrm{eff}}$ & Magnetic & $T_{\textrm{C}}$, $T_{\textrm{N}}$ or $\theta_p$ & Moment & $\mu_{\textrm{eff}}$\\
  & order & (K) & ($\mu_B$/RE ion) & ($\mu_B$) & order & (K) & ($\mu_B$/RE ion) & ($\mu_B$) \\
\hline
LaN  &  &   &  & & & & & \\
CeN  & no order\footnote{Hulliger 1979 ~\cite{hullig79} and refs. therein}&   &  &    & & & & \\
PrN  & VV PM & $-11$-0\footnote{Hulliger 1978 ~\cite{hullig78} and refs. therein} & 0.24\footnotemark[2] & 3.57-3.7\footnotemark[2]    & & & &  \\
NdN  & FM & 27.6-35\footnotemark[1] &  1.8-3.1\footnotemark[1] & 3.65-3.70\footnote{Vogt and Mattenberger 1993 ~\cite{VogtMatten93} and refs. therein}    & & & &  \\
SmN  & AFM & $\textless$2, 3, 18\footnotemark[1]& &     & FM & 20-30\footnote{Preston et al.~2007 ~\cite{Preston07}}$^,$\footnote{Meyer et al.~2008 ~\cite{Mey2008}}  & &  \\
EuN  & VV PM\footnotemark[3] & & &     & VV PM\footnote{Ruck et al.~2011 ~\cite{Ruck2011}}& & &  \\
GdN  & FM & 62-72, 90\footnotemark[1]$^,$\footnotemark[3] & 6.6-7.26\footnotemark[1]$^,$\footnotemark[3] & 8.15-8.6\footnotemark[1]$^,$\footnotemark[3] &     FM & 25 (GdN/NbN)\footnote{Xiao and Chien 1996 ~\cite{xiao96}}, 58\footnote{Li et al.~1997~\cite{dxli97}} & 6.88\footnotemark[8], 6.8\footnotemark[12] & 7.92\footnotemark[8]\\
     &    &  & & & &  30-60 (GdN/W/NbN)\footnote{Osgood III et al.~1998 ~\cite{osgood98}} & 7.5\footnotemark[14], 6.0\footnotemark[15] &7.0\footnote{Plank et al.~2011 ~\cite{plank10}} \\
     &    &  & & & & 61\footnote{Nakagawa et al.~2004~\cite{nakagawa04a}}, 59\footnote{Leuenberger et al.~2005 ~\cite{Leuenberger05}}, 50\footnote{Si et al.~2007~\cite{si07}} & 
& \\
      &    &  & & & & 70\footnote{Scarpulla et al.~2009 ~\cite{scarpulla09}}, 65\footnote{Natali et al.~2010 ~\cite{natali10}},  60\footnote{Senapati et al.~2011 ~\cite{senapati11a}}$^,$\footnote{Thiede et al.~2011 ~\cite{thiede11}} & & \\
 & & & &  & & & & \\

TbN  & FM & 34-42\footnotemark[1]$^,$\footnotemark[3] & 6.7-7.0\footnotemark[1]$^,$\footnotemark[3] & 9.3-10\footnotemark[1]$^,$\footnotemark[3]  & FM & 48\footnote{Wachter et al.~1998 ~\cite{wachter98}}, T$_N$=31 (AFM)\footnotemark[18], 44\footnote{Yamamoto et al.~2004~\cite{yamamoto04}} & & 8.5\footnotemark[18]      \\
DyN  & FM & 17-26\footnotemark[1]$^,$\footnotemark[3]& 4.8, 7.4\footnotemark[1]$^,$\footnotemark[3] & 10.5\footnotemark[3]    & FM & 21\footnotemark[11], 25\footnotemark[4] & &   \\
HoN  & FM & 13.3-18\footnotemark[1]$^,$\footnotemark[3]& 6.0, 8.9\footnotemark[1]$^,$\footnotemark[3] & 10.8\footnotemark[3]    & FM & 18\footnotemark[19]
& & \\
ErN  & FM & 3.4-6\footnotemark[1]$^,$\footnotemark[3] & 3, 5.5-6\footnotemark[1]$^,$\footnotemark[3]& 9.4\footnotemark[3] & FM & 7.5\footnote{Nakagawa et al.~2006~\cite{nakagawa06b}}, 6.3\footnote{Meyer et al.~2010 ~\cite{meyer10}}  & 6.0\footnotemark[22]  & 9.0\footnotemark[22]     \\
TmN  & PM &  $\textless$1.3\footnotemark[3] & 7.6\footnotemark[3]    & & & & \\
YbN  & AFM & 0.73, $\textless$2\footnotemark[1]$^,$\footnotemark[3]$^,$\footnote{Degiorgi et al.~1990~\cite{degiorgi90}}& 0.39\footnotemark[3] & 4.44\footnotemark[22], 4.8\footnotemark[3]    & AFM & $\sim$0.5\footnote{Kasuya and Li 1997~\cite{kasuya97}}  & & \\
LuN  &  &  & &    & & & & \\


\end{tabular}
\end{ruledtabular}
\label{tabcomparegd}
\end{table}
\end{center}
\end{widetext}

\begin{figure}
\includegraphics[width=8cm]{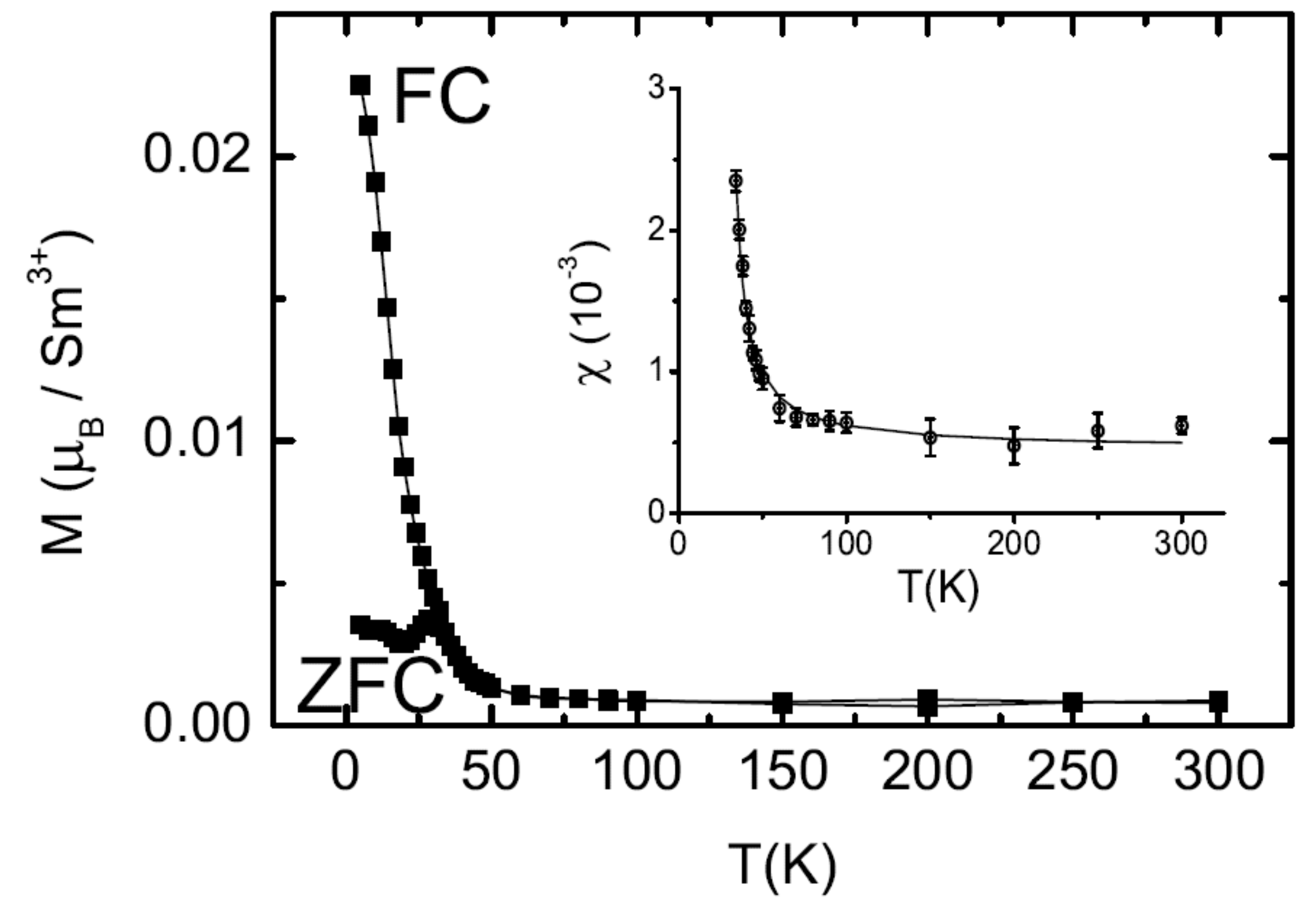}

\caption{Temperature dependence of the magnetisation of SmN after cooling without (ZFC) and with (FC) an applied field of 5 kOe.  The inset shows the result of a fit of the FC susceptibility ($\chi=M/H$) in a Van Vleck approach. From Ref.~\onlinecite{Mey2008}.
\label{SmN_mag}}
\end{figure}

In agreement with previous studies performed on bulk polycrystals, \cite{VogtMatten93,hullig79} semiconducting DyN in thin film form was demonstrated to undergo a ferromagnetic transition below 25~K.\cite{Preston07} In the case of ErN thin films\cite{meyer10} the $T_C$ of 6~K is also in agreement with previous investigations, however more details have been established owing to the (111) texture of the films. The magnetisation
curves of the textured thin films measured in an applied magnetic field parallel and
perpendicular to the plane of the films are not compatible with
the original assumption that the Er magnetic moments would be
perpendicular to the (111) direction. Neither is the magnetisation compatible
with the moments being aligned along the (111) direction, which
would be in agreement with calculations of the single ion anisotropy.\cite{Tram63} Instead, the magnetisation mostly agrees with an alignment of the magnetic moments along the (100) direction, similar to the magnetic moment arrangements found in HoN and DyN.\cite{child63,VogtMatten93}

Among the light RENs (RE=Ce, Pr, Nd, Sm, Eu), SmN and EuN are of special
interest because of the
proximity of excited states.  SmN involves the Sm$^{3+}$ ion, which in the free ion state
($^6H_{5/2}$) presents a magnetic moment $\mu=gJ=0.71~\mu_B$. However,
due to the small energy separation between the ground and lowest excited
state multiplets, the angular momentum $J$ is in most cases not a
good quantum number and the magnetic moment has to be defined as
$\mu=-(<L_z> + 2<S_z>)~\mu_B$. The spin and orbital angular momenta
are equal in magnitude with opposite sign, so that they cancel
each other giving rise to a zero total magnetic moment. A zero total moment is expected
in SmN, and any magnetic order is indeed difficult to evidence.
Previous experiments on bulk SmN were rather in favour of
antiferromagnetic ordering instead of a ferromagnetic one with a small
magnetic moment (see Table \ref{tabcomparegd}).~\cite{busch,stutius,VogtMatten93} 
The recent experiments on thin films definitely show that the
compound undergoes a ferromagnetic transition below
30~K, shown in Figure \ref{SmN_mag}.\cite{Mey2008} Above this temperature, the magnetisation curves
and low field susceptibility are in agreement with a Van Vleck
paramagnetism. In the ordered state, the magnetic hysteresis loop
exhibits an increasingly large coercivity when lowering the
temperature while the magnetic moment remains very small. This is
reminiscent of the well-known behaviour of ferrimagnetic
materials close to their compensation point
. Due to the polycrystalline structure of the
films, there is no saturation of the moment at high field because
of the huge anisotropy of the Sm ion. However, the analysis of the
data within a ferromagnetic scenario is undisputable. In addition, the
nearly zero moment is in agreement with theoretical calculations.
\cite{Larson07} The compound SmN is thus of great interest as a
ferromagnetic material without a fringe field, as has already been
pointed out for the metallic alloy
(Sm$_{1-x}$Gd$_x$)Al$_2$.\cite{adachi99,avisou08} However, unlike (Sm$_{1-x}$Gd$_x$)Al$_2$, SmN is a semiconductor.

\begin{figure}
\includegraphics[width=8cm]{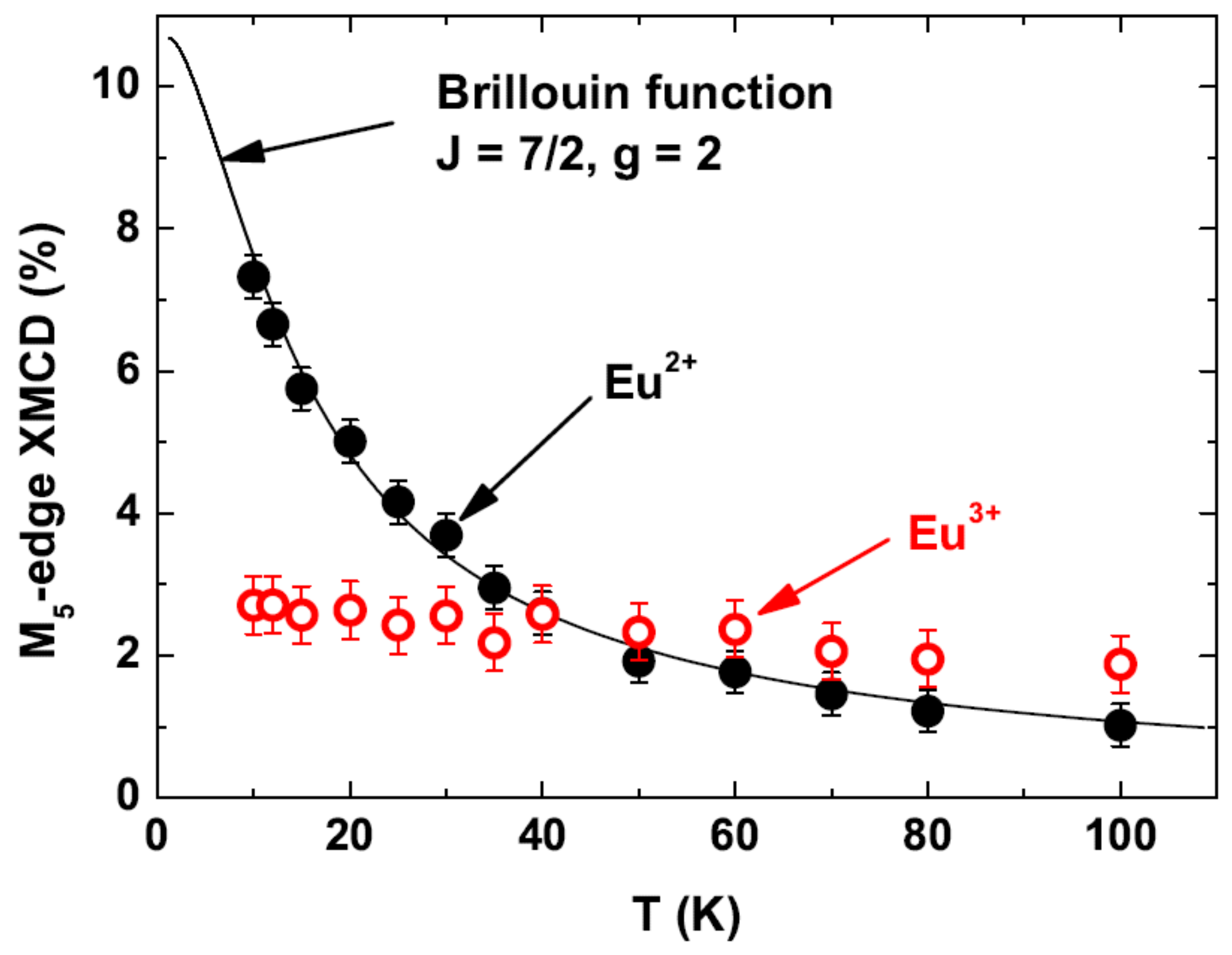}
\caption{Eu $M_5$-edge amplitudes at 1131 eV (black filled circles) and 1134 eV (open red circles) are measures of the moments on the Eu$^{2+}$ and Eu$^{3+}$ ions in EuN, respectively.  The black line is a fit to the Eu$^{2+}$ moment by the Brillouin function appropriate for a half-filled 4$f$ shell in a field of 50 kOe. From Ref.~\onlinecite{Ruck2011}.
\label{EuN_M_edge}}
\end{figure}

\begin{figure}
\includegraphics[width=8cm]{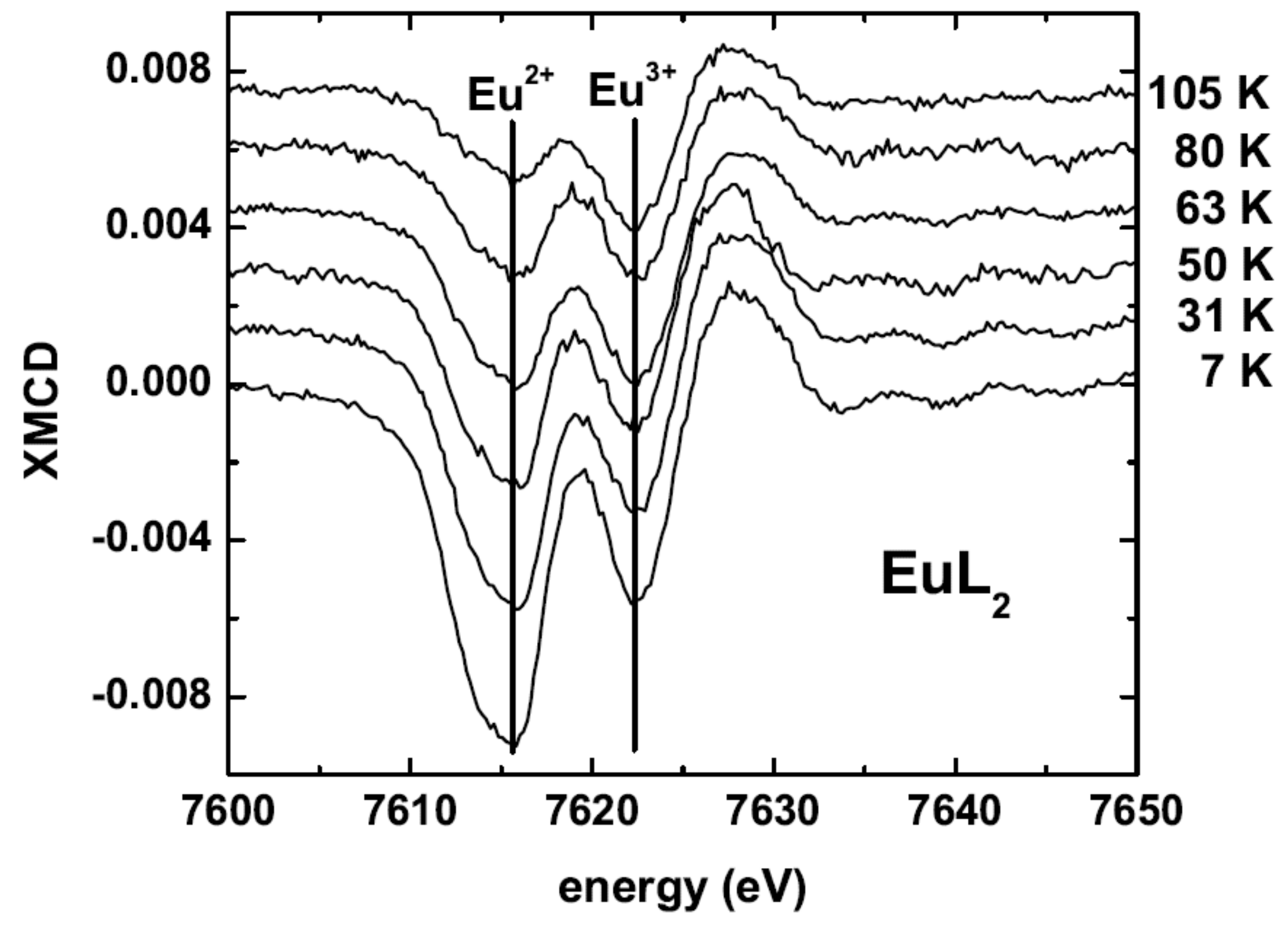}
\caption{Eu $L$-edge amplitudes vs temperature in EuN. From Ref.~\onlinecite{Ruck2011}.
\label{EuN_L_edge}}
\end{figure}

EuN also involves an irregular RE becauseEu in its
free ion trivalent state has no magnetic moment ($J=0$). EuN is thus expected to be a Van Vleck paramagnet, with the Zeeman interaction
mixing several multiplets of Eu$^{3+}$ owing to the weak separation
energy of 460~K between the $J=0$ ground state and the $J=1$ excited
state. Therefore, a constant magnetic susceptibility is expected at
low temperature. Old data were never conclusive because of the
presence of O impurities.~\cite{VogtMatten93} Recent experiments were achieved on epitaxial thin films.\cite{Ruck2011}  Unexpectedly, the magnetisation data
show a Curie-like temperature behaviour of the susceptibility. The
explanation is provided by XMCD experiments, using a combination of
the data from two absorption edges of Eu. The $M_{4,5}$
absorption edge directly probes the magnetic state of the 4$f$
orbitals, while the $L_{2,3}$ absorption edge probes the empty
states of the 5$d$ band, which is expected to be polarised through
the intra-atomic $f-d$ interaction. A small amount of Eu$^{2+}$  is
detected in the $M$-edge XAS spectra, corresponding to about 1\% of
the total Eu signal. The XMCD spectra are dominated by the
divalent impurity signal, owing to the intrinsic high paramagnetic
moment of Eu$^{2+}$ (7$\mu_B$). Figure \ref{EuN_M_edge} shows the temperature
dependence of the $M$-edge XMCD signal for the two valence states, resembling a
Brillouin function for Eu$^{2+}$ and quasi-constant for
Eu$^{3+}$. The origin of these two contributions is found in the
$L$-edge experiments on the same sample. Figure \ref{EuN_L_edge} reproduces the temperature variation of the $L_2$ XMCD spectra. In
agreement with the $M$-edge results, the $d$-orbital divalent
contribution strongly decreases with increasing temperature, while
the trivalent contribution remains practically constant. An
\textit{ab-initio} calculation of the XAS and XMCD spectra was undertaken to
understand this behaviour. The comparison between the experimental
spectra and the simulated one allowed the relationship
between the Eu$^{3+}$ and the Eu$^{2+}$ 5$d$ signal amplitude to be evaluated. Clear evidence for a linear dependence is established, suggesting
that part of the magnetic polarisation of the 5$d$ orbitals in EuN
is due to the Eu$^{2+}$ impurities through interatomic 5$d$-5$d$
exchange.

It is interesting to note a small number of early studies that were reported on N-containing RE compounds with elevated $T_C$ values owing to the inclusion of a significant fraction of dopants replacing the N ions.  GdN$_{1-x}$C$_x$ was reported to be ferromagnetic with a $T_C$ value of 190~K, and material with an additional deficiency of N was reported with a $T_C$ as high as 340~K.~\cite{gambino68,rhyne72}  In addition, replacing up to 30\% of O in EuO with N augments $T_C$ from 69.5~K to as much as 77~K.\cite{chevalier78} This latter result was attributed to a decreasing RE-RE distance with an increasing N concentration, resulting in a stronger ferromagnetic exchange interaction between the RE cations. These reportsfurther support the possibility of enhancing $T_C$ values in the RENs through appropriate control over the strain or carrier concentration by way of doping, similar also to the case of Gd-doping in EuO.\cite{altendorf12}

\subsection{Theories of Magnetism in RENs}\label{magnetism}
In this section, we discuss further in detail studies of the ferromagnetism
in RENs, mostly on GdN because this compound has received the bulk of the recent attention. The magnetism of GdN is usually studied together
with the rest of the Gd pnictide series.

Before we get into a detailed discussion of the LSDA+$U$ results, we want to discuss the question
of why the ferromagnetism of GdN is of special interest.
First, it is somewhat surprising that the compounds which
are definitely semimetallic in the Gd pnictide series - GdP, GdAs, GdSb, GdBi - are all antiferromagnetic,
while GdN, the only one which is now believed to be a semiconductor, is ferromagnetic. Usually,
one finds antiferromagnetism in insulators and ferromagnetism in metals. Second, as pointed out
by Kasuya and Li,\cite{KasuyaLi} the similarity of the ferromagnetic properties of GdN to those of EuO is
surprising. Although both systems have a half filled $f$ shell (because Eu in EuO is divalent),
the position
of the energy levels is quite different. In EuO the $4f$ bands lie above the O-$2p$
bands and thus form the VBM, while in GdN, as we have seen, the majority spin $4f$
bands lie well below the N-$2p$ bands. In a simple model of the ferromagnetism
the exchange interaction between nearest neighbour Gd in second order perturbation theory is given by
\begin{equation}
J_1^{2}=2I_{df}\frac{t_{df}^2}{S\Delta E_{df}}.
\end{equation}
Here $I_{df}$ is an intra-atomic $d-f$ exchange coupling, while $t_{df}$ is a hopping parameter
from $d$ to $f$ orbitals on neighbouring sites, $S$ is the spin and $\Delta E_{df}$ is the
splitting of the $d$ and $f$ levels. This model thus would predict a much smaller exchange
coupling and hence smaller $T_C$ in GdN than in EuO. It is therefore quite surprising that both materials
have similar $T_C$, $\sim$69~K for EuO and 58-70~K for GdN. This observation led Kasuya and Li
to propose a novel exchange coupling scheme\cite{KasuyaLi} arising from 4th-order perturbation theory
and described by
\begin{equation}
J_1^{(4)}=-4 I_{df}\frac{t_{pf}t_{fd}t_{dp}}{S\Delta E_{pf}\Delta E_{pd}^2},
\end{equation}
involving $p-d$ as well as $f-d$ hopping.
Instead of using such intricate perturbation coupling schemes, spin density functional
theory attempts to obtain the exchange couplings directly from the total energy calculations,
as we now describe.

\begin{widetext}
\begin{center}
\begin{table}
\caption{Comparison of exchange interactions in Gd-pnictides (GdX) from different papers.}
\begin{ruledtabular}
\begin{tabular}{lcccc|ccccc}
GdX & \multicolumn{4}{c}{$J_1$ (meV)} & \multicolumn{4}{c}{$J_2$ (meV)} \\
    & Larson\footnote{Larson and Lambrecht\cite{Larson06}:FP-LMTO,$U_f+U_d$} & Duan\footnote{Duan et al.~\cite{duanapl06}:FLAPW,$U_f$ only} & Mitra \footnote{Mitra and Lambrecht \cite{mitra08mag}:FP-LMTP, $U_f+U_d$} & Li\footnote{Li et al.~\cite{dxli97}: expt.} & Larson$^a$ & Duan$^b$ & Mitra$^c$ & Li$^d$ \\ \hline
GdN & 1.84   & 0.86 & 0.42 & 1.74 & $-0.87$ & $-0.14$ & $-0.36$ &  \\
GdP & 1.13   & $-0.17$ & 0.34 & 0.60 & $-1.35$ & $-0.74$ & $-0.82$ & $-0.92$ \\
GdAs& 0.98   & $-0.22$ & 0.12 & 0.22 & $-1.52$ & $-0.91$ & $-1.03$ & $-0.95$ \\
GdSb& 0.80   & $-0.51$ & 0.15 & 0.14 & $-1.95$ & $-1.13$ & $-1.22$ & $-1.63$ \\
GdBi& 0.69   & $-0.66$ & 0.01 & $-0.11$ & $-2.45$ & $-1.37$ & $-1.44$ & $-1.71$ \\
\end{tabular}
\end{ruledtabular}
\label{tabcomparegdn}
\end{table}
 \end{center}
\end{widetext}

The common approach used in several of these papers is to map a generalised Heisenberg Hamiltonian
\begin{equation}
H=-\sum_{ij}J_{ij}{\bf S}_i\cdot{\bf S}_j, \label{eqheis}
\end{equation}
to the total energies of a series of different magnetic configurations, for example
ferromagnetic (FM) order, antiferromagnetic ordering along [001] (alternating layers of
up spin and down spin along a [001] direction,  or so-called AFM-I), antiferromagnetic
ordering along [111] (AFM-II), and other hypothetical ordering schemes.  Within a
model with a certain number of parameters, say first and second-nearest neighbours, one can
then easily write down expressions for the total energy in each magnetic configuration.
Calculating these energy differences from
first principles using LSDA+$U$, one then extracts the exchange parameters from the model. From these exchange parameters,
the Curie or N\'eel temperatures are estimated using either the mean field approximation or some
more sophisticated statistical method, for example the Monte Carlo approach.

To demonstrate the convergence of the model, one typically tries to show that other
configurations beyond the ones used to extract the model parameters are also well reproduced.
For example, Larson et al.~\cite{Larson06} consider also the AFM-III ordering, which consists of alternating 2 layers of up spin with 2 layers of down spin along [001] and show that its energy difference from
the FM case is well reproduced if they determine the nearest and second-nearest neighbour parameters
$J_1$ and $J_2$ from the AFM-I-FM and AFM-II-FM
energy differences. Duan et al.~\cite{duanapl06} include a third-nearest neighbour interaction $J_3$
instead but find it to be small.
There are slight variations to this approach, as some authors apply the method to a quantum
Heisenberg Hamiltonian with $S=7/2$, while others use a classical Heisenberg Hamiltonian.
At the mean field level these map exactly to each other, so it is straightforward to translate
the exchange interactions from one study to another even if the original papers use slightly different
definitions. In the discussion below we will refer to classical Heisenberg spins and assume the
sum is over all sites $i$ and $j$ in Eq. \ref{eqheis}, i.e., each pair is counted twice.

Before we discuss the differences between the results of the different studies, let us
point out that, apart from GdN, all agree on the lowest ground state for the Gd-pnictides to be AFM-II ordering, in agreement with experiment, and that all find GdN to be ferromagnetic,
also in agreement with experiment.

Duan et al.~\cite{duanprl05} emphasized the volume dependence of the properties of GdN. They noted
that GdN in their approach comes out very close to a metal-insulator transition and hence
carefully studied the volume dependence. At the calculated equilibrium lattice constant, they
obtained a semimetallic band structure, but at slightly expanded lattice constant a band gap was obtained. They noted the strong dependence of the exchange interactions on the lattice constant.
As mentioned above, this suggestion was picked up by Leuenberger et al.,\cite{Leuenberger06} who studied strained
GdN experimentally and indeed found a behaviour close to a metal-insulator transition.

In a second paper, Duan et al.~\cite{duanapl06} studied the exchange interactions up to third nearest neighbour interactions in the
Gd-pnictide series and explained the increasingly stronger antiferromagnetic second nearest
neighbor coupling parameter in terms of superexchange via the intervening N. They suggested that the
nearest neighbour exchange interaction is RKKY in nature and obtained a ferromagnetic interaction
for GdN and antiferromagnetic coupling for the other pnictides.

Larson and Lambrecht \cite{Larson06} independently obtained similar results but
used an additional $U_d$ as already mentioned, which turns GdN into a semiconductor even without an
increase in lattice constant. Mitra and Lambrecht \cite{mitra08mag} improved on the results of
Larson  with a more careful method of extracting the exchange interactions.  The key here was
to recalculate the ferromagnetic/antiferromagnetic energy differences in exactly
the same unit cell so that systematic computational errors cancel. They obtained
much better agreement for the exchange interactions than Duan et al.~\cite{duanapl06}
A comparison between these three calculations is given in Table \ref{tabcomparegdn}.

\begin{widetext}
\begin{center}
\begin{table}
\caption{Comparison of N\'eel and Curie-Weiss temperatures obtained by different groups.}
\begin{ruledtabular}
\begin{tabular}{lcccc|cccc}
GdX & \multicolumn{4}{c}{$T_N$ (or $T_C$ in GdN) (K)} & \multicolumn{4}{c}{$T_{CW}$ (K)} \\
    & Larson\footnote{Larson and Lambrecht\cite{Larson06}:FP-LMTO,$U_f+U_d$} & Duan\footnote{Duan et al.~\cite{duanapl06}:FLAPW,$U_f$ only} & Mitra \footnote{Mitra and Lambrecht \cite{mitra08mag}:FP-LMTP, $U_f+U_d$} & Li\footnote{Li et al.~\cite{dxli97}: expt.} & Larson$^a$ & Duan$^b$ & Mitra$^c$ & Li$^d$ \\ \hline
GdN  & 47\footnote{0.7$T_{CW}$} & 34 \footnote{Monte Carlo simulation} & 8$^e$ & 58 & 67 & 37 & 11 & 81 \\
GdP & 22\footnote{0.7 $T_N^{MF}$} & 12 & 13$^g$ & 15.9 & 21 & $-25$ & $-3^f$ & 4.0 \\
GdAs & 24 & 14 & 17 & 18.7 & 10 & $-31$ & $-18$ & $-11.8$ \\
GdSb & 31 & 20 & 20 & 23.4 & $-8$ & $-50$ & $-22$ & $-31.3$ \\
GdBi & 39 & 22 & 23 & 25.8 & $-25$ & $-62$ & $-32$ & $-34.0$ \\
\end{tabular}
\end{ruledtabular}
\label{tabtc}
\end{table}
\end{center}
\end{widetext}

Although different codes were used by the Lambrecht and Duan groups, respectively a full-potential
linearized muffin-tin orbital (FP-LMTO) method and a full-potential linearized augmented plane wave (FLAPW) method, this is not the main origin of the differences.
The band structure methods give essentially converged results for a given choice of
Hamiltonian.  The remaining differences for GdN in particular can thus be clearly traced to
the use of a $U_d$ parameter opening up the band gap, as used by the Lambrecht group.

Besides the N\'eel temperature $T_N$, one can also calculate the Curie-Weiss temperature $T_{CW}$ within
the mean field model and compare both to the experiments. Arguably, the calculations of Mitra and Lambrecht obtain
the trend in both $T_N$ and $T_{CW}$ slightly more accurately.
The predictions of the different groups for the critical temperatures and $T_{CW}$
are compared in Table \ref{tabtc}. It is important
to note here that the experimental values for the exhange parameters were in fact obtained
by Li et al.~\cite{dxli97} from experimental values for $T_{CW}$ and $T_N$ analysed within
mean field theory. The exchange parameters are found using
\begin{eqnarray}
T_N&=&-J_2/k_B, \nonumber \\
T_{CW}&=&(4J_1+2J_2)/k_B
\end{eqnarray}
respectively within the mean field approximation and thus determine directly $J_2$
and $J_1$.
As the $T_{CW}$ values are obtained from the high temperature
susceptibility, the mean field theory is better justified for them than for $T_N$.
The results of Duan et al.~\cite{duanapl06} from both Monte Carlo and mean field methods show that the
actual critical temperatures are typically underestimated by about 30 \% by the mean field theory.
Overall, one can say that the magnetic exchange interactions in the series of Gd pnictides
is fairly well understood on the basis of the LSDA+$U$ method. Similarly, the same approach
was also used for the Eu-chalcogenides\cite{Larsonjpcm06} and in that system also reproduced the trend
in $T_N$ and $T_{CW}$ in good agreement with experiment.
Again, one finds a switch from ferromagnetic to antiferromagnetic AFM-II behaviour in the series,
although the theory predicts the switch to occur between EuO and EuS, while in the experiment
the switch occurs between EuS and EuSe.

Ghosh et al.~\cite{ghosh05} calculated magnetic moments of the Gd monopnictides but not the exchange interactions.
The magnetic moments also show small differences between the different groups.
For example, for GdN, Ghosh et al.~\cite{ghosh05}
report a spin (orbital) moment of 6.79 (0.13) $\mu_B$ on Gd and $-0.14$ $\mu_B$ on N. Duan et al.~\cite{duanapl06} report 7.038 $\mu_B$ as the total magnetic moment on Gd and Larson et al.~\cite{Larson07} give
 an $f$ contribution to the Gd spin moment of 6.93 $\mu_B$, a $d$ contribution to the spin moment of
0.081 $\mu_B$ and a N moment of $-0.083$ $\mu_B$. Mitra and Lambrecht \cite{mitra08mag} report
only Gd-$d$ and N induced magnetic moments of $0.08$ and $-0.08$ $\mu_B$, respectively.
Of course, all these calculations result in approximately 7 $\mu_B$ per Gd. However, different authors have emphasised different
ways of decomposing the total magnetic moment. In principle, one expects a zero orbital moment
for Gd$^{3+}$ in agreement with Larson et al.~\cite{Larson07}
Mitra and Lambrecht \cite{mitra08mag} emphasized the exact cancellation of the induced magnetic moments of the Gd-$d$ orbitals with the N moment in the ferromagnetic but not the antiferromagnetic state. They thus
coined the description of GdN as an \textquotedblleft antiferromagnet in disguise\textquotedblright. In fact, the
Gd-$d$ moments are surrounded by N moments in perfect antiferromagnetic alignment, and vice-versa.

In fact, the basic picture that emerges from all these calculations is that the exchange interactions
in GdN arise from the small induced moments on Gd-$d$ and N-$p$ orbitals which are coupled to the
large $4f$ moment on Gd through intra-atomic exchange coupling. The $4f$ moments themselves are
too localised to have any direct coupling between each other. There are different ways
to describe the effective exchange couplings depending on how one maps these on an effective
Hamiltonian. The common approach is to map them on a classical Heisenberg model with spins only on the Gd atoms. As already mentioned, it is within this approach
that Li and Kasuya \cite{KasuyaLi} introduced
a phenomenological model involving various orders of perturbation theory
and which is also used in the LSDA+$U$ work discussed above.  Unfortunately,
no direct correlation between this perturbation theory and the LSDA+$U$ calculations
has been found, in particular not for the high order theory proposed for GdN.
On the other hand, Duan et al.~\cite{duanapl06} find some perturbation theory
justification for the trend with anion diameter in $J_2$, which
they describe in terms of a superexchange via the anions. The larger the anion, the larger
the overlap and hence the increasing trend in magnitude of the $J_2$ with increasing anion size, which is in fact
observed in all LSDA calculations. Unfortunately, the trends in $J_1$ are less
well understood and there is poorer agreement between the different calculations
of $J_1$.

However, one may also calculate the exchange interactions in a model
including spins on both Gd and N. In fact, Mitra and Lambrecht \cite{mitra08mag} analyse this
different approach using the exchange interactions calculated using Liechtenstein's
linear response approach.\cite{Liechtensteinjmm87} Because this model is evaluated within the
atomic sphere approximation (ASA) to LMTO, it includes empty spheres on the interstitial
sites (introduced to obtain a better space filling with spherical potentials).
In many ways, this is a more direct approach. It calculates the exchange interactions
directly from the Green's functions for a reference magnetic configuration,
usually the ferromagnetic ground state. It considers small transverse fluctuations
from this model to define the exchange interactions, rather than the total energy differences between different collinear spin arrangements. This model makes no \textit{a priori} assumptions on what is
the correct Hamiltonian or which spins to include.  The analysis of Mitra and Lambrecht \cite{mitra08mag} shows that this model justifies the simpler approach including only Gd spins
for all Gd pnicitides other than the nitride. For the nitride, one finds exchange interactions
of similar magnitude between Gd, N and empty spheres. Thus, they suggest, one
perhaps should view GdN as a ferrimagnet rather than a ferromagnet.

Interestingly, these different viewpoints lead to similar conclusions about the critical
temperature in GdN, namely that it is very small.
Within a mean field approach to the ferrimagnetic model (not incuded in that
paper but worked out later by the author), one obtains
19~K as the critical temparature, whereas the spin on Gd only model gave 11~K in Mitra and
Lambrecht's work. While the improved computational details eventually gave a convergence
between the different group's results for the exchange interactions in the other
Gd pnictides, the differences became more obvious for GdN and in fact worsened the
agreement with experiment. The estimate of Larson \cite{Larson06} gave a $T_C$
of 67~K, in good agreement with the experiments which gave values typically around 70~K,
although Li et al.~\cite{dxli97} place $T_C$ at 58~K.  Using a more accurate Monte Carlo approach,
Duan et al.~\cite{duanapl06} obtained $T_C$=34~K. Even if we apply a Monte Carlo vs.~mean field reduction factor of
0.7 to Larson's results we obtain $T_C$=47 K. In Mitra's results, we obtain only 11~K in mean field
or 8~K if we include the Monte Carlo reduction factor. Clearly the difference between the calculations of Mitra and
Duan must result from the fact that in Duan's case GdN has a semimetallic band structure
while in Mitra's case the band structure is semiconducting. However, we know that a semiconducting state
is in agreement with experiments.\cite{trodahl07,Granville06}  While an RKKY
carrier-mediated explanation for the ferromagnetic nearest neighbour interaction makes some sense
in a semimetal, it does not in a strictly insulating material. Even so, it is clear that the
interactions must come from the $d$ electrons. One can thus think of a sort of virtual
excitation to the $d$ like conduction band which then mediates the interaction between the
localised $4f$ moments. This is pretty much the vision incorporated in Kasuya and Li's
perturbation theory.\cite{KasuyaLi}
We are faced with the problem that improving the methodology led to better agreement with experiments for the
other pnictides but worse agreement for the nitride! For this reason Mitra and Lambrecht \cite{mitra08mag} also checked the case of
Gd metal with the linear response method and arrived at $T_C$=285~K,
(including the 0.7 correction factor)
in quite good agreement with the experimental value of $T_C$=297~K.
This led the authors to the conclusion that pure GdN must have a much lower $T_C$
and that the disagreements with experiment might result from extrinsic factors such
as $V_N$.

A first attempt at studying the latter was reported by Punya et al.~\cite{Punyamrs}
This study showed that $V_N$ indeed increases the nearest neighbour interaction and
reduces the second-nearest neighbour antiferromagnetic interaction, leading overall to an
increase in $T_C$ by about a factor 2. This, however, does not yet take into
account the possibility of longer range, more itinerant interactions from the presence of
free carriers in the Gd $d$-band.

Sharma and Nolting \cite{Sharma10} present a separate study of the free carrier contribution to the
exchange interactions in GdN. Their many-body Green's function formalism is
somewhat intricate and its relation to the Liechtenstein linear response formalism is not entirely
clear. For example, simply shifting the Fermi level in the latter approach to simulate
free carriers did not lead to a strong increase in $T_C$.\cite{mitra08mag}
In Sharma's approach the Gd $f$ electrons are represented as localised spins only
and the band structure of the remaining electrons is extracted from an LMTO Hamiltonian
without explicit inclusion of the $f$ electrons.\cite{Sharma08}
The coupling to the localised spins is then treated
with a more advanced Green's function method than the simple linear response. Nonetheless, this work presents some evidence that the discrepancy between theory for perfect insulating GdN and
experiment may indeed stem from the carrier-mediated interactions.

The result of these studies is to bring us almost full circle to the assertions from the early work on GdN by
Wachter and Kaldis,\cite{WachterKaldis} who claimed that \textquotedblleft semiconducting GdN (n/Gd$\ll$ 10$^{-3}$), if it were possible, would be an antiferromagnet in zero field. However, the lowest carrier
concentration obtained until now makes GdN a semimetal\textquotedblright.  While the recent theory work on pure
GdN which best agrees with the known experimental data, namely that it is in fact a semiconductor,
still gives a ferromagnetic rather than antiferromagnetic state, it is one with a very low
$T_C$.

Evidence for carrier mediated ferromagnetism and even a second antiferromagnetic
phase was recently reported by Senapati et al.~\cite{senapati11a} in films with $T_C$=60~K.
On the other hand, in another recent experimental study,\cite{yoshitomi11}
a much lower $T_C$ of 37~K was reported for GdN. This value was based on using Arrott plots
and on fairly thin films of 30-95~nm thick. They attributed the differences to the other
recent studies by Granville et al.~\cite{Granville06} and Leuenberger et al.~\cite{Leuenberger05} to differences
in stoichiometry, film thickness and grain size.

In a recent hybrid functional calculation (using the HSE functional\cite{HSE03,HSE06}), somewhat higher $T_C$ values
of 55~K in the mean field approximation and 42~K in a random phase approximation were reported.\cite{Schlipf2011}
So perhaps there is finally some convergence between theory and experiment on the value of
$T_C$. The exchange interactions $J_1$ and $J_2$ as defined earlier in the HSE calculation
are found to be 1.09~eV and 0.17~eV respectively. So, in contrast to the LSDA+$U$ results,
they find a positive $J_2$ and somewhat larger $J_1$, but we do not yet know how effective HSE is for the
other Gd pnictides or EuO.

Other magnetic properties are also of interest and in this context it is important to mention
the magnetocrystalline anisotropy energy (MAE) determined in the recent study of Abdelouahed and Alouani\cite{alouani09}
for Gd, GdN and GdFe$_2$. They find a different easy axis in fcc Gd (along [111]) than in GdN
(along [001]). They find the right order of magnitude for the MAE in hcp Gd and show that
GGA+$U$, GGA and GGA with open core calculations lead to results that differ by orders of magnitude.
Although they did not extract explicitly the anisotropy parameters, the difference between the
hard and easy axes appears to be of order 30~$\mu$eV in hcp Gd, 5~$\mu$eV in fcc Gd and only
0.5~$\mu$eV in GdN. They thus predict a rather large tunability of the magnetic anisotropy
in GdN$_x$ with various amounts of N.

As for the other RENs, a systematic study of ferromagnetic versus antiferromagnetic
ordering has not yet been done. Johannes and Pickett\cite{Johannes05} studied the exchange couplings
for EuN and EuP. Their perspective is that Eu$^{3+}$ is a nominally non-magnetic ion with $J=0$,
as discussed earlier. Their LSDA+$U$ band structure is similar to the ones
recently obtained in the GW approximation.\cite{richter11} In agreement with the SIC calculations
of Horne et al.,\cite{Horne} they find a trivalent solution, but in disagreement with
Larson et al.,\cite{Larson07} they find a way to satisfy Hund's rules with maximum $L_z$.
As in the case for the other RENs, they find a semimetallic band structure for EuN when no explicit corrections
are made to the RE-$d$ bands. They find the ferromagnetic energy to lie 4.9~meV below the AFM-II
configuration. This is comparable to the results for GdN obtained by Larson.\cite{Larson06}
The paper by Johannes and Pickett \cite{Johannes05} on EuN and EuP includes an interesting
discussion of the magnetic exchange interactions, including a more complex effective Hamiltonian
than the Heisenberg Hamiltonian, which explicitly includes spin-orbit coupling terms.

\section{Electronic and optical properties}\label{optics}

It is in the determination of the electronic behaviour, the electronic band structures, where the recent thin-film studies of the RENs have made the most important advances. As discussed in the introduction the reasonable agreement about the magnetic states of most RENs found in 20th century studies was in direct contrast to the uncertainties about their electronic band structures and transport properties. In Section \ref{theory} the most recent band structure calculations were presented, and here we review the experimental evidence for those band structures and for the conducting states; are they semiconductors or semimetals? In view of both ambient-temperature electronic and low-temperature spintronic potential, it is of interest to answer these questions for both the magnetically ordered and the paramagnetic phases, and to establish the effects of magnetic order on the transport properties. Thus in this section we review more recent transport studies probing the states near the Fermi level and optical and X-ray spectroscopy investigations used to probe the band gaps and states deeper into the valence and conduction bands.

Historically there was an enormous variation in transport measurements. It was realised that defects played a defining role, which itself is the signature of a semiconductor. Furthermore, essentially all results found in early studies \cite{Sclar,hullig79} showed low-energy, presumably free-carrier absorption and an enhanced absorption at higher energy interpreted as signalling interband edges in the 0.7-2.9~eV range. There was substantial disagreement even among measurements on any one of the RENs as regards the band gaps as well as the density of free carriers. The gaps reported are shown in Table \ref{optgaps}, based on an assortment of experimental works.

\begin{table}
\caption{Interband edges in the paramagnetic phase ($T\textgreater$\textit{}$T_C$)}
\begin{ruledtabular}
\begin{tabular}{l | c | c }
REN & \multicolumn{2}{c}{Direct gap (eV)} \\
& pre-1994 & 1994-present \\

LaN & 0.82\footnote{Hulliger 1979 ~\cite{hullig79} and refs. therein} &     \\
CeN &      & 1.76\footnote{Xiao and Takai 1998~\cite{xiao98}}\\
PrN & 1.03\footnotemark[1] &     \\
NdN & 0.80\footnotemark[1] &     \\
PmN &      &     \\
SmN & 0.70\footnotemark[1] &     \\
EuN & 0.76\footnotemark[1] & 0.95\footnote{Richter et al.~2011~\cite{richter11}}    \\
GdN & 0.98\footnotemark[1],0.85\footnotemark[1] & 4.1\footnote{Shalaan and Schmidt 2006~\cite{shalaan06}},1.3\footnote{Trodahl et al.~2007~\cite{trodahl07}}$^,$\footnote{Yoshitomi et al.~2011~\cite{yoshitomi11}} \\
TbN & 0.80\footnotemark[1]$^,$\footnote{Bommelli et al.~1995~\cite{bommeli95}}$^,$\footnote{Wachter et al.~1998~\cite{wachter98}}   &  0.80\footnote{Bommelli et al.~1995~\cite{bommeli95}}$^,$\footnote{Wachter et al.~1998~\cite{wachter98}}   \\
DyN & 0.91\footnotemark[1],0.95\footnotemark[1],2.6\footnote{Sclar 1964~\cite{Sclar}} & 1.2\footnote{Azeem et al.~2012~\cite{Azeem2012}} \\
HoN & 1.05\footnotemark[1],0.73\footnotemark[1]     & 1.88\footnotemark[7], 1.48\footnote{Brown et al.~2012~\cite{brown12}} \\
ErN & 1.2\footnotemark[1],1.3\footnotemark[1]  &  2.4\footnotemark[7] \\
TmN & 1.10\footnotemark[1]  &     \\
YbN & 1.03\footnotemark[1],1.5\footnotemark[1],1.4\footnote{DeGiorgi et al.~1990~\cite{degiorgi90}} &  \\
LuN & 1.55\footnotemark[1],1.6\footnotemark[1]     &  \\
\end{tabular}
\label{optgaps}
\end{ruledtabular}
\end{table}

In 1990 there was a thorough investigation of YbN,\cite{degiorgi90} suggested at the time to promise a heavy fermion state. The results were interpreted to indicate YbN is a semimetal with an optical gap of 0.1~eV, now with a low carrier concentration absorbing relatively weakly at lower energies. The study in this case was extended to well above the gap, reporting several higher-energy interband transitions. Even to date this is the only of these materials to have anything more than an optical band gap reported. It is interesting that the same group reported metallic conduction in TbN, with an absorption onset near 0.75 eV that they interpret as a plasma edge.\cite{wachter98} 

We start in the next section with a review of transport studies, which investigate the excitations at low energy where the differentiation between semiconductors and semimetals is most apparent. Following that discussion we review the few optical studies that have been reported, mostly to determine optical band gaps. XAS/XES investigations of the full band structure, delineating the partial densities of filled and empty states, are then reviewed in Section \ref{xray}. Although these do not have the resolution to signal a clear gap between valence and conduction bands, they show considerable detail in the overall density of states and offer a more complete comparison with gross behaviour of the calculated band structures.

\subsection{Transport and the optical band gap}
\subsubsection{GdN}

\begin{figure}
\includegraphics[width=8cm]{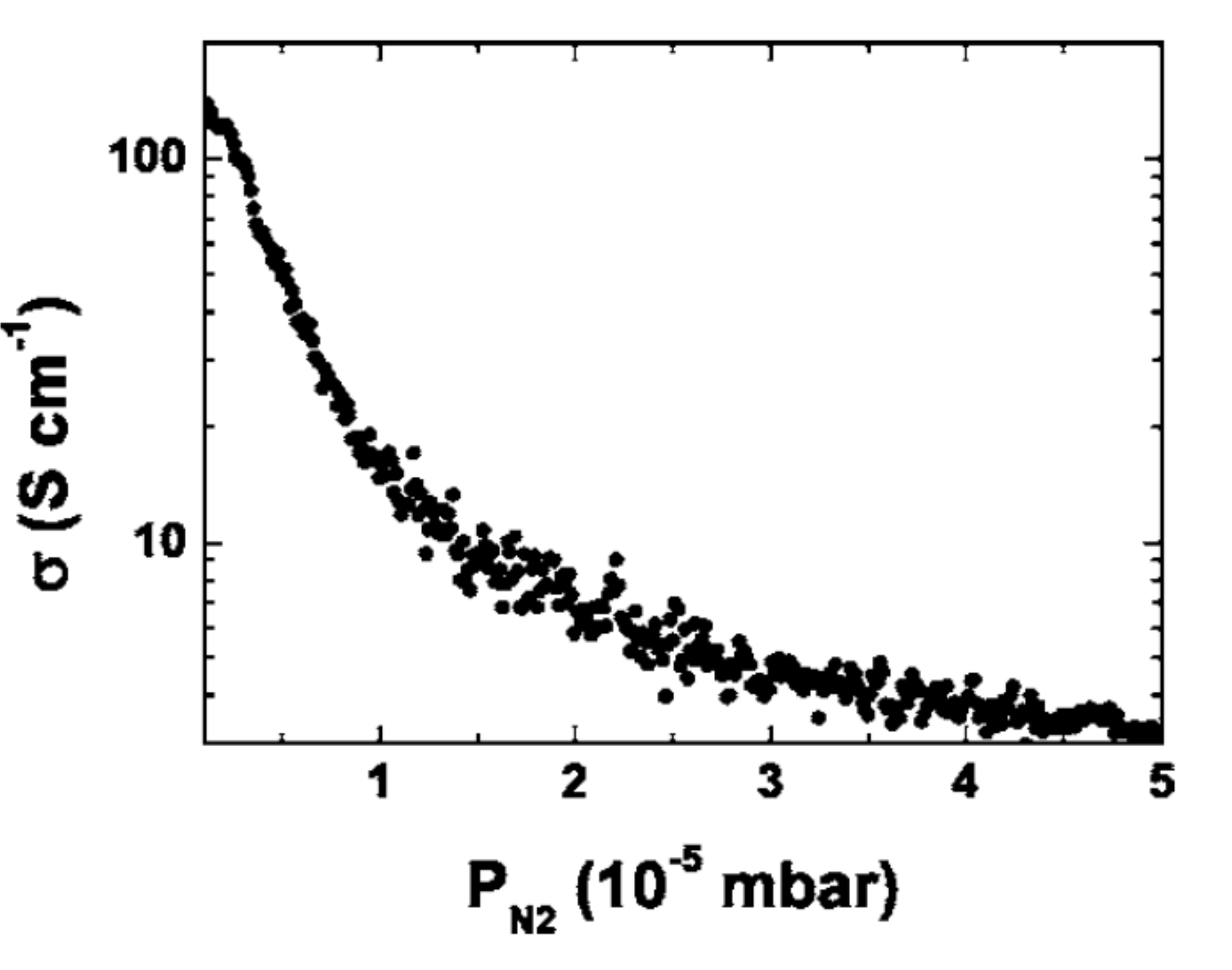}

\caption{The conductance of GdN as a function of the pressure of N$_2$ during deposition, recorded during a growth under a continuously diminishing pressure. From Ref.~\onlinecite{Granville06}.
\label{conduc_press}}
\end{figure}

Recent conductivity measurements of GdN still show considerable variation among different growth conditions, but nonetheless the results are within a somewhat smaller range than historical data. Reported ambient-temperature values for GdN range from 10$^{-4}~\Omega$cm in high-temperature grown epitaxial films\cite{ludbrook09,scarpulla09,natali10} to 1~$\Omega$cm in polycrystalline films grown at lower temperatures.\cite{Granville06,plank10} The dominant dopants are $V_N$, as has been clearly demonstrated by the dependence of the conductivity on the partial pressure of N$_2$ during growth (Figure \ref{conduc_press}). It is expected that each vacancy will bind either one or two electrons, thus acting as either single or double donors at all temperatures, with the carrier concentration rising as the bound electrons are released at higher temperature.\cite{Leuenberger05,Granville06,plank10} The few Hall measurements that are reported confirm that the carriers are electrons in the conduction band.\cite{ludbrook09,scarpulla09,natali10} The reaction at the surface of deposited Gd is slow, so that the ratio of N$_2$/Gd flux must exceed 100 to achieve conductivities as low as 10 Scm$^{-1}$. Such low conductivities suggest carrier concentration of order 10$^{18}$-10$^{19}$ cm$^{-3}$.\cite{plank10} It is tempting to assign that also to the order of $V_N$ density, though the possible existence of traps would prevent accumulation of all donated electrons in the conduction band. Epitaxial films are grown at higher temperature, for which the low formation energy of $V_N$, along with the likely reduced concentration of electron traps, ensure they are more heavily doped than ambient-temperature grown polycrystalline films.\cite{Punyamrs} Hall effect data on the more conductive films show electron concentrations of order 10$^{20}$-10$^{21}$ cm$^{-3}$, with at most a very weak temperature dependence that is all but masked by the field dependence of the extraordinary Hall effect.\cite{ludbrook09,natali10} There appear to be no Hall effect data on less conductive films, but evidence for carrier freeze-out comes from strongly activated conductivities in the paramagnetic state.\cite{plank10} 

\begin{figure}
\includegraphics[width=8cm]{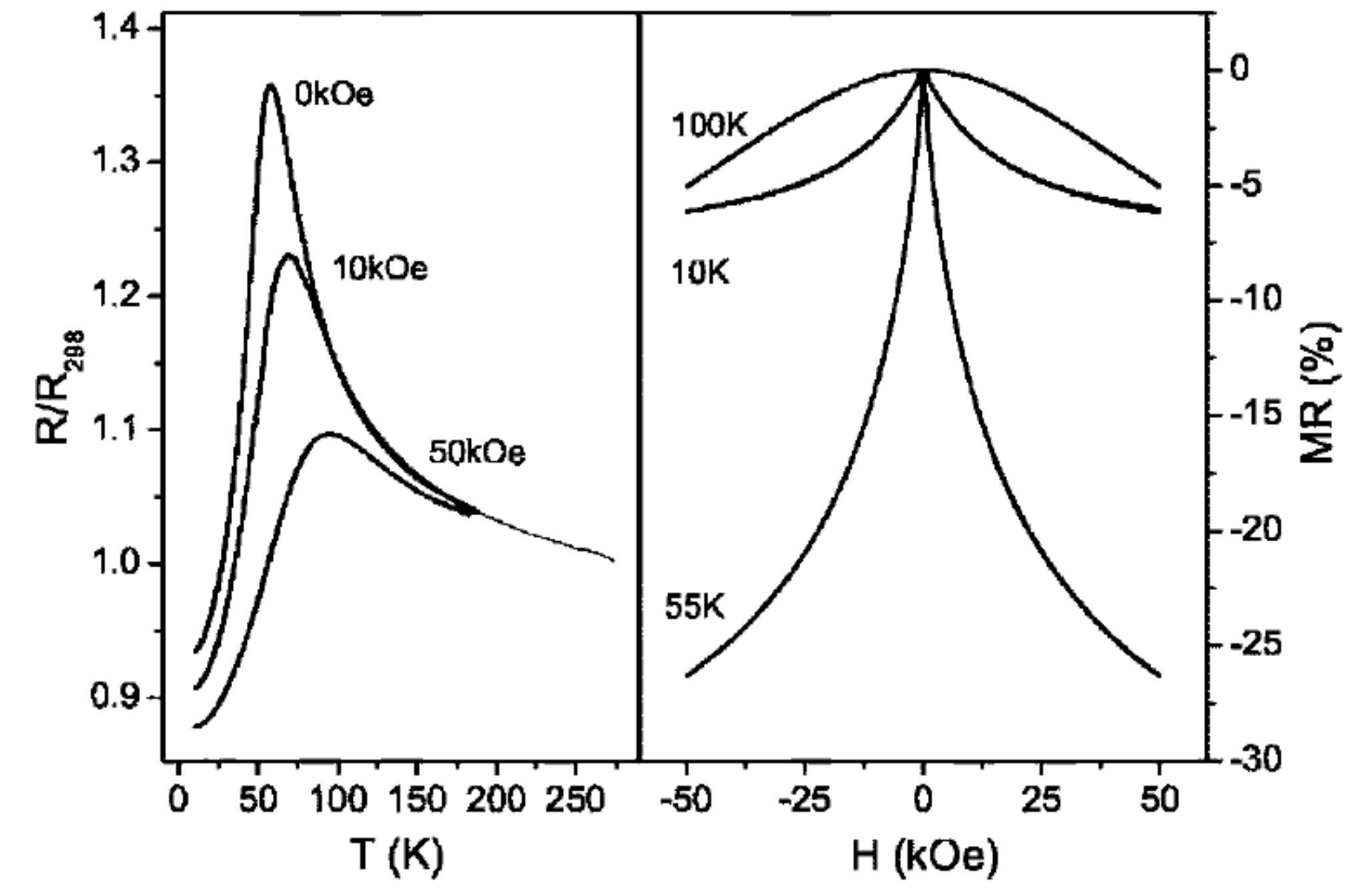}
\caption{Resistance (normalised to the value at 298 K) of a 1000 \AA~GdN film as a function of temperature in different magnetic fields (left panel). The resistivity at 298 K is $\sim$10~m$\Omega$~cm.  Right panel: Magnetoresistance ($R(H)-R(0))/R(0)$ vs magnetic field at temperatures below, near, and above $T_C$. From Ref.~\onlinecite{Leuenberger05}.
\label{GdN_MR}}
\end{figure}

The temperature coefficient of resistance (TCR) varies from weakly positive to strongly negative as the ambient-temperature conductivity falls, signalling thermally activated transport. However, conventional semi-metallic behaviour has also been reported in films with high carrier density.\cite{ludbrook09,scarpulla09,plank10} 
A well-established resistive anomaly forms a peak at $T_C$, as seen in Figure \ref{GdN_MR}, with the resistivity falling in the ferromagnetic phase by up to an order of magnitude.\cite{Leuenberger05,Granville06} Both metallic and semiconductor signs of TCR are reported below $T_C$.\cite{Leuenberger05,Granville06,ludbrook09,scarpulla09,natali10,plank10} The data, and especially their sensitivity to the growth temperature and N$_2$ pressure, suggest that GdN is a semiconductor in both the paramagnetic and ferromagnetic phases. 

A sizeable magnetoresistance has been observed in GdN by two groups.~\cite{Leuenberger05,ludbrook09} At 100~K, the non-saturating magnetoresistance reaches 5\% in a field of 50 kOe. As shown in Figure~\ref{GdN_MR}, with an applied field the anomalous peak in the resistivity seen at $T_C$ is reduced and shifted to higher temperatures. The result is an enhanced magnetoresistance near $T_C$ which approaches 35\% in epitaxial films.\cite{ludbrook09}

\begin{figure}
\includegraphics[width=8cm]{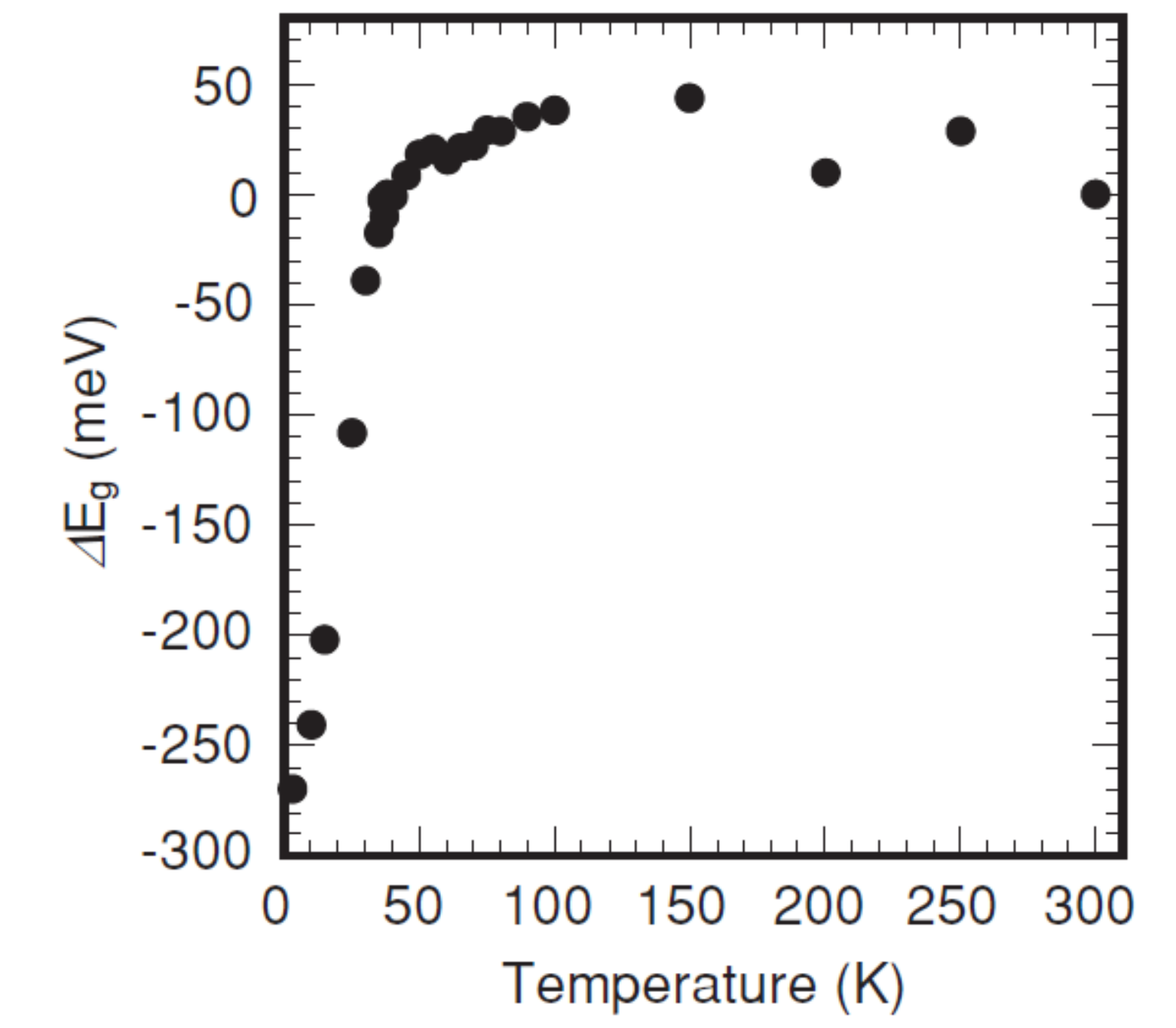}
\caption{Temperature dependence of the direct band-gap energy of 95-nm thick GdN. $\Delta$%
$E_g$ corresponds to the change in the band-gap energy versus the room-temperature value. From Ref.~\onlinecite{yoshitomi11}.
\label{GdN_gap}}
\end{figure}

A clear ambient-temperature (paramagnetic) optical absorption edge of 1.3~eV in GdN has been reported by two groups\cite{trodahl07,yoshitomi11} in the past few years. Both studies also found a red shift of the gap to 0.9-1.0~eV deep in the ferromagnetic phase, (Figure \ref{GdN_gap}), a result of the narrowed majority-spin gap in the ordered spin state. Very recently, evidence was found also for a red shift of both majority-spin and minority-spin gaps in an AlN/GdN/AlN heterostructure.~\cite{Vidyasagar2012} It is likely this narrowing leads to an enhancement of carrier concentration as evidenced by the resistive anomaly at $T_C$. The paramagnetic and ferromagnetic gaps have combined to tune the value of $U_d$ in a LSDA+$U$ band-structure calculation. The refined spin-ordered band structure then shows a finite but relatively small absolute gap of about 0.4~eV between the majority-spin VBM at $\Gamma$ and the CBM at $X$.\cite{trodahl07} Sub-gap absorption in both optical studies seems to be small, implying carrier densites of less than 10$^{21}$ cm$^{-3}$, in agreement with a doped semiconductor or very low-carrier-concentration semimetal. Interestingly, the gap is close to those measured in early studies, but in enormous disagreement with the 4.1 eV claimed more recently for GdN on films that had not been protected by a passivating cap.\cite{shalaan06}

\subsubsection{Other RENs}

There appear to be only four other RENs for which recent optical or transport measurements have been reported. An optical gap of 0.95 eV was measured for EuN,\cite{richter11} which has been of theoretical interest for its possible divalency and for a suggested hidden spin order in the more likely trivalent state.\cite{Johannes05} The existence of the interband edge and the relatively weak free carrier absorption below the gap are strong evidence that EuN is non-metallic, in direct disagreement with some of the calculated band structures.\cite{richter11} CeN is reported to be metallic,\cite{lee03} but in contrast it has also been reported to be a low-conductivity semiconductor with an optical gap of 1.76 eV.\cite{xiao98} HoN has been reported to show optical gaps in the 1.48 eV to 3.32 eV energy range, but with substantial absorption below the gap suggesting a very heavy doping, and possibly even metallic carrier densities.\cite{joshandjames}  Most recently, DyN has been shown to have an optical gap of 1.2~eV with a blue-shift where $V_N$ are present associated with the Moss-Burstein effect.~\cite{Azeem2012}

Recent temperature-dependent resistivities have been reported for SmN,\cite{Preston07} DyN\cite{Preston07} and ErN.\cite{meyer10} All three show resistivities typical of heavily doped semiconductors, and negative TCRs with anomalies at their $T_C$ that are very much weaker than found in GdN.

Sumarising the transport/optical evidence, then, there is strong evidence that SmN, GdN, DyN and ErN are semiconductors. EuN shows a clear interband edge at 0.9 eV, with sufficiently weak subgap absorption to suggest that it is semiconductor. HoN shows a clear interband absorption edge charateristic of a semicondutor but the free carrier absorption at lower energy in the one sample that has been reported leaves its conduction state uncertain. The conducting state of CeN is uncertain. There are as yet no clear data for the rest of the series. 

\subsubsection{Theoretical advances, band gaps}

We have already to some extent discussed the band gaps in the previous sections. Here
we resummarize the results with this focus.

The SIC calculations of Aerts et al.~\cite{Aerts2004} predicted both half-metallic (PrN-GdN), insulating
(TbN-HoN) and semimetallic (ErN-YbN) behavior would occur in the REN series. As already
mentioned, most of the LSDA+$U$ calculations find that when only $U_f$ is included, the RENs
have a small band overlap and are thus semimetals. This continues the trend of the other
RE-V, in particular the Gd pnictides and ErAs.\cite{Petukhov96,Larson07}
Based on the early experimental work,\cite{Wachter79} Larson et al.~\cite{Larson06}
pointed out the need for a $U_d$ parameter or shift of the $d$ bands to open a gap.
This seemed also justified based on Lambrecht's analysis\cite{Lambrecht2000} of scaling the shifts inversely
proportional to the dielectric constant from known Fermi surface properties of the semimetallic
pnictides. Using this approach, Larson et al.~\cite{Larson07} found a systematic upward trend of the
direct band gap at $X$ from LaN to LuN both in the experimental data and in the calculations. This correlated with the
decreasing lattice constants. However, the fluctuations of the individual values around this
general behavior are fairly large and it is not clear if these are due to experimental
uncertainties or whether the LSDA+$U$ level of theory is able to capture these details.
There is certainly no good agreement in the fluctuations of theory and experiment around the
general increasing trend.
One of the main successful predictions of the theories including $f$ electrons explicitly
and in contrast to the earlier models treating $f$ electrons as core electrons, is the
prediction of a gap red-shift in GdN, similar to that occuring in EuO. The clear experimental confirmation of this redshift came  from Trodahl et al.~\cite{trodahl07}
and led to a refinement of the $U_d$ parameter for GdN. It is not clear however how
good the approximation is to keep this $U_d$ parameter constant across the REN series.
The commonly made approximation of averaging the majority and minority spin gap
to obtain the effective gap above $T_C$ was explicitly tested with
non-collinear spin calculations with randomly pointing spins by Mitra. These results were
included in Ref.~\onlinecite{trodahl07}.
The HSE \cite{Schlipf2011} and QSGW calculations\cite{Chantis07} both predict a very small gap
in GdN when applied at the experimental lattice constant.

The general band structure picture is thus well accepted. An indirect gap very close to zero
between majority-spin valence bands at $\Gamma$ and majority-spin conduction bands at $X$
is followed by a larger minority-spin gap. In some calculations this gap is slightly negative
and in others slightly positive depending on the details of the method.
The optical absorption however is dominated by the smallest direct gap at $X$. Above the magnetic
ordering temperature, one presumably observes an average of the smaller majority-spin and higher minority-spin gaps. In earlier work that absorption threshold was placed
at 0.98~eV in GdN, while in the more recent measurements it is found to be at about 1.3~eV.
Below $T_C$, the spin-polarization results in a smaller gap between
the majority-spin states. The theory and experiment agree pretty well on this redshift which
is found to be 0.3-0.4~eV. These predictions are common to the LSDA+$U_f+U_d$ theory, the
HSE and the GW. For the other RENs, there are fewer detailed studies available
to check experiment and theory. However, the anomalous band structure predicted by LSDA+$U$
for EuN with a band crossing the Fermi level is not supported by the experiment or by the
GW calculations.

It would be quite interesting to find direct confirmation in the optical spectra of these
different spin-up and spin-down bands but this has not yet been possible experimentally.

\subsubsection{Theoretical advances, interband optical response functions}

Higher energy interband optical transitions have been measured for only YbN,\cite{degiorgi90} but a theoretical treatment calculating the optical response functions for the RENs has been reported by Mitra.\cite{mitra08opt} These could provide detailed information
on the band structure when spectroscopic ellipsometry data become available in the future.
In particular, they analyse group theoretical selection rules and provide the only
fully symmetry-labelled band structure for GdN.

A study of optical response functions was also
carried out by Ghosh et al.~\cite{ghosh05} The two papers however
differ rather strongly in details. Ghosh used two slightly different LSDA+$U$ implementations.
One simply adds $U_f$. The other includes a smaller $U_f$
but additional shifts of the $d$ and $f$ bands adjusted separately the position of the
empty and filled $f$ states with respect to the $d$-bands. This latter calculation gives a small
gap in GdN but no attempt was made to adjust $U_d$ to reproduce the experimental band gap.
 They report calculated optical conductivity, dielectric functions and reflectity for the entire series of Gd pnictides. For GdN, their optical conductivity shows a broad
spectrum peaking at about 8~eV, similar to the results of Mitra and Lambrecht,\cite{mitra08opt} which however, peak
at 10~eV.  They identify a doublet of peaks at 3-4~eV with N-$p$ to Gd-$d$ transitions and a
peak at 7~eV with occupied Gd-$f$ to empty Gd-$d$ transitions.
In contrast, Mitra and Lambrecht provide a much more
detailed discussion of the fine structure of the dielectric function. They instead
assign a peak at 7~eV in the dielectric function to a transition from the valence band to unoccupied $f$
bands because they only see this peak in the minority-spin channel. On the other hand, they
find no strong optical transitions from the occupied $f$ bands to the conduction band.

The reflectivity spectra look very different, because in Ghosh et al.~\cite{ghosh05}
these are dominated by the strong peak at low energy due to the zero gap in their calculation.
Although not very clearly stated, it appears that they include the intraband Drude
contribution in these spectra which leads to a high reflectivity of nearly 100 \%, dropping down to about
25 \% within less than 1~eV.  In contrast, Mitra's results, which do not include
intraband transitions and has a finite
gap, show a much lower reflectivity at zero energy, also near 25\%, but then shows the
details of this peak at higher energies. In Ghosh's figures, these details are not visible
in their results. The results of the reflectivity of GdAs and GdSb by Ghosh et al.~\cite{ghosh05}
agree well with experiment. Given that in real GdN there is always a large free carrier
concentration, their calculated spectra may still be quite relevant for the
overall expected behavior of the reflectivity.

\subsubsection{Magneto-optics}
Interestingly, Ghosh et al.~\cite{ghosh05} also discuss the magneto-optic Kerr effect in
the Gd compounds. Remarkable magneto-optic results were earlier discovered in
CeSb\cite{Pittini96} and CeS\cite{Pittini97} and successfully addressed with LSDA+$U$ calculations by
Liechtenstein et al.,\cite{Liechtenstein94} but remain to be studied experimentally in the nitrides.  The optical properties of CeN were studied by Delin et al.~\cite{Delin97} and provide
evidence for the $4f$ band formation in CeN within the LSDA+$U$ method.

\subsection{X-ray spectroscopies}\label{xray}
\subsubsection {Experimental results}

In view of the considerable uncertainties of the theory, and the need to accurately locate the 4$f$ electron energies, validation of the higher-energy electronic structure by experiment is important. Here again the majority of available X-ray photoelectron (XPS), absorption (XAS) and emission (XES) spectroscopic data are for GdN. XPS results by the G\"{o}ttingen group provided the first information about the filled states, locating the Gd-4$f$ levels about 8~eV below the Fermi level.\cite{Leuenberger05} The same study investigated the N-$p$ projected density of empty states by XAS at the N $K$-edge, and the Gd-4$f$ empty states by Gd $M$-edge absorption. The XMCD signals at both edges were also investigated, as discussed above. The $K$-edge data were compared with computed results by Aerts et al.,\cite{Aerts2004} providing the very first theory-experiment confrontation for the higher-energy structure in the conduction band density of states, with the results serving to drive further experimental and theoretical work in the search for close agreement between the two. The $M$-edge data reflect primarily the atomic multiplet structure in the 4$f$ shell, showing no discernable solid-state effects. The Gd $L_{2,3}$ spectra were measured by Leuenberger et al.~\cite{Leuenberger05,Leuenberger06} and compared with theory by Abdelouahed and Alouani\cite{alouani07} and Antonov et al.~\cite{Antonov07}

Preston et al.~reported XAS/XES spectra from epitaxial GdN films.\cite{preston:032101} The spectra were compared with N-$p$ like partial densities of states from LSDA+$U$ calculations tuned to the measured optical gap by Larson et al.,\cite{Larson07} obtaining reasonable agreement between the main features in both the filled and empty levels. The effects of core-holes were also investigated. The Gd $M_{4,5}$-edges were also studied in both absorption and emission. This combination provided the first measurement of the 4$f$ filled-to-empty level splitting as about 12.5~eV, which, in conjunction with the XPS measurement of the occupied 4$f$ levels' location, places the empty 4$f$ levels at about 5~eV above the conduction band edge.	

\begin{figure}
\includegraphics[width=8cm]{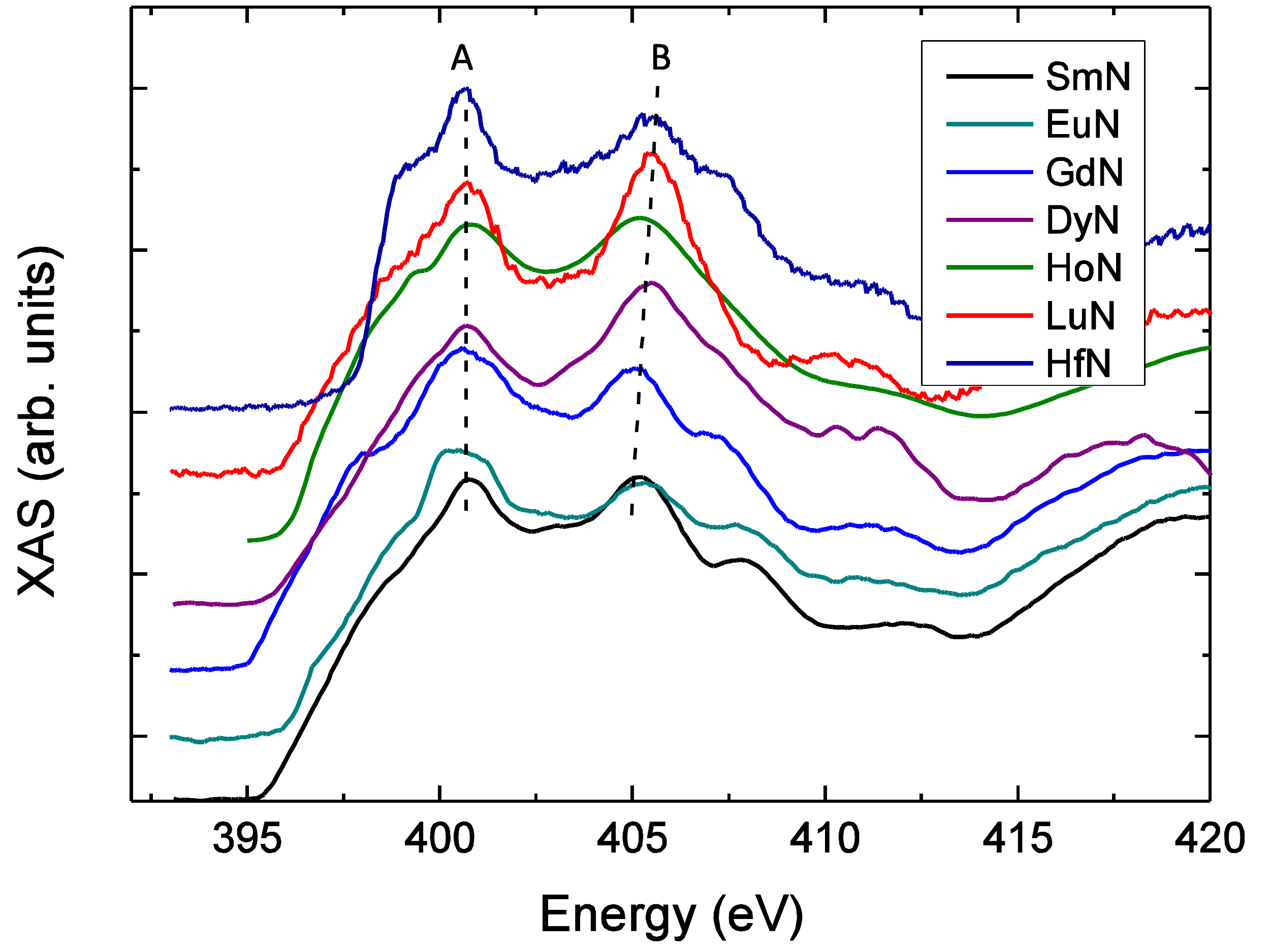}

\caption{Nitrogen $K$-edge X-ray absorption spectra from a series of REN films (from Refs.~\onlinecite{Preston07,preston:032101,richter11,brown12}) along with metallic HfN (Ref.~\onlinecite{farrell}). The peaks near 401~eV and 405~eV are associated with crystal field split RE $5d$ $t_{2g}$ and $e_g$ states, respectively. Spectra offset for clarity.
\label{REN_XAS}}
\end{figure}

More detailed XAS/XES spectra for SmN and DyN were reported by Preston et al.~\cite{Preston07}~and compared to LSDA+$U$ densities of states. The valence-to-conduction band gap was suggested by the results to lie close to 1.5 and 1~eV for DyN and SmN, respectively, though the estimate is problematic due to the uncertain core-hole influence on the absolute energy scale of the XAS spectra. A full set of XPS, XAS, and XES spectra from HoN have been recently reported to investigate $p-f$ hybridisation.\cite{brown12} Further XAS data have been reported for LuN\cite{preston:032101} and EuN.\cite{richter11} Figure~\ref{REN_XAS} provides a comparison between the N $K$-edge XAS data from the set of six RENs, along with metallic HfN.\cite{farrell} Band structure calculations and symmetry considerations show that the largest features, labelled A and B, correspond to crystal field split Gd-5$d$ $t_{2g}$ and  $e_g$ levels, respectively. This assignment is supported by the trend across the series, where the $e_g$ peak shifts upwards in energy as the lattice constant decreases reflecting the fact that these orbitals are oriented directly towards the nearest neighbour N ions.\cite{preston:032101}

\subsection{Phonon spectra}\label{Raman}
The phonon properties of the RENs have also received some attention from
theorists. In  the rocksalt structure, there is no first order allowed Raman spectrum because of
the presence of an inversion center. Nonetheless, Raman spectra were observed.\cite{degiorgi90,degiorgi93,Granville09} They were shown to
be disorder induced Raman spectra and dominated by the $LO$-phonons at the Brillouin zone $L$ points.
This vibrational mode corresponds to a breathing mode and is thus most strongly excited
with above band gap resonant Raman excitation. The role of the various phonons was
first elucidated in ScN, a transition metal cousin of the RENs by means of full phonon
density of states and phonon band structure calculations.\cite{Paudel09}
In the RENs, pseudopotential calculations were not found to be successful and hence a frozen
phonon method using the LSDA+$U$ band structure approach\cite{Granville09}
 for the total energy was used. Phonons
were calculated at $\Gamma$, $X$ and $L$ and the $LO$-phonons at $L$ were found to account
well for the observed Raman peaks. Their trends with lattice constant were explained and
agree with the observed behavior of lattice constant as a function of the RE atomic number.
Jha et al.~\cite{Jha95} studied the phonon dispersion and density of states in Yb pnictides,
including YbN using a semi-empirical three-body potential.

\section{Summary}\label{summary}

The RENs were first investigated 50 years ago, but with results that were recognised to be subjected to both $V_N$ and reaction with oxygen. However, recent advances in UHV-based thin-film growth and passivation have permitted substantially enhanced reliability of the data concerning their electronic and magnetic properties, with results that suggest devices for both spintronics and as conventional narrow-gap semiconductors. The experimental work is still very incomplete; only a few of the series have even reasonably well-established conducting states, semiconducting or metallic. However, even within those few there are some clear opportunities to form heterojunctions that present potentially exploitable properties; indeed there is already demonstrated a GdN-based spin filter.\cite{GdNspinfilter}
In concert with experimental investigation these compounds have been studied even more thoroughly within various theoretical treatments. In that context they are attractive as a test for treatments of strong correlations in the presence of a relatively simple NaCl crystal structure. Thus much of this review summarises exactly the various treatments used in electronic and magnetic structure calculations, outlining the relative strengths and weaknesses in both their fundamental underpinning and their agreement with experimental results. The points of contact between theory and experiment lie primarily in the measured Curie temperatures, the magnitude of optical band gaps and the delineation of higher-energy excitation with X-ray absorption and emission spectroscopies.

Recent thin film growth has been reported by physical vacuum deposition (CeN, SmN, EuN, GdN, DyN, HoN, ErN, LuN), sputtering (GdN) and CVD (GdN, DyN). In most cases the films have been subjected to magnetic measurements, and for a smaller selection simple transport studies have indicated heavily doped ferromagnetic semiconductor behaviour. The most fundamental magnetic parameters (Curie temperature, saturation moment) are in substantial agreement with the historical data, and are now supported by XMCD results for at least a small selection of the series. That agreement is in stark contrast to their electronic behaviour, with most so far investigated (SmN, GdN, DyN, ErN, and possibly EuN, HoN) showing clearly semiconductor-like temperature dependences of resistivity. Optical interband absorption and weak free-carrier absorption corroborates that conclusion in at least some of the REN series.

Epitaxial film growth has so far been demonstrated for only three of the series; GdN, EuN and SmN. Even with this very limited set the recent data show a striking contrast; GdN has a huge magnetic moment and a very weak coercive field whereas SmN is a near-zero moment ferromagnet with a coercive field some 2-3 orders of magnitude larger. Coupled with their similar semiconductor electronic structure the combination of the two suggests potential as a magnetoresistance memory element.

The RE elements, including its end members (La, Lu), comprise 15 species. One, Pm, has no stable isotopes, making it inappropriate for any condensed matter investigation. Nonetheless, there has been no more than a very dilute start at investigating the materials within the recent thin-film decade. Even for the most thoroughly studied, GdN, there remains a vexing theoretical-experimental discrepancy concerning its Curie temperature. There is a clear need for continued work to settle these and many other questions.

\begin{acknowledgments}
We acknowledge funding from the NZ FRST (Grant No. VICX0808), the Marsden Fund (Grant No. 08-VUW-030), and the MacDiarmid Institute for Advanced Materials and Nanotechnology, funded by the New Zealand Centres of Excellence Fund. Special thanks are due to A. Svane, W. Felsch and T. Kita for giving permission to use the original figures from their publications and the authors acknowledge insightful discussions with Eva Anton.

\end{acknowledgments}  

\bibliography{REreview}
\end{document}